\documentclass[useAMS,usenatbib, usegraphicx]{mn2e}

\usepackage{amsmath}
\usepackage{amssymb}
\usepackage{natbib}
\usepackage{times}
\usepackage{multirow}
\usepackage{psfrag}
\usepackage{color}

\def\beq{\begin{equation}}
\def\enq{\end{equation}}

\def\bar{\begin{eqnarray}}
\def\ear{\end{eqnarray}}

\title[Impact of uncertainties on the timing of pulsars in binary systems]
{Impact of the orbital  uncertainties on the timing \\ of pulsars in binary systems}

\author[G. A. Caliandro, D. F. Torres, \& N. Rea]{G. A. Caliandro$^{1}$\thanks{andrea.caliandro@ieec.uab.es}, 
D. F. Torres$^{1,2}$, \& N. Rea$^1$ \\
$^{1}$Institut de Ci\`encies de l'Espai (IEEC-CSIC)
Campus UAB, Fac. de Ci\`encies, Torre C5, parell, 2a planta
08193 Barcelona,  Spain\\
$^{2}$Instituci\'o Catalana de Recerca i Estudis Avan\c{c}ats (ICREA) Barcelona, Spain}
\begin{document}

\date{Draft version \today}


\maketitle

\label{firstpage}

\begin{abstract}

The detection of pulsations from an X-ray binary is an unambiguous signature of the presence of a neutron star in the system.
When the pulsations are missed in the radio band, their detection at other wavelengths, like X-ray or gamma-rays, requires orbital demodulation, since the length of the observations are often comparable to, or longer than the system orbital period. 
The detailed knowledge of the orbital parameters of binary systems plays a crucial role in the detection of the spin period of pulsars, since any uncertainty in their determination translates into a loss in the coherence of the signal during the demodulation process. 
In this paper, we present an analytical study aimed at unveiling how the uncertainties in the orbital parameters might impact on periodicity searches. We find a correlation between the power of the signal in the demodulated arrival time series and the uncertainty in each of the orbital parameters. This correlation is also a function of the pulsar frequency. We test our analytical results with numerical simulations, finding good agreement between them. Finally, we apply our study to the cases of LS 5039 and LS I +61 303 and consider the current level of uncertainties in the orbital parameters of these systems and their impact on a possible detection of a hosted pulsar.
We also discuss the possible appearance of a sideband ambiguity in real data. The latter can occur when, due to the use of uncertain orbital parameters, the power of a putative pulsar is distributed in frequencies lying nearby the pulsar period.  Even if the appearance of a sideband is already a signature of a pulsar component, it may introduce an ambiguity in the determination of its period. We present here a method to solve the sideband issue. 
\end{abstract}

\begin{keywords}
stars: neutron, pulsars, gamma-rays: observations
\end{keywords}

\section{Introduction}

A few High Mass X-ray Binaries (HMXBs) have been detected to emit  GeV and TeV photons. 
LS I +61 303 (see \citealt{LSIfermi}, \citealt{LSImagic}, \citealt{LSIveritas}, \citealt{LSIhess}), LS 5039 (see \citealt{LSfermi}, \citealt{LShess}), PSR B1259-63/LS 2883 (see \citealt{B1259fermi}, \citealt{B1259hess}), HESS J0632+057 (see \citealt{hess0632}, \citealt{J0632}), and the most recently discovered 1FGL J1018.6-5856 \citet{J1018.6} are all examples of these systems. These former systems are generally referred to as gamma-ray binaries. Other objects like Cyg X-1 and Cyg X-3 have been observed to flare in gamma-rays, but their emission is neither dominant nor persistent at these energies (see \citealt{CygX-1},  \citealt{CygX-3}).

These HMXB systems are composed by a massive OB or Be star and a compact object, the nature of which is in general unknown.
In the case of Cyg X-1 (and less securely of Cyg X-3) it is likely, however, that their compact objects are
black holes surrounded by an accretion disk filled by matter captured from the massive star.
On the other hand, among the gamma-ray emitting HMXBs, 
the compact object of PSR B1259-63/LS 2883 is  a radio pulsar with a spin period of 48 ms \cite{Johnson1999}. After \citet{Matre}, e.g., 
\citet{Dubus2006}, \citet{Sier1}, and
\citet{Zdziarski2010}, among others, proposed detailed theoretical models to explain the emission of these gamma-ray binaries as due to the interaction of the relativistic particle wind from the pulsar with the wind of the massive star, or via processes in the pulsar wind zone directly. Recently, following the detection of a very short, magnetar-like burst coming from the direction of LS I +61 303, \citet{TorresRea2012} 
developed on a model based on assuming the existence of a high magnetic field - long period pulsar in this system.
The dichotomy on the nature of gamma-ray binaries is a trending topic of high-energy astrophysics (see, e.g., \citealt{Mirabel2006}). 
Obviously, the detection of pulsations from them would lead to a clear and unequivocal solution.

Deep searches for pulsations in radio frequencies have been performed especially for LS I +61 303 and LS 5039 (e.g. \citealt{McSwain2011}),  without success. The lack of radio pulsation can be explained by the dense environment of the massive star. Indeed, for PSR B1259-63/LS 2883 (the largest system, with a period of $\sim 4$ years) the radio pulsations disappear at the periastron distance (\citealt{Johnson1999}, \citealt{Johnson2005}). Since the orbits of LS I +61 303 and LS 5039 are much smaller than that of PSR B1259-63, it is natural to expect the radio emission of their compact object being always affected by free-free absorption and dispersion.
Searches for pulsations were performed also with X-ray data for LS I+61 303, and LS 5039 and the deepest upper limits on the pulse fraction were put in the works by \citet{Rea2010}, \citet{Rea2011}, and \citet{Rea2011b}. Note again that for the only firmly established TeV binary containing a pulsar, PSR B1259-63, X-ray pulsations were not detected either \citep{Cher09}, pointing to an X-ray emission being dominated by wind-wind or intra-wind shocks. 

The \textit{Fermi} satellite, and its main instrument on board, the Large Area Telescope, is continuously surveying the sky since its launch in June 2008 \citep{Atwood2009}. This experiment offers a good opportunity to perform searches for gamma-ray pulsations from binaries. However, given the dim character of gamma-ray fluxes, and the paucity of counts, large integration times are needed for pulsation searches. The event arrival times need orbital demodulation. An uncertain knowledge of the orbital parameters, in this gamma-ray case, or at any other frequencies or systems where demodulation is needed, would lead to lose the coherence of the pulsed signal. The most accurate measurements of the orbital parameters for LS 5039 and LS I+61 303 are derived fitting the Doppler shift of optical spectral lines emitted by the massive star (see 
\citealt{Casares2005a}, \citealt{Casares2005b}, \citealt{Grundstrom2007}, \citealt{Aragona2009}). This technique led so far to uncertainties of the order of 1\% to 10\%.
Inverting the problem, one may ask how well should a certain orbital parameter be known in order to secure that a pulsed signal is not lost through the demodulation process.

The purpose of this work is to analytically study how much the uncertainties on the orbital parameters affect
the power spectrum of a putative pulsed signal from an X-ray binary. 

The paper is organized as follows. Section 2 presents our approach to the problem, using perturbation theory. Section 3 introduces the perturbation functions, and a new set of variables that simplify the treatment of the problem. In Section 4 and 5, we compute the probability density function of the perturbed emission times, and for each orbital parameter, the impact of their uncertainty in the power spectrum. Section 6 provides constraints on the uncertainty over each of the orbital parameters such that the loss in the power of the pulsar signal is smaller than a given value. Finally, the discussion applies our results to LS 5039 and LS I 61 303, together with providing numerical simulations that validate our results and conclusions. Numerical validation of our conclusions is given via simulations and subsequent timing analysis of pulsed signals from different binaries, where the knowledge of the orbital parameters is blurred ad-hoc.

\section{The problem of pulse extraction}

\subsection{Numerical problems for a blind search approach}

We start showing how a blind search approach for pulsars in binary systems is generally doomed.

For a blind search of isolated pulsars, a FFT should be performed with a number of frequency bins ($N_F$)
equal to $T_{\rm obs} F_{\rm MAX}$, where $T_{\rm obs}$ is the viewing period, and $F_{\rm MAX}$ the highest frequency searched.
Many trials are needed
to correct for the first derivative of the frequency, $F1$.
The step size in $F1$ should not be larger than
$1/T_{\rm obs}^2$ in order to keep the
signal power within a single frequency bin.
The number of $F1$ trials, $N_{F1}$ (that is proportional to $T_{\rm obs}^2$) would be enormous for 
viewing periods lasting few years.
Then, even for isolated pulsars, performing a blind search is
computational demanding. To face this issue different techniques were proposed, the most successful so far being the 
time-differencing of \cite{Atwood2006}, and the method of \cite{Pletsh}.

In the case of a pulsar in a binary system, many trials would be needed also for
each of the orbital parameters.
We could arbitrary choose to cover the uncertainty ranges of the orbital parameters (we consider just 5 parameters)
with an equal number of trials $N_p$ each.
The total number of trials would be $N_T=N_p^5 \times N_{F1}$.

To have an idea on the numbers involved, we can consider a blind search similar to that in \cite{SazPark}, but
applied to binary systems. In that case, pulsations are searched in one year of \textit{Fermi}-LAT data using the time-differencing technique 
with time windows of 6 days, up to 64 Hz frequency. The number of frequency bins in the FFT were $N_F = 2^{26}$, and the trials in the frequency derivative were $N_{F1} = 2000$.
In the case of binary systems, an hypothesis of just 10 trials for each orbital parameter leads to a total number of trials equal to $N_T = 2 \times 10^8$. 

And yet,  is $N_p=10$ a guarantee that the pulsar detection does not get lost, regardless of its frequency? How can we understand whether the uncertainty range is oversampled or not by the choice of the number of trials? Is it possible to optimize the trials in order to minimize the required CPU time to run the analysis? 
It is not possible to answer all these questions without studying how the parameter uncertainties affect the results of the periodicity search.

\subsection{The analytical approach}

\begin{figure}
\center
\includegraphics[width=0.45\textwidth]{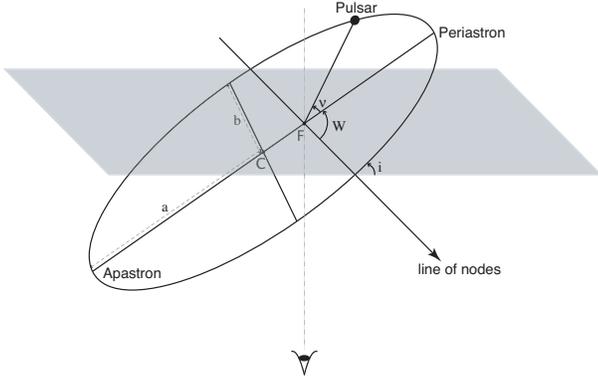}
\caption{Schematic view of an orbit and its parameters. The gray plane is the reference plane perpendicular to the observer line of sight, and $i$ is the inclination of the plane of the orbit respect to it. {\sf a} and {\sf b} are the semi-major and the semi-minor axis, respectively. {\sf F} is the focus of the orbit that hosts the baricenter of the binary system, while {\sf C} is its center. The distance {\sf CF} is equal to the semi-major axis times the eccentricity (${\sf a} \cdot e$). $W$ is the longitude of the periastron, and $\nu$ is the true anomaly.
\label{Orbit}}
\end{figure}

The photons emitted by a pulsar in a binary system experience several delays travelling towards the observer. These delays are: the dispersive delay due to the Interstellar Medium ($\Delta_{IS}$), the propagation and relativistic delays within the Solar System ($\Delta_{\odot}$), and the corresponding delays accounting for the geometry in the binary system itself ($\Delta_{B}$).
The so-called timing formula correlates the photon emission time in the pulsar reference frame ($t_e^{psr}$) with the photon 
arrival time to the observer ($t_a^{obs}$) as
\begin{equation}
t_e^{psr} = t_a^{obs}-\Delta_{\odot}-\Delta_{IS}-\Delta_{B}.
\label{2.1}
\end{equation}
The delay introduced by the motion of the pulsar around its companion star ($\Delta_{B}$) is mainly due to the R\"{o}mer delay ($\Delta_R$),
\bar
&& \Delta_{B}  \simeq  \Delta_R = \nonumber \\ &&
 A\left[ {\rm sin}W \, ({\rm cos}E-e) + \sqrt{1-e^2} \cdot {\rm cos}W \, {\rm sin}E \right], 
\label{2.2}
\ear
where $A$ is the projection of the semi-major axis on the plane perpendicular to the observer's line-of-sight, $W$ is the longitude of periastron, $e$ is the eccentricity, and $E$ is the eccentric anomaly (see for example \citealt{BT76}, or \citealt{Camenzind}).
Post-Newtonian effects will be neglected in this work.
A schematic view of an orbit and its parameters is shown in Figure \ref{Orbit}, while Table \ref{VarTable} lists the main variables.

Assuming that a pulsar is part of a binary system, the detection of periodic signals from it can be achieved by calculating the power spectrum of the emission time series in Eq.~(1). In practice, however,  one is faced with the problem that the uncertainties in the estimation of the delays lead to a wrong calculation of the emission times ($t_{ew}$), affecting the results of the analysis. 
In this work, we focus on studying the impact of the uncertainties in the orbital parameters on the ability for detecting pulsations. 
For this reason,
it will be assumed that the delays due to the interstellar medium ($\Delta_{IS}$), and to the propagation in the solar system ($\Delta_{\odot}$) are exactly known (or rather, that their uncertainty is much smaller in comparison to $\Delta_R$). The R\"{o}mer delay in the binary system (Eq. 2), instead, is considered 
wrongly estimated ($\Delta_{Rw}$).

Comparing the timing formula (Eq. 1) in the cases of a correct and a wrong calculation of the R\"{o}mer delay ($\Delta_R$ and $\Delta_{Rw}$ respectively), the relation between the correct ($t_{e}$) and wrong ($t_{ew}$) emission times turns out to be
\begin{equation}
 t_{ew} = t_e - \delta \Delta_R,
\label{2.3}
\end{equation}
where $\delta \Delta_R =  \Delta_{Rw} - \Delta_{R}$.
%
A mathematical interpretation of Eq.~(\ref{2.3}) is simply that  the times $t_{ew}$ are 
the correct emission times perturbed by $\delta \Delta_R$. 
At first order approximation, the impact of the errors in the orbital parameters on the R\"{o}mer delay calculation, can then be taken into account by writing the perturbation factor ($\delta \Delta_R$) as
\begin{equation}
\delta \Delta_R = \sum_p \frac{\partial \Delta_R}{\partial p} dp , 
\label{2.5}
\end{equation}
where the sum is over the orbital parameters $p = \{A,W,e,P_{\rm orb}, T_0\}$, and $T_0$ is the epoch of the periastron.
Eq.~(\ref{2.5})  allows us to factorize the problem, and to evaluate how the timing analysis is affected by each orbital parameter separately from the others. With this aim we introduce the perturbation functions, $f_p$, relative to each of the orbital parameter, $p$, as
\begin{equation}
f_p = \frac{\partial \Delta_R}{\partial p} dp .
\label{2.6}
\end{equation}

\begin{table*}
 \caption{Meaning of the variables used in the paper}
 \label{VarTable}
 \begin{tabular}{@{}ll}
  \hline
  Variable & Meaning \\
  \hline
    $ t_e^{psr}$ or $ t_e $ & Photon emission time in the pulsar reference frame    \\
    $ t_a^{obs}$ &    Photon arrival time to the observer\\
    $ t_{ew} $ & Perturbed emission time: photon emission time in the pulsar reference frame affected by the errors on the orbital parameters \\
    $ \Delta_R$ & R\"{o}mer delay    \\
    $\Delta_{Rw}$  & Evaluation of the R\"{o}mer delay affected by the errors on the orbital parameters \\
    $A$ & Projection of the semi-major axis on the plane perpendicular to the observer line of sight \\
    $W$ & Longitude of periastron \\
    $e$ & Eccentricity \\
    $P_{\rm orb} $ & Orbital period    \\
    $ \Omega_{\rm orb} = 2\pi/P_{\rm orb} $ & Angular frequency of the binary system    \\
    $E$ & Eccentric anomaly \\
    $ T_0 $ & Epoch of the periastron    \\
    $ f_p$ or $ f $ & Perturbation function, defined in Eq.~(\ref{2.6}), and described in Section 3    \\
    $ \{\delta k, \phi, \psi, \delta c\}$ &    Set of transformed orbital parameters. $\phi$ is defined in Eq.~(\ref{3.3}), the other three in Eq.~(\ref{3.6})\\
    $ T_{\rm obs} $ & Duration of the observation    \\
    $ \omega_0 = 2\pi/P_{psr} $ & Angular pulsar frequency    \\
  \hline
 \end{tabular}
\end{table*}

\section{The perturbation functions}

In Eq.~(\ref{2.6}) the perturbation functions are defined as partial derivatives of the R\"{o}mer delay ($\Delta_R$), which  is 
in turn expressed in Eq.~(\ref{2.2}). 
It is possible to rewrite this latter formula in a more compact form, using the following trigonometric identity:
\begin{align}
a \cdot {\rm cos}\,x + b \cdot {\rm sin}\,x = c \cdot {\rm sin}(x+\phi), \nonumber \\
\phi = {\rm atan}(a/b), \label{3.1} \\
c = \sqrt{a^2 + b^2} . \nonumber
\end{align}
Applying it to cos$\,E$ and sin$\,E$ in Eq.~(\ref{2.2}), the R\"{o}mer delay is then:
\begin{equation}
\Delta_R = M(A,e,W)\cdot{\rm sin}(E+\phi(e,W)) - Q(A,e,W),
\label{3.2}
\end{equation}
where
\begin{align}
M = A\sqrt{1-e^2{\rm cos}^2W} , \nonumber \\
\phi = {\rm atan}({\rm tan}(W)/\sqrt{1-e^2})) , \label{3.3}\\
Q = e\,A\,{\rm sin}W .\nonumber
\end{align}
To evaluate the perturbation functions, we calculate the partial derivatives of Eq.~(\ref{3.2}) for a generic orbital parameter $p$,
\bar
\frac{\partial \Delta_R}{\partial p} = \frac{\partial M}{\partial p} \cdot{\rm sin}(E+\phi) \hspace{3.5cm}  \nonumber \\
+ M\cdot{\rm cos}(E+\phi) \cdot \left( \frac{\partial E}{\partial p}  + \frac{\partial\phi}{\partial p} \right) - \frac{\partial Q}{\partial p}.
\label{3.4}
\ear
Here again, we can apply the trigonometric identity in Eq.~(\ref{3.1}) to sin$(E+\phi)$ and cos$(E+\phi)$, obtaining:
%
%
%
\begin{equation}
\frac{\partial \Delta_R}{\partial p} = K\cdot{\rm sin}(E+\phi+\psi) - C,
\label{3.5}
\end{equation}
from which we can define
\begin{align}
\delta k = Kdp = \sqrt{\left[ M\left( \frac{\partial E}{\partial p}  + \frac{\partial\phi}{\partial p}\right) \right]^2 
+ \left[ \frac{\partial M}{\partial p} \right]^2 } \hspace{0.1cm} dp , \nonumber \\
\psi = {\rm atan}\left( \tfrac{M\left(  \frac{\partial E}{\partial p}  + \frac{\partial\phi}{\partial p}\right)} {\frac{\partial M}{\partial p}} \right) , \label{3.6}\\ 
\delta c = Cdp = \frac{\partial Q}{\partial p} dp . \nonumber
\end{align}
Finally, a compact formula for the perturbation function is
\begin{equation}
f = \frac{\partial \Delta_R}{\partial p}dp = \delta k\cdot{\rm sin}(E+\phi+\psi) - \delta c.
\label{fE}
\end{equation}
Since this formula is valid for all the orbital parameters, we omitted the sub-index $p$ in Eq.~(\ref{fE}). 
From a mathematical point of view, we have done a transformation from the canonical orbital parameters $\{A,W,e,P_{\rm orb},T_0\}$ and their errors $\{dA,dW,de,dP_{\rm orb}, T_0\}$, to the parameters $\{\delta k, \phi, \psi, \delta c\}$. 

The behavior of the perturbation function, at each instant, is given by its dependence on the eccentric anomaly $E$,
and it is connected to the emission times $t_e$ by the relation
\begin{equation}
E - e \cdot {\rm sin}E = \Omega_{\rm orb} (t_e - T_0),
\label{Ete}
\end{equation}
where $\Omega_{\rm orb} = 2\pi / P_{\rm orb}$ is the system frequency. This equation provides the emission times as a function of $E$. 
The inverse function ($E=E(t_e)$) can not be expressed analytically using Eq.~(\ref{Ete}). However, it can be approximated, and in order to express the perturbation function dependence with $t_e$ and do further analytical steps, 
we will simply assume that $E \sim \Omega_{\rm orb} t_e$. This implies, 
\begin{equation}
f(t_e) \simeq \delta k\cdot{\rm sin}(\Omega_{\rm orb} t_e+\phi+\psi) - \delta c.
\label{fte}
\end{equation}
Appendix \ref{App1} gives an assessment of the approximation made to reach the latter formula.
Note that  $\delta k$ and $\psi$ can also be functions of $t_e$, because their defining formulae (see Eq. \ref{3.6}) contain the partial derivatives ${\partial E}/{\partial p}$. By using Eq.~(\ref{Ete}), we can see that the partial derivative is null when the orbital parameter $p$ incarnates into $A$, or $W$. In these two cases, $\delta k$ and $\psi$ are constants. In contrast, the partial derivative ${\partial E}/{\partial p}$ is a function of $E$ when $p$  is either $e$ or $P_{\rm orb}$, since $\delta k$ and $\psi$ are function of $E$, and consequently, also of $t_e$. This difference in the dependence of $\delta k$ and $\psi$ will lead to a different treatment of the problem, as described below in Section 6.

\section{The probability density function of the perturbed emission times}

The expectation value of the power spectrum can be evaluated using the probability density function (pdf) 
of the phases assigned to each photon (see appendix \ref{App0}).
In order to evaluate the power spectrum of the perturbed emission times $t_{ew}$, we shall calculate their pdf ($P_{t_{ew}}$), as well as that of the phases assigned to them ($P_{\theta}$).

Since the times $t_{ew}$ are correlated with the correct emission times ($t_e$) by Eq.~(\ref{2.3}), which we rewrite here as
\begin{equation}
 t_{ew} = t_e - f(t_e),
\label{4.1}
\end{equation}
$P_{t_{ew}}$ can be calculated if $P_{t_e}$ is known.
Appendix \ref{App2} shows that for all realistically observable binary systems, $t_{ew}$ is a monotonic increasing function of $t_e$. This makes the calculation of its pdf easier. Indeed,   its pdf is:
\begin{equation}
P_{t_{ew}}(t_{e}) = U\frac{P_{t_e}(t_e)}{1-f^{\prime}(t_e)},
\label{4.4}
\end{equation}
where here (and hereafter)  $U$ indicates a normalization factor, and a ${\prime}$ represents the first derivative respect to the emission time $t_e$. 
Finally, to compute the power spectrum of the perturbed time series $t_{ew}$ at a frequency $\omega$ one has to calculate the phases, defined as
\begin{equation}
\theta = \omega t_{ew}.
\label{4.5}
\end{equation}
Since $\omega$ acts like a constant in Eq.~(\ref{4.5}) , the pdf of the phases has the same form of Eq.~(\ref{4.4}), i.e., 
\begin{equation}
P_{\theta}(t_{e}) = U\frac{P_{t_e}(t_e)}{1-f^{\prime}(t_e)}
\label{4.6}
\end{equation}

The fundamental features of the power spectrum of the perturbed time series $t_{ew}$ can be derived assuming that the signal emitted by the pulsar in its reference frame is sinusoidal with frequency $\omega_0$, 
\begin{equation}
 P_{t_e}(t_e) = 1+{\rm sin}(\omega_0 t_e).
\label{4.7}
\end{equation}
Substituting the latter in Eq.~(\ref{4.6}) we get
\begin{equation}
P_{\theta}(t_{e}) = U\frac{1+{\rm sin}(\omega_0 t_e)}{1-f^{\prime}(t_e)}.
\label{4.8}
\end{equation}
In order for Eq.~(\ref{4.8}) to be useful for our purposes, we shall apply some approximations,  expressing it as a function of the phases $\theta$.
Appendix \ref{App2} also shows that for not unreasonably large values of the uncertainties, 
the first derivative of the perturbation function is $f^{\prime}(t_e) \ll 1$. Thus, it can be ignored in the denominator of Eq.~(\ref{4.8}).
The argument of the sine in Eq.~(\ref{4.8}) has also to be expressed in terms of $\theta = \omega t_{ew}$. 
Substituting Eq.~(\ref{4.1}) we obtain:
\begin{equation}
\theta = \omega t_{ew} = \omega t_e - \omega f(t_e) \Longrightarrow t_e = \theta/\omega + f(t_e),
\label{4.9}
\end{equation}
and at a first order approximation we can set
\begin{equation}
t_e = \theta/\omega + f(\theta/\omega).
\label{4.10}
\end{equation}
Even if we do not give now a direct estimation of this approximation, we shall realize that it is effectively safe when in Section 6 our analytical results will be compared with simulations.
Putting it all together, the approximated pdf of the phases ($\theta$) is
\begin{equation}
P_{\theta}(\theta) \simeq U \left[ 1+{\rm sin}\left( \frac{\omega_0}{\omega} \theta + \omega_0 f(\theta/\omega)\right)  \right]  ,
\label{4.11}
\end{equation}
and substituting in Eq.~(\ref{fte}) $t_e$ by $\theta/\omega$, the perturbation function expressed in terms of $\theta/\omega$ is equal to
\begin{equation}
f(\theta/\omega) \simeq \delta k\cdot{\rm sin}\left( \frac{\Omega_{\rm orb}}{\omega} \theta+\phi+\psi \right)  - \delta c .
\label{4.12}
\end{equation}

\section{How the perturbation function affects the power spectrum}

Once the pdf $P_{\theta}$ of the phases perturbed by the errors on the orbital parameters has been evaluated, 
we can study how the power spectrum is affected.
In the Appendix \ref{App0} we briefly introduce a method by which the power spectrum is directly expressed as a function of the pdf $P_{\theta}$.
There, we find that a key role is played by the terms
\begin{equation}
\sum_{i=0}^{N-1} P_{\theta}(2\pi i + \theta),
\label{5.1}
\end{equation}
which correspond to the pdf $P_{\theta}$ folded in $2\pi$. Indeed, in Eq.~(\ref{5.1}),  
$0 \leq \theta<2\pi$, the folding is due to the sum over the term $2\pi i$, and $N$ is the number of rotations made by the plausible neutron star during the whole observation $T_{\rm obs}$ 
\begin{equation}
N = \frac{\omega_0 T_{\rm obs}}{2\pi}.
\label{5.2a}
\end{equation}

\begin{figure*}
\center
\includegraphics[width=0.45\textwidth]{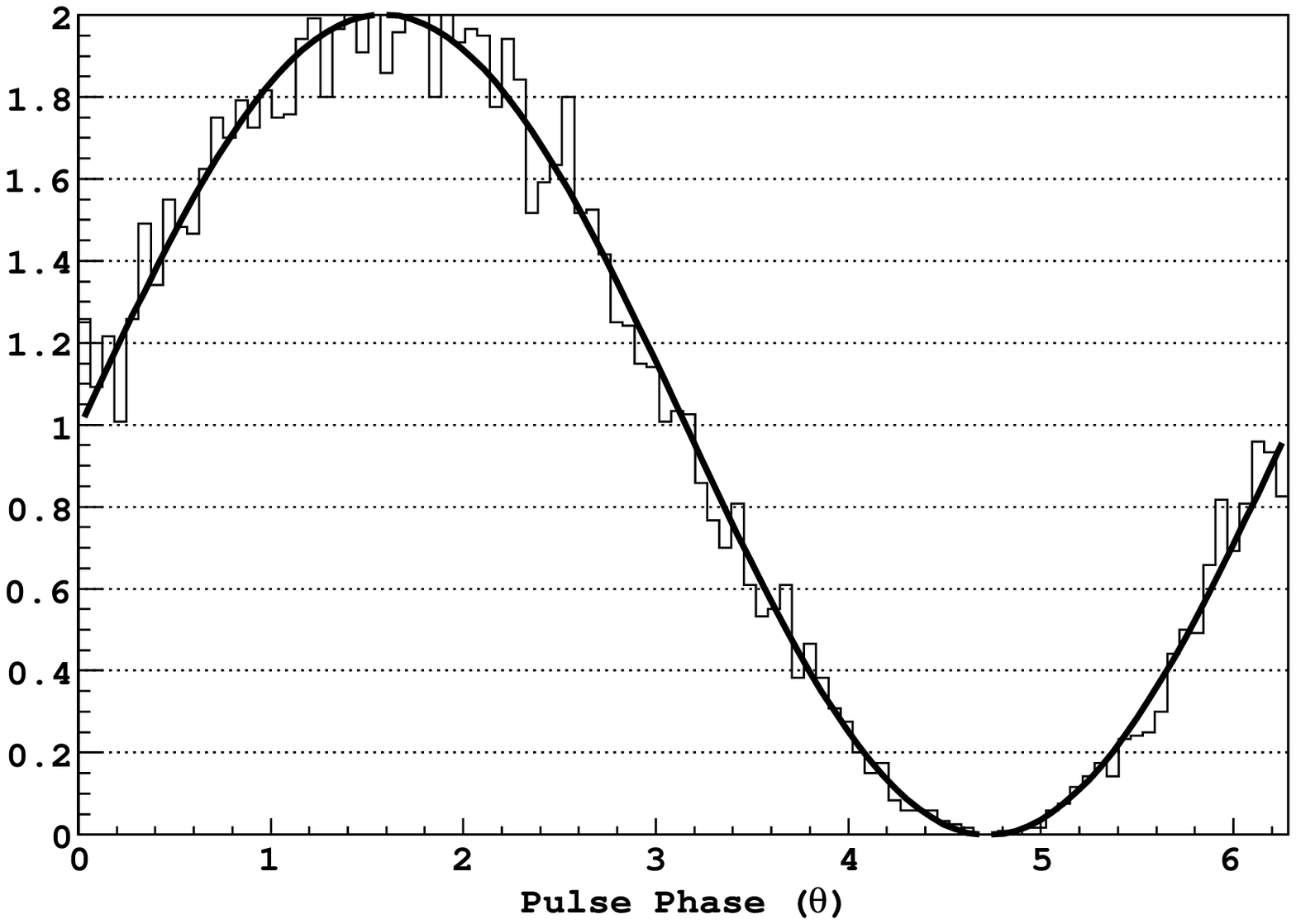}
\hspace{0.5 cm}%
\includegraphics[width=0.45\textwidth]{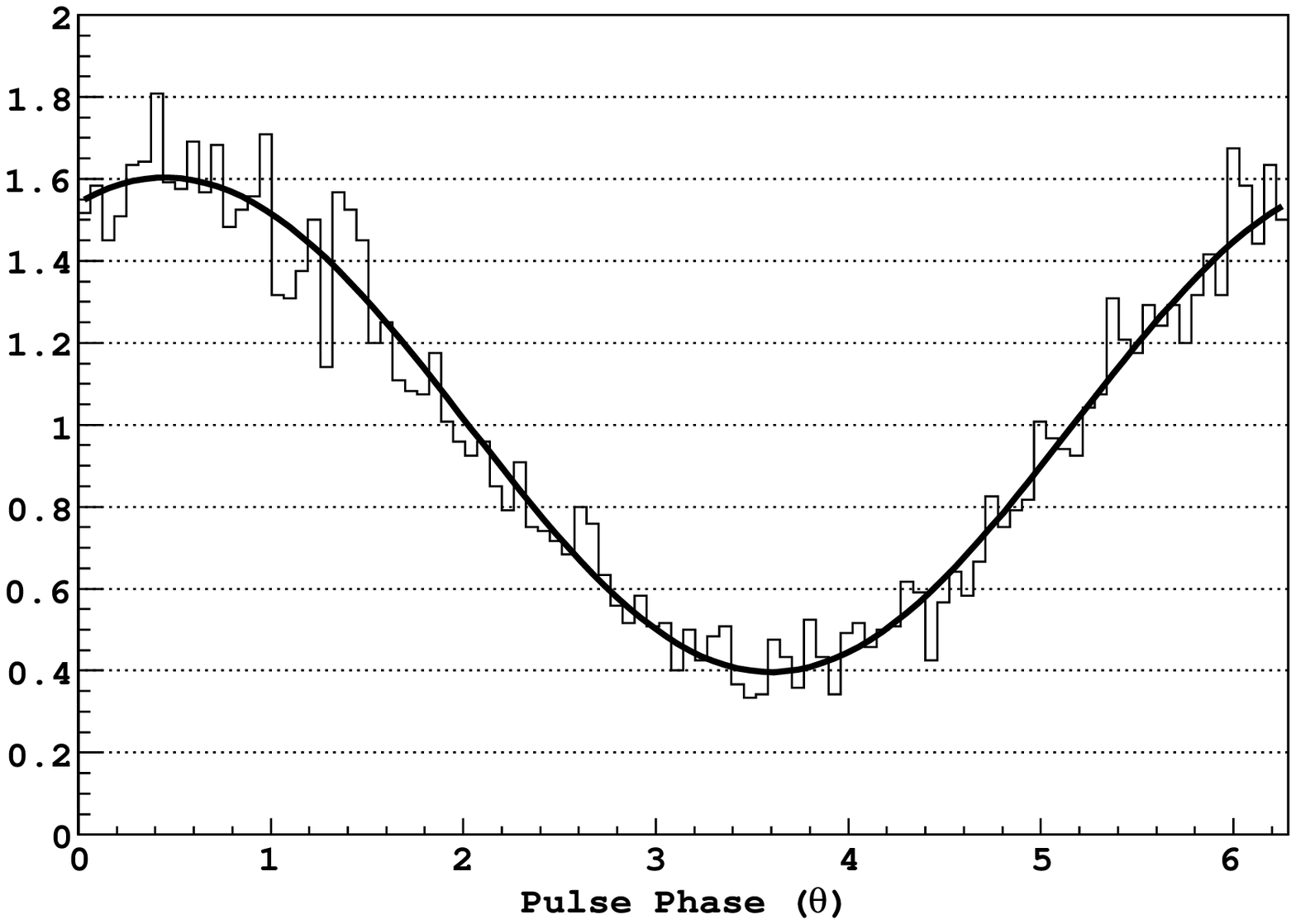}
\caption{The histograms in both panels are the normalized folded phase distribution of a periodic signal emitted by a pulsar in a binary system with the parameters listed in Table \ref{Tab1}. 
The emission is simulated as described in section 6.1.1, assuming a pure sinusoidal signal with a rate of 300 counts/day. The histograms are fitted by a sine function (black lines).
\textit{Left}: the true values of the orbital parameters are used to calculate the R\"{o}mer delay. The amplitude of the sine function is equal to 1.0.
\textit{Right}: a value of the projection of the semi-major axis $A$ different from the true one by 0.075 lt-s has been used to calculate the R\"{o}mer delay. This causes the reduction of the amplitude of the sine function, that in this case is equal to 0.6.
\label{Pthetas}}
\end{figure*}

Figure \ref{Pthetas} shows the folded $P_{\theta}$ due to a 100\% sinusoidal signal emitted by the pulsar, assuming two different sets of orbital parameters. In the left panel, the {\it correct} set is used to calculate the R\"{o}mer delay, so the perturbed function is null and 
\begin{equation}
\frac{1}{N}\sum P_{\theta}(2\pi i + \theta) = 1+{\rm sin}\theta,
\label{5.2}
\end{equation}
where the factor $1/N$ cancels out setting $U=1$ in Eq.~(\ref{4.11}).
In the right panel, a value of the projected semi-major axis $A$ slightly different from the true one is assumed. We can note that the folded $P_{\theta}$ is still sinusoidal, but the effect of the wrong value of $A$ is to reduce its amplitude so that
\begin{equation}
\frac{1}{N}\sum_{i=0}^{N-1} P_{\theta}(2\pi i + \theta) = 1 + \varepsilon \cdot {\rm sin}(\theta + \alpha).
\label{Pfold}
\end{equation}
Similarly, the power spectrum calculated at the signal frequency $\omega_0$ is maximum when the right value of $A$ is used (so when $\varepsilon = 1$ in Eq.~(\ref{Pfold})). Using a different value of $A$, the power is reduced by a factor equal to $\varepsilon^2$ 
(the square happens because of Eq. (\ref{pow2}) in Appendix \ref{App0}). 
This effect is not  only related to the semi-major axis, but it is valid for all the orbital parameters, as we  demonstrate below.

At the signal frequency $\omega_0$, the term in the sum of Eq.~(\ref{Pfold}) is explicitly equal to
\begin{equation}
P_{\theta}(2\pi i + \theta) = 1 + {\rm sin}\left( \theta + \omega_0 \cdot f\left( \frac{2\pi i + \theta}{\omega_0}\right) \right) .
\label{5.3}
\end{equation}
This is obtained substituting the variable $\theta$ with $2\pi i + \theta$, and setting $\omega = \omega_0$ in Eq.~(\ref{4.11}).
The dependence of the perturbation function from $\theta$ is negligible in Eq.~(\ref{5.3}). Indeed, making the same substitutions in Eq.~(\ref{4.12}) we have
\begin{equation}
f\left( \frac{2\pi i + \theta}{\omega_0}\right) = \delta k\cdot{\rm sin}\left(  \frac{\Omega_{\rm orb}}{\omega_0}2\pi i + \frac{\Omega_{\rm orb}}{\omega_0}\theta +\phi+\psi \right)  - \delta c.
\label{5.4}
\end{equation}
Since here $0 \leq \theta<2\pi$, $\Omega_{\rm orb}/\omega_0 \ll 1$, and $({\Omega_{\rm orb}}/{\omega_0}) \theta$ is always very small, while the term $({\Omega_{\rm orb}}/{\omega_0})2\pi i$ can be as large as $\Omega_{\rm orb} T_{\rm obs} = 2\pi T_{\rm obs}/P_{\rm orb}$ for $i=N$. Similarly, the dependence from $\theta$ is negligible in  $\delta k$ and $\psi$. In what follows, the perturbation function will be labelled as $f_i$, indicating that it depends only on the index $i$.

With all this, the folded pdf becomes
\begin{equation}
\frac{1}{N}\sum_{i=0}^{N-1} P_{\theta}(2\pi i + \theta) = 1 + \frac{1}{N}\sum_{i=0}^{N-1}{\rm sin}(\theta + \omega_0  f_i),
\label{5.5}
\end{equation}
where the sum in the right hand is of a set of several sines with the same periodicity. 
The trigonometric identity of Eq.~(\ref{3.1}) is a particular case of a more general theorem stating that a sum of sines and cosines with equal periodicity, but different amplitudes and offset phases is equal to a single sine with same periodicity, and with
amplitude and offset phase depending on those in the sum. This implies that the sum in Eq.~(\ref{5.5}) is equal to $\varepsilon \cdot {\rm sin}(\theta+\alpha)$, proving the equivalence with Eq.~(\ref{Pfold}).

Summarizing: the effects of the errors of the orbital parameters ($dp$) on the power spectrum calculated at the signal frequency $\omega_0$ are described by the single factor $\varepsilon$, so that $P(\omega_0, dp) = \varepsilon^2 P(\omega_0, 0)$.


With a few steps of extra algebra  we shall 
find a useful formula to evaluate the factor $\varepsilon^2$. The right hand side of Eq.~(\ref{Pfold}) can be written as
\begin{equation}
1 + \varepsilon \cdot {\rm sin}(\theta + \alpha) = 1 + \varepsilon \left[ {\rm sin}\theta\,{\rm cos}\alpha + {\rm cos}\theta\,{\rm sin}\alpha\right] .
\label{5.6}
\end{equation}
Similarly, the right hand side of Eq.~(\ref{5.5}) is equal to
\begin{eqnarray}
1 + \frac{1}{N}\sum_{i=0}^{N-1}{\rm sin}(\theta + \omega_0  f_i) = \nonumber \\
 1 + \frac{1}{N}\sum_{i=0}^{N-1}\left[ {\rm sin}\theta\,{\rm cos}(\omega_0f_i) + {\rm cos}\theta\,{\rm sin}(\omega_0f_i)\right] =
\label{5.7} \\ 
1 + {\rm sin}\theta\, \frac{1}{N}\sum_{i=0}^{N-1}{\rm cos}(\omega_0f_i) + {\rm cos}\theta\, \frac{1}{N}\sum_{i=0}^{N-1}{\rm sin}(\omega_0f_i). \nonumber
\end{eqnarray}
Comparing Eqs.~(\ref{5.6}) and (\ref{5.7}), we get
\begin{eqnarray}
\sum_{i=0}^{N-1}{\rm cos}(\omega_0f_i) = N \varepsilon {\rm cos}\alpha,
\label{5.8} \\
\sum_{i=0}^{N-1}{\rm sin}(\omega_0f_i)= N \varepsilon {\rm sin}\alpha.
\label{5.9}
\end{eqnarray}
And squaring and adding Eqs.~(\ref{5.8}) and (\ref{5.9}) we obtain
\begin{equation}
\left[ \sum_{i=0}^{N-1}{\rm sin}(\omega_0f_i) \right]^2 + \left[ \sum_{i=0}^{N-1}{\rm cos}(\omega_0f_i) \right]^2 = N^2 \varepsilon^2 .
\label{5.10}
\end{equation}
%

\section{Constraints on the parameters}

So far we have commented on the way in which 
the perturbation function affects the power spectrum. In this Section, we aim 
to constrain the uncertainties ($dp$) in the parameters ($p$) in order to maintain the ability to detect pulsations.
In practice, this reduces in searching for a formula that allows to state that if the uncertainty is smaller than a given value, $dp <  x$, 
then $\varepsilon^2 > y$. The larger is $\varepsilon$ the better, until for $\varepsilon=1$ there is no loss introduced by imprecise knowledge of the orbit. 
If one aims to search for pulsations from a compact object in a binary system, for which orbital parameters are known just to an indicative level, this study will provide
the maximum steps in the sampling so that the signal detection is secure at a certain level.

Developing the squares of the two sums in Eq.~(\ref{5.10}), it becomes
\begin{eqnarray}
\left[ \sum {\rm sin}(\omega_0f_i) \right]^2 + \left[ \sum {\rm cos}(\omega_0f_i) \right]^2  = \nonumber \\
 N + 2{\rm cos}(\omega_0f_1){\rm cos}(\omega_0f_2) + 2{\rm sin}(\omega_0f_1){\rm sin}(\omega_0f_2) + ...  = \nonumber \\
N + 2\left[ {\rm cos}(\omega_0 (f_1 - f_2) ) + ... \right] = \label{6.1} \\
 N + 2\left[ \sum_{i=1}^{N} \sum_{j=i+1}^{N} {\rm cos}\left( \omega_0 (f_i - f_j) \right)  \right] . \nonumber 
\end{eqnarray}
To continue further, we need now to discern the cases for which the partial derivative of the eccentric anomaly ($\partial E / \partial p$) is null, from those for which it is still a function of $E$. Indeed, in the former case (for the orbit parameters $A$, and $W$), $\delta k$ and $\psi$ in Eq.~(\ref{5.4}) are constants that do not depend on the sub-index $i$ of the perturbation function.

\subsection{Cases for which $\delta k$, and $\psi$ are constants}

We consider the double sum in the square brackets of the last equation.
The difference of the perturbation functions $(f_i - f_j)$ is equal to
\begin{eqnarray}
f_i - f_j =  \nonumber \\
=\delta k \left[ {\rm sin}\left(  \frac{\Omega_{\rm orb}}{\omega_0}2\pi i +\phi+\psi \right) - {\rm sin}\left(  \frac{\Omega_{\rm orb}}{\omega_0}2\pi j +\phi+\psi \right) \right]  \nonumber \\
= \delta k \left[ K_i - {\rm sin}\left(  \frac{\Omega_{\rm orb}}{\omega_0}2\pi (i+\Delta n) +\phi+\psi \right) \right] \label{6.1.1} \\
= \delta k \left[ K_i - {\rm sin}\left(  \frac{\Omega_{\rm orb}}{\omega_0}2\pi \Delta n + \Lambda_i \right) \right],   \nonumber
\end{eqnarray}
where $K_i$ has values between $-1$ and 1, and we set $j=i+\Delta n$. In this way, for each fixed value of the index $i$, the difference $(f_i - f_j)$ is function of $\Delta n$, and the sum over $j$ in Eq.~(\ref{6.1}) becomes
\begin{equation}
\sum_{\Delta n = 1}^{N-i}  {\rm cos}\left( \omega_0 (f_i - f_{i+\Delta n}) \right).
\label{6.1.2}
\end{equation}
Furthermore, $(f_i - f_{i+\Delta n})$ is periodic, with period $\Delta n = \omega_0/\Omega_{\rm orb}$. This feature allows us to solve, with a good approximation, Eq.~(\ref{5.10}), and evaluate $\varepsilon^2$. 

The argument of the sum in Eq.~(\ref{6.1.2}) has the same periodicity of $(f_i - f_{i+\Delta n})$. Then, we can approximate it as
\bar
\sum_{\Delta n = 1}^{N-i}  {\rm cos}\left( \omega_0 (f_i - f_{i+\Delta n}) \right) \approx \hspace{3cm} \nonumber \\
\frac{N-i}{\omega_0/\Omega_{\rm orb}} \sum_{\Delta n = 1}^{\omega_0/\Omega_{\rm orb}}  {\rm cos}\left( \omega_0 (f_i - f_{i+\Delta n}) \right).
\label{6.1.3}
\ear
In the right hand side of Eq.~(\ref{6.1.3}) the sum is over one cycle, while the term $({N-i})/({\omega_0/\Omega_{\rm orb}})$ is the number of cycles of the function $(f_i - f_{i+\Delta n})$. This approximation is good when the number of cycles is large. Actually, we will show that it is still good when the full observation includes just one cycle or more, whereas it starts to be annoyingly imprecise when less than one cycle is observed.

Setting $x=\Delta n \cdot 2 \pi \Omega_{\rm orb}/\omega_0$, Eq.~(\ref{6.1.3}) can be further approximated as
\begin{eqnarray}
\frac{N-i}{\omega_0/\Omega_{\rm orb}} \sum_{\Delta n = 1}^{\omega_0/\Omega_{\rm orb}}  {\rm cos}\left( \omega_0 (f_i - f_{i+\Delta n}) \right)
\hspace{2cm} \nonumber \\
\approx \frac{N-i}{\omega_0/\Omega_{\rm orb}} \cdot \frac{\omega_0}{2 \pi \Omega_{\rm orb}} \int_{0}^{2 \pi} {\rm cos}\left( \omega_0 \cdot \delta k [K_i - {\rm sin}(\Lambda_i + x)] \right) dx \nonumber \\ 
= \frac{N-i}{2\pi} C_i ,
\label{6.1.4}
\end{eqnarray}
where the integral has been referred to as $C_i$, indicating that it depends only on the index $i$. 
Note that since the integral is over one cycle ($0$, $2\pi$), the phase $\Lambda_i$ does not have any effect, so we can safely set $\Lambda_i=0$. Then, $C_i$ is a function of the term $K_i$, that we already noticed has  $-1$ and 1 as its minimum and maximum values, respectively.
Considering $K_i$ as a continuous function, the average value of $C_i$ is equal to the double integral
\beq
\overline {C_i} = \frac{1}{2}\int_{-1}^{1}  \int_{0}^{2 \pi} {\rm cos}\left( \omega_0 \cdot \delta k [K_i - {\rm sin}\,x] \right) dx dK_i .
\label{6.1.5}
\enq
The order of the integrals can be inverted, and  an analytical solution for the integral on $dK_i$ can be found. 
\begin{eqnarray}
 \int {\rm cos}\left( \omega_0 \delta k \cdot [K_i - {\rm sin}\,x] \right) dK_i \nonumber \\
 = \frac{{\rm sin}(\omega_0 \delta k \cdot K_i) \cdot {\rm cos}(\omega_0 \delta k \cdot {\rm sin}\,x )}{\omega_0 \delta k},
\label{6.1.6} \\
\Rightarrow
\int_{-1}^{1} {\rm cos}\left( \omega_0 \delta k \cdot [K_i - {\rm sin}\,x] \right) dK_i =  \nonumber \\
\frac{2 {\rm sin}(\omega_0 \delta k) \cdot {\rm cos}( \omega_0 \delta k \cdot {\rm sin}\,x)}{\omega_0 \delta k} .
\label{6.1.7}
\end{eqnarray}
Finally, the average value of $C_i$ is equal to
\beq
\overline {C_i} = \frac{ {\rm sin}(\omega_0 \delta k) }{\omega_0 \delta k} \int_{0}^{2 \pi} {\rm cos}( \omega_0 \delta k \cdot {\rm sin}\,x) dx .
\label{6.1.8}
\enq
Summarizing, the sum in Eq.~(\ref{6.1.2}) is equal to Eq.~(\ref{6.1.4}):
\beq
\sum_{\Delta n = 1}^{N-i}  {\rm cos}\left( \omega_0 (f_i - f_{i+\Delta n}) \right) = \frac{N-i}{2\pi} C_i .
\label{6.1.9}
\enq
Substituting it in Eq.~(\ref{6.1}), and then in Eq.~(\ref{5.10}) we get
\beq
N^2\varepsilon^2 = N + 2 \left[ \sum_{i=1}^{N}  \frac{N-i}{2\pi} C_i \right] .
\label{6.1.10}
\enq
Taking into account the average value of $C_i$, this equation  approximately becomes
\begin{eqnarray} 
\hspace{1cm}
N^2\varepsilon^2 &=& N +  2 \left[ \sum_{i=1}^{N} (N-i) \right] \frac{\overline {C_i}}{2\pi}  \nonumber \\
&=&  N +  2 \left[ \frac{N^2 - N}{2} \right] \frac{\overline {C_i}}{2\pi}  \nonumber \\
&=& N^2\frac{\overline {C_i}}{2\pi} + N\left(1-\frac{\overline {C_i}}{2\pi}\right).
\label{6.1.11}
\end{eqnarray}
In conclusion, we have found that for $N \gg 1$, the factor $\varepsilon^2$ that measures the loss in 
the power spectrum when values of the orbital parameters are not precisely known is approximately 
\beq
\varepsilon^2 = \frac{1}{2\pi} \cdot \frac{ {\rm sin}(\omega_0 \delta k) }{\omega_0 \delta k} \int_{0}^{2 \pi} {\rm cos}( \omega_0 \delta k \cdot {\rm sin}\,x) dx.
\label{6.1.12}
\enq

\begin{figure}
\center
\includegraphics[width=0.49\textwidth]{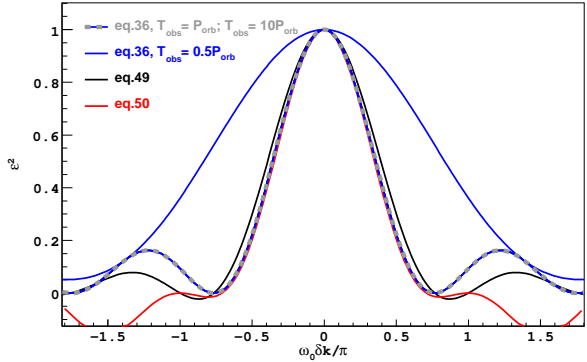}
\caption{The factor $\varepsilon^2$ calculated with different approximations in the case described in Section 6.1. See the text for the details.\label{fig2}}
\end{figure}

This result gives a clear description of the power spectrum around $\delta k = 0$, from which the following 
conclusions can be derived:

\begin{itemize}
 \item The factor $\varepsilon^2$ is function only of $\omega_0 \delta k$. The other parameters ($\phi, \psi, \delta c$) do not affect the power spectrum. \\

\item Eq.~(\ref{6.1.12}) does not depend on the duration of the observation $T_{\rm obs}$, at least within our approximations. Figure \ref{fig2} 
shows the factor $\varepsilon^2$ as a function of $\omega_0 \delta k$, as expressed in Eq.~(\ref{6.1.12}). Also in Figure \ref{fig2} we show 
the values of $\varepsilon^2$ calculated by the non-approximated formula of Eq.~(\ref{5.10}), for $T_{\rm obs} = 10 P_{\rm orb}$, $T_{\rm obs} = P_{\rm orb}$, $T_{\rm obs} = 0.5 P_{\rm orb}$. As we can notice from the plot, for $T_{\rm obs} \geq P_{\rm orb}$, Eq.~(\ref{6.1.12}) is a good approximation of $\varepsilon^2$, whereas it underestimates the values of $\varepsilon^2$ for $T_{\rm obs} < P_{\rm orb}$.
This feature is particularly advantageous for pulsation searches. Indeed, the maximum errors on the orbital parameters needed to avoid washing out the periodic signal are unaffected by the duration of the observation, which, on the other hand, the longer it is, the higher is the signal to noise ratio of the power spectrum, which is proportional to the total number of events (Scargle 1982).
We can deduce from Figure \ref{fig2} that reducing the observation time to $T_{\rm obs} \ll P_{\rm orb}$, the factor $\varepsilon^2$ remains unaffected. This is what most commonly happen in radio observations, which are so short that the orbital motion has negligible effects on the pulsation search.\\

\item The integral in Eq.~(\ref{6.1.12}) can not be solved analytically, but its behavior is very similar to the \textit{sinc} function. Indeed, we have empirically found that a very good approximation for $\varepsilon^2$ is 
\beq
\varepsilon^2 = \left[ \frac{ {\rm sin}(\omega_0 \delta k) }{\omega_0 \delta k}\right]^2 \left[1-2\left( \frac{\omega_0 \delta k}{\pi}\right) ^2 \right] .
\label{6.1.13}
\enq
As shown in Figure \ref{fig2}, Eq.~(\ref{6.1.13}) is even better than Eq.~(\ref{6.1.12}) 
to describe the central peak of $\varepsilon^2$, but it fails in the side lobes.
\end{itemize}

Clearly, these considerations are valid for the case analyzed in this section, i.e., when the parameters $\delta k$ and $\psi$ have no dependence on the eccentric anomaly $E$. This happens when the derivative of the eccentric anomaly with respect to the canonical orbital parameter ($\partial E/\partial p$) is null, as is the case for the projection of the semi-major axis $A$, and the longitude of periastron $W$.

\subsubsection{Semi-major axis A and simulations}

When we take into account the error on $A$, $\omega_0 \delta k$ is explicitly equal to (see Eqs. \ref{3.3} -- \ref{3.6}) 
\beq
\omega_0 \delta k = \omega_0 dA \sqrt{1-e^2 \cos(W)^2} .
\label{6.1.1.1}
\enq
In order to maintain the factor $\varepsilon^2$ higher than a certain level (say $\varepsilon^2>0.4$) the term $\omega_0 \delta k/\pi$ has to be lower than the inverse value ($\omega_0 \delta k/\pi < \left[ \varepsilon^2\right]^{-1}(0.4) \sim  0.4$), meaning that
\beq
dA < \frac{\pi \left[ \varepsilon^2\right]^{-1}}{\omega_0 \sqrt{1-e^2 \cos(W)^2}}.
\label{6.1.1.2}
\enq
Clearly, the error $dA$ is inversely proportional to the pulsar frequency ($\omega_0$). It is interesting to notice that high eccentricities are less constraining for $dA$, even though the term $\sqrt{1-e^2 \cos(W)^2}$ change very slowly with the eccentricity. For example it is lower than 0.5 only for $e \gtrsim 0.85$, and cos$(W) \sim 1$.

\begin{table}
  \caption{Parameters of the simulated pulsar.}
  \begin{center}
  \begin{tabular}{@{}lr@{}}
  \hline
   Parameters   &   value  \\
 \hline
$P_{psr}$       &      300 ms\\
$P_{\rm orb}$       &      4.0 days\\
$T_0$       &      54587.00 MJD\\
$A   $    &      2.5 lt-s\\
$W   $    &      44.39$^{\circ}$\\
$e   $   &      0.61\\
$T_{\rm obs}$       &      40 days\\
\hline
\end{tabular}
\label{Tab1}
\end{center}
\end{table}

In order to check these results, we simulated the barycentred arrival time series from a pulsar in a binary system with the features described in Table \ref{Tab1}. Then we demodulate it, but modifying one of the orbital parameters (in this case $A$). In this way we obtain the demodulated arrival time series perturbed by the variation of the orbital parameter,  called hereafter {\it perturbed time series}. Its power spectrum is expected to follow the analytical description described above.

The simulation consists of two steps. First, the emission time series in the pulsar reference frame has been created. All the time stamps are taken as random numbers in the range $[0; \, 2\pi/\omega_0]$, following a pure sinusoidal distribution
\begin{equation}
 P_t(t) = 1 + {\rm sin}(\omega_0 t),
 \label{6.1.1.3}
\end{equation}
where $\omega_0$ is the assumed frequency of the pulsar.
Then, in order to cover the full duration of the observation, each time stamp has been randomly delayed adding a value $({2\pi}/{\omega_0}) n$, where $n$ is a random integer number uniformly distributed in $[0, \omega_0 T_{\rm obs}/2\pi$].
Finally, to pass from the emission times in the pulsar reference frame ($t_e$), to the barycentred arrival time ($t_a$), the R\"{o}mer delay $\Delta_R(t_e)$ is added to each $t_e$. The R\"{o}mer delay is calculated taking into account the orbital parameters.
The demodulation is performed by means of the algorithm described in \citet{Rea2011}, which makes use of the program TEMPO2 (\citet{Hobbs-Edwards-Manchester2006}).

\begin{figure*}
\centering
\begin{minipage}[b]{0.9\textwidth}
\centering
\includegraphics[width=1.0\textwidth]{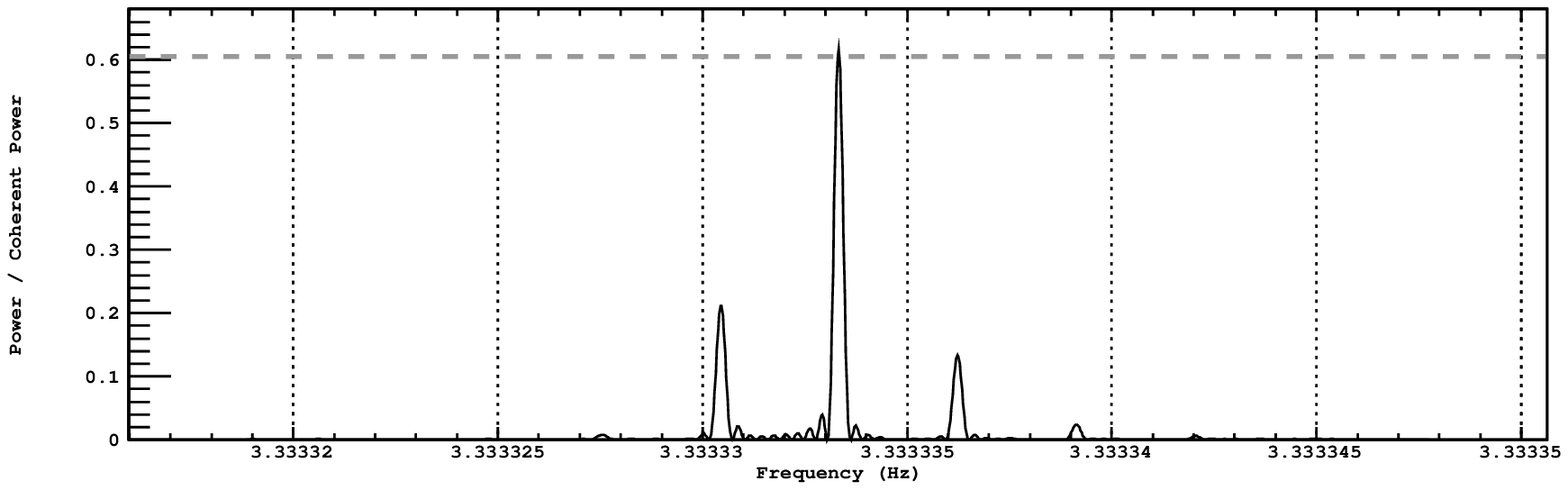}
\end{minipage}
\begin{minipage}[b]{0.9\textwidth}
\centering
\includegraphics[width=1.0\textwidth]{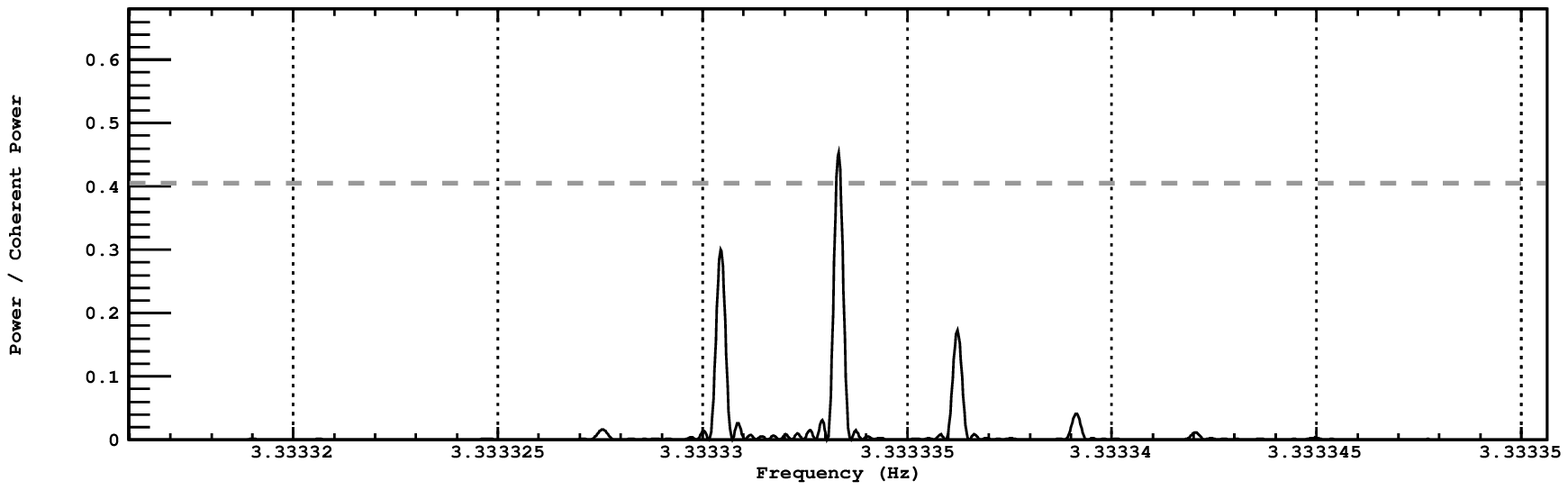}
\end{minipage}
\begin{minipage}[b]{0.9\textwidth}
\centering
\includegraphics[width=1.0\textwidth]{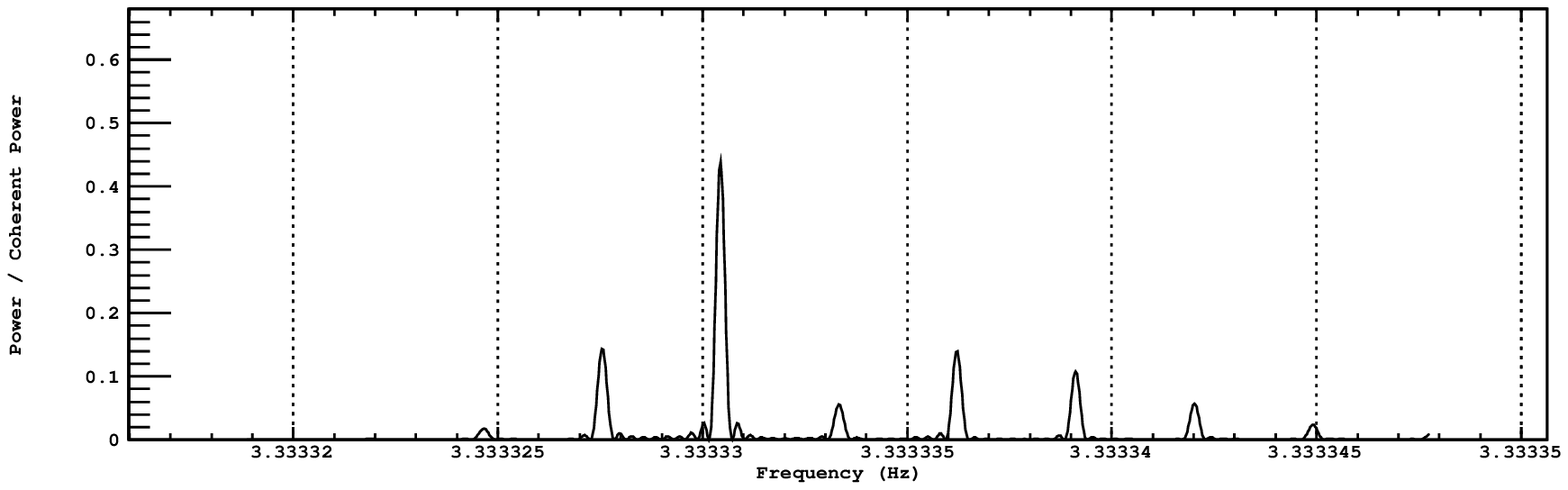}
\end{minipage}
\caption{Power spectra of the simulated time series, demodulated varying the projection of the semi-major axis by three different amounts $dA$. \textit{Top panel}: $dA=0.053$ lt-s correspond to $\omega_0 \delta k / \pi = 1/\pi$, and $\varepsilon^2 \sim 0.6$ is expected (dashed line). 
\textit{Middle panel}: $dA=0.067$ lt-s, correspond to $\omega_0 \delta k / \pi = 0.4$, and to an expected $\varepsilon^2 \sim 0.4$ (dashed line).
\textit{Bottom panel}: $dA=0.125$ lt-s , imply $\omega_0 \delta k / \pi = 0.75$, a null power is expected. The so-called sidebands dominate in this case.
The \textit{Coherent Power} at the denominator of the $y$ axis is the power calculated for an unperturbed demodulation  ($dA=0$) at the pulsar frequency $\omega_0$.}
\label{fig.dA}
\end{figure*}

The power spectra shown in Figure \ref{fig.dA} are calculated from time series demodulated when varying the projection of the semi-major axis by three different amounts $dA$. In the top panel $dA=0.053$ lt-s correspond to $\omega_0 \delta k / \pi = 1/\pi$. From Figure \ref{fig2} or Eq.~(\ref{6.1.13}), for this value of $\omega_0 \delta k / \pi$ the power calculated at the pulsar frequency $\omega_0$ is expected to be $\sim 0.6$ times the coherent power resulting from an unperturbed demodulation. 
This is confirmed by the power spectrum in the top panel in figure \ref{fig.dA}.
In the middle panel $dA=0.067$ lt-s, corresponding to $\omega_0 \delta k / \pi = 0.4$, and $\varepsilon^2 \sim 0.4$.
In the bottom panel $dA=0.125$ lt-s, implying $\omega_0 \delta k / \pi = 0.75$, for which a null power is expected. Indeed, the peak at the pulsar frequency is very suppressed, while sidebands dominate in this case. 
The ratio of the power to the coherent power of the central peak in the plots of Figure \ref{fig.dA} can be directly compared with the $\varepsilon^2$ evaluated in Figure \ref{fig2} at the corresponding values of $\omega_0 \delta k / \pi$. We thus find a good agreement between the analytical procedure and the simulations.

\subsubsection{Sidebands ambiguity and a method for removing it}

As shown in the bottom panel of Figure \ref{fig.dA}, it is possible that for a certain value of $dA$, the peak at the true pulsar frequency is strongly suppressed, while the highest sideband peak is detected above the noise level. How can we realize in the course of a real observation whether the peak detected is a sideband or the pulsar itself? Before answering to this question, it is important to clarify that the presence of the sideband structure in the power spectrum is a clear signature of the presence of a pulsar in the system, since otherwise we would see only noise.

\citet{Ransom2001} studied the sidebands by a not-demodulated arrival time series. The number and the height of the sideband peaks depend on the orbital parameters. Here we summarize some important features of the sidebands, which can help solving the ambiguity described above.
\begin{itemize}
 \item The difference in frequency of two adjacent peaks is equal to the orbital frequency, so that with respect to the pulsar frequency, the sideband peaks are located at
\begin{equation}
\omega_{sb} = \omega_0 \pm n \Omega_{\rm orb} ,
 \label{6.1.1.4}
\end{equation}
where $n$ is an integer number, and $\omega_{sb}$ the frequency of the $n$'th sideband peak. \\  

\item The sum of the power of all the sideband peaks is equal to the total coherent power of the pulsar.  \\

\item The larger is the orbit, more numerous are the sideband peaks, because the modulation in the arrival time series is stronger. 

\end{itemize}

In our case, the sidebands appear in the power spectra of the perturbed time series because they still have a residual modulation. 
This means that larger is $dA$, the higher the residual modulation will be, and because of the third bullet commented above, more numerous sideband peaks will appear. We can deduce that the highest probability to detect a sideband peak, rather than the pulsar one is when $\varepsilon^2$ is close to the minimum in Figure \ref{fig2}. Indeed, when it happens, the pulsar peak is almost at zero power and consequently the  sideband peaks get stronger. In this case, their number is small because we are close to the true value of the orbital parameters. Therefore, at least one sideband peak could have a power higher than the noise level.

Thus one can think of a method to solve the ambiguity introduced by the possible presence of sidebands as follows;
\begin{enumerate}
\item When a significant peak is detected at $\omega_d$ in the power spectrum, we can assume it to be at the minimum of the $\varepsilon^2$ curve, regardless on whether this is true or not. \\

\item  With this assumption we can get an estimation of $dA$ (or more in general of $\delta k$) by means of Eq.~(\ref{6.1.1.2}) setting in it $\varepsilon^2=0$, and $\omega_0 = \omega_d$ (since $\Omega_{\rm orb} \ll \omega_0$, for Eq.~(\ref{6.1.1.4}) it is always true that $\omega_d \simeq \omega_0$). \\

\item 
We can use the so-estimated $dA$ to define a sub-sampling of the parameter space in order to explore the profile in Figure \ref{fig2}, and to understand if we really are in its minimum, or on the top of the peak. Using the new sampling, and if in the former case, it is expected that a frequency peak with higher power with respect to the detected one will appear at a frequency $\omega = \omega_d \pm n \Omega_{\rm orb}$. This new peak is then the true pulsar frequency. 
In contrast, if the new peak does not appear, or its power is lower than the first detected one, the true pulsar frequency remains $\omega_0 = \omega_d$, as originally detected.
\end{enumerate}

Since the presence of the sidebands in the power spectrum means that there is a pulsar in the system, their detection will immediately solve the nature of the compact object if such is unknown. A technique dedicated to the detection of sidebands was formulated by \citet{Ransom2001} (see also \citealt{Ransom2003}). In brief, because of the first feature listed above, the sideband peaks appear as a short series of regular pulsations  in the power spectrum of the arrival times series. Then, the detection of the sidebands is possible by taking the Fourier transform of this short section of the power spectrum, which is expected to have a peak at the orbital frequency of the system. 
This technique was originally formulated for the detection of radio binary systems with short period $P_{\rm orb} < T_{\rm obs}$. Since this condition is common in gamma-ray observations, this technique can be easily adapted to the perturbed time series studied in this work.

Whether the first or the second method described above is more appropriate, strongly depends on the signal to noise ratio, as well as on the pulsed fraction of the signal.
For weak pulsed signals the first method should be more appropriate, because 
most of the sideband peaks would be lower than the noise level. 
On the other hand, when the method proposed by \cite{Ransom2001} is applied to calculate the significance of the signal one should take into account also the trials in Fourier transforming several short section of the power spectrum.

\subsubsection{Longitude of the periastron W and simulations}

For the longitude of the periastron, the expression of $\omega_0 \delta k$ is
\beq
\omega_0 \delta k = \omega_0 A dW \sqrt{\frac{1-e^2+e^4{\rm cos}^2W {\rm sin}^2W}{1-e^2 \cos^2W}} .
\label{6.1.16}
\enq
The error $dW$ is inversely proportional to both the pulsar frequency and the projection of the semi-major axis $A$. 
Its dependence on $W$ and $e$ is in the square root. For $W=\pi$, this term is equal to 1, so it does not affect $dW$. In contrast, for $W=\pi/2$, or $W=\frac{3}{2}\pi$ the square root is equal to $\sqrt{1-e^2}$, so also in this case high eccentricities are less constraining for $dW$.

\begin{figure*}
\centering
\begin{minipage}[b]{0.9\textwidth}
\centering
\includegraphics[width=1.0\textwidth]{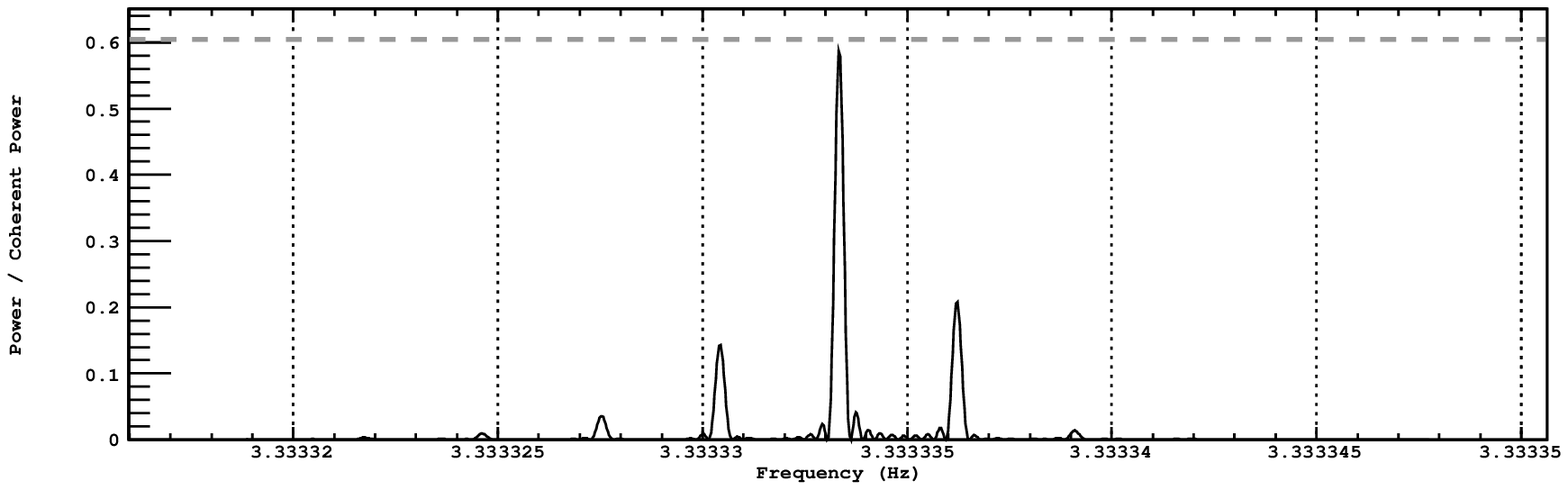}
\end{minipage}
\begin{minipage}[b]{0.9\textwidth}
\centering
\includegraphics[width=1.0\textwidth]{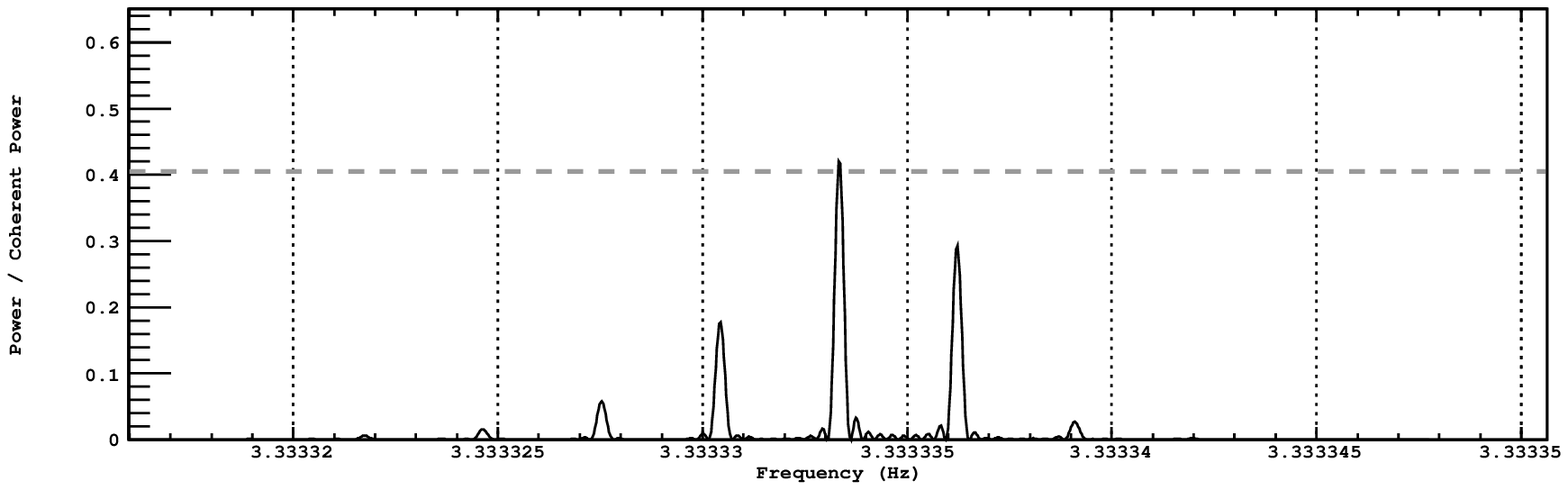}
\end{minipage}
\begin{minipage}[b]{0.9\textwidth}
\centering
\includegraphics[width=1.0\textwidth]{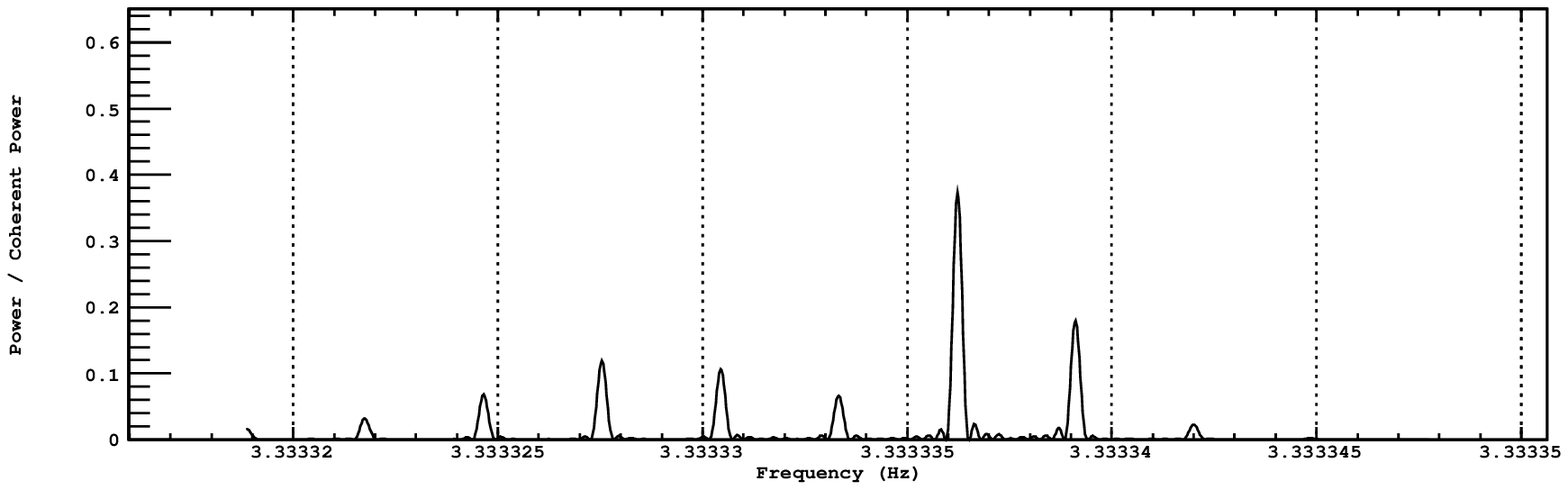}
\end{minipage}
\caption{Power spectra of the simulated time series, demodulated varying the longitude of the periastron by three different amounts $dW$. \textit{Top panel}: $dW=1.34^{\circ}$ correspond to $\omega_0 \delta k / \pi = 1/\pi$, and $\varepsilon^2 \sim 0.6$ is expected (dashed line). 
\textit{Middle panel}: $dW=1.68^{\circ}$, correspond to $\omega_0 \delta k / \pi = 0.4$, and to an expected $\varepsilon^2 \sim 0.4$ (dashed line).
\textit{Bottom panel}: $dW=3.15^{\circ}$, imply $\omega_0 \delta k / \pi = 0.75$, a null power is expected.
The \textit{Coherent Power} at the denominator of the $y$ axis is the power calculated for an unperturbed demodulation ($dW=0$) at the pulsar frequency $\omega_0$.  
}
\label{fig.dW}
\end{figure*}

We have demodulated the arrival time series of the pulsar simulated in the previous Section by varying $W$ in three different amounts, $dW$. Figure \ref{fig.dW} shows the power spectra of the perturbed time series. In the top panel $dW = 1.34^{\circ}$ implies $\omega_0 \delta k = 1$, and  $\varepsilon^2 \sim 0.6$ by Eq. \ref{6.1.13}. In the middle panel $dW = 1.68^{\circ}$, $\omega_0 \delta k /\pi = 0.4$, and  $\varepsilon^2 \sim 0.4$ (Eq. \ref{6.1.13}, see also Figure \ref{fig2}). In the bottom panel $dW = 3.15^{\circ}$, $\omega_0 \delta k /\pi = 0.75$, and $\varepsilon^2 \sim 0$. Also in this case the sidebands dominate in the bottom panel, while the peak at the pulsar frequency is prominent in the top one.

\subsection{Cases for which $\delta k$ and $\psi$ are functions of the eccentric anomaly $E$}

Similarly to the previous Section, we would like to find a solution for $\varepsilon^2$ when the parameter known with the higher uncertainty is the eccentricity $e$, or the epoch of the periastron $T_0$, or the orbital period $P_{orb}$. These cases are more complicated because the partial derivatives of the eccentric anomaly $E$ with respect to these parameters are not null. 
This implies that $\delta k$ and $\psi$ are not constants, but rather, functions of the eccentric anomaly $E$.
In the following subsections we analyze these three cases one by one.

\subsubsection{Eccentricity $e$ and simulations}

The differential of Eq.~(\ref{Ete}) with respect to the eccentricity $e$ leads to the following expression for the partial derivative of $E$ respect to the eccentricity 
\beq
\frac{\partial E}{\partial e} = \frac{{\rm sin}E}{1-e \, {\rm cos}E} \sim \frac{{\rm sin}(\Omega_{\rm orb} t_e)}{1-e \, {\rm cos}(\Omega_{\rm orb} t_e)}.
\label{6.2.1}
\enq
Since ${\partial E}/{\partial e}$ is periodic, so are $\delta k$ and $\psi$. 
In order to find a solution for $\varepsilon^2$, we need to solve the double sum in the square brackets of Eq.~(\ref{6.1}), where $f_i - f_j$ is in this case
\begin{eqnarray}
\label{6.2.2}
f_i - f_j =  
 \delta k_i \cdot {\rm sin}\left(  \frac{\Omega_{\rm orb}}{\omega_0}2\pi i +\phi+\psi_i \right) -
\\
 \delta k_j \cdot {\rm sin}\left(  \frac{\Omega_{\rm orb}}{\omega_0}2\pi j +\phi+\psi_j \right) \nonumber  \\
=   K_i - \delta k_{i+\Delta n} \cdot {\rm sin}\left(  \frac{\Omega_{\rm orb}}{\omega_0}2\pi (i+\Delta n) +\phi+\psi_{i+\Delta n} \right). \nonumber 
\end{eqnarray}
Note that $\delta k_i$ and $\psi_i$ contain the partial derivative of $E$ as function of $(2\pi i + \theta)/\omega_0$, which is equal to
\begin{eqnarray}
\frac{\partial E}{\partial e}\left( \frac{2\pi i + \theta}{\omega_0}\right)  = \frac{{\rm sin}(\frac{\Omega_{\rm orb}}{\omega_0}2\pi i + \frac{\Omega_{\rm orb}}{\omega_0}\theta)}{1-e \, {\rm cos}(\frac{\Omega_{\rm orb}}{\omega_0}2\pi i + \frac{\Omega_{\rm orb}}{\omega_0}\theta)} 
\nonumber \\
\sim \frac{\frac{\Omega_{\rm orb}}{\omega_0}2\pi i)}{1-e \, {\rm cos}(\frac{\Omega_{\rm orb}}{\omega_0}2\pi i)} .
\label{6.2.3}
\end{eqnarray}
Since this is periodic in $\Omega_{\rm orb}/\omega_0$, then also $f_i - f_j = f_i - f_{i+\Delta n}$ is periodic with period $\Delta n = \Omega_{\rm orb}/\omega_0$. This is the same feature we have found in the previous Section. Hence, we can follow the same reasoning and integrate the argument of the double sum ($\cos(\omega_0(f_i - f_j))$) in one cycle, as we did in Eq.~(\ref{6.1.4}), to get
\begin{eqnarray}
\frac{N-i}{\omega_0/\Omega_{\rm orb}} \sum_{\Delta n = 1}^{\omega_0/\Omega_{\rm orb}}  {\rm cos}\left( \omega_0 (f_i - f_{i+\Delta n}) \right)
\approx \nonumber \\
\approx \frac{N-i}{\omega_0/\Omega_{\rm orb}} \cdot \frac{\omega_0}{2 \pi \Omega_{\rm orb}} \times  \nonumber \\ 
\int_{0}^{2 \pi} {\rm cos}\left( \omega_0 \cdot  [K_i - \delta k(x) \cdot {\rm sin}(x + \psi(x))] \right) dx \nonumber \\
= \frac{N-i}{2\pi} C_i 
\label{6.2.4}
\end{eqnarray}
Here $K_i = \delta k_i \cdot {\rm sin}\left(  \frac{\Omega_{\rm orb}}{\omega_0}2\pi i +\phi+\psi_i \right)$ is a periodic function that oscillates between the values $K_{min}$ and $K_{max}$.
Following the reasoning of the previous Section, from Eqs.~(\ref{6.1.5}) and (\ref{6.1.12}) we have in this case that
\begin{eqnarray}
\varepsilon^2 = \frac{1}{2\pi(K_{max}-K_{min})} \int_{K_{min}}^{K_{max}} \times \hspace{2cm} \nonumber \\  
\int_{0}^{2 \pi}  {\rm cos}\left( \omega_0 \cdot  [K_i - \delta k(x) \cdot {\rm sin}(x + \psi(x))] \right) dx dK_i .
\label{6.2.5}
\end{eqnarray}

The integral over $dK_i$ can be analytically solved, obtaining
\begin{eqnarray}
\varepsilon^2 = \frac{1}{2\pi}\frac{{\rm sin}(\omega_0 K_{max})-{\rm sin}(\omega_0 K_{min})}{\omega_0(K_{max}-K_{min})}     \times
\hspace{1.3cm}
\nonumber \\ 
\hspace{1.3cm}
\int_{0}^{2 \pi}  {\rm cos}\left( \omega_0 \cdot \delta k(x) \cdot {\rm sin}(x + \psi(x))] \right) dx .
\label{6.2.6}
\end{eqnarray}
Here the values $K_{min}$ and $K_{max}$ are not symmetric. Hence the expression outside the integral can not be further simplified to a \textit{sinc} function. Furthermore, $K_{min}$ and $K_{max}$ cannot be expressed by an analytic formula. 
However, some features can be found analyzing the function $K_i$:
\begin{eqnarray}
 K_i = \delta k_i \cdot {\rm sin}\left(  \frac{\Omega_{\rm orb}}{\omega_0}2\pi i +\phi+\psi_i \right) = 
\nonumber \\
 A \, de   \left(\frac{}{}     (1-e^2{\rm cos}^2W) \times  \right.   \nonumber \\
 \left.
 \left[ \frac{{\rm sin}\,x}{1-e \, {\rm cos}\,x} + \frac{e \, {\rm tan}(W)}{1-e^2+{\rm tan}(W)}\frac{1}{\sqrt{1-e^2}}  \right]^2 + \right.
 \nonumber \\
 \left.
  \frac{e^2{\rm cos}^4W}{1-e^2{\rm cos}^2W} \right)^{1/2} \times 
\label{6.2.7}   \\
 {\rm sin}\left( x + {\rm atan}\left( \frac{{\rm tan}(W)}{\sqrt{1-e^2}}\right)  +  {\rm atan}\left( -\frac{1-e^2{\rm cos}^2W}{e \, {\rm cos}^2W} 
 \right. \right.  \nonumber \\
 \left. \left.
 \left[  \frac{{\rm sin}\,x}{1-e \, {\rm cos}\,x} + \frac{e \, {\rm tan}(W)}{1-e^2+{\rm tan}(W)}\frac{1}{\sqrt{1-e^2}} \right]   \right)   \right)  ,  \nonumber 
\end{eqnarray}
where we have set $x=({\Omega_{\rm orb}}/{\omega_0}) 2\pi i$. 
Eq.~(\ref{6.2.7}) can be more easily read as $K_i = A\,de \, g(x; e, W)$, so that $K_{max}=A\, de \, g_{max}$ and $K_{min}=A\, de \, g_{min}$. Substituting in Eq.~(\ref{6.2.6}) it is clear that $\omega_0 \, A\, de$ has a key role in this case, from which we can conclude that the maximum error of the eccentricity ($de$) that is allowed to avoid washing out the pulsed signal is inversely proportional to the pulsar frequency ($\omega_0$), and to the projection of the semi-major axis ($A$).

\begin{figure}
\begin{minipage}[b]{0.5\textwidth}
\centering
\includegraphics[width=1\textwidth]{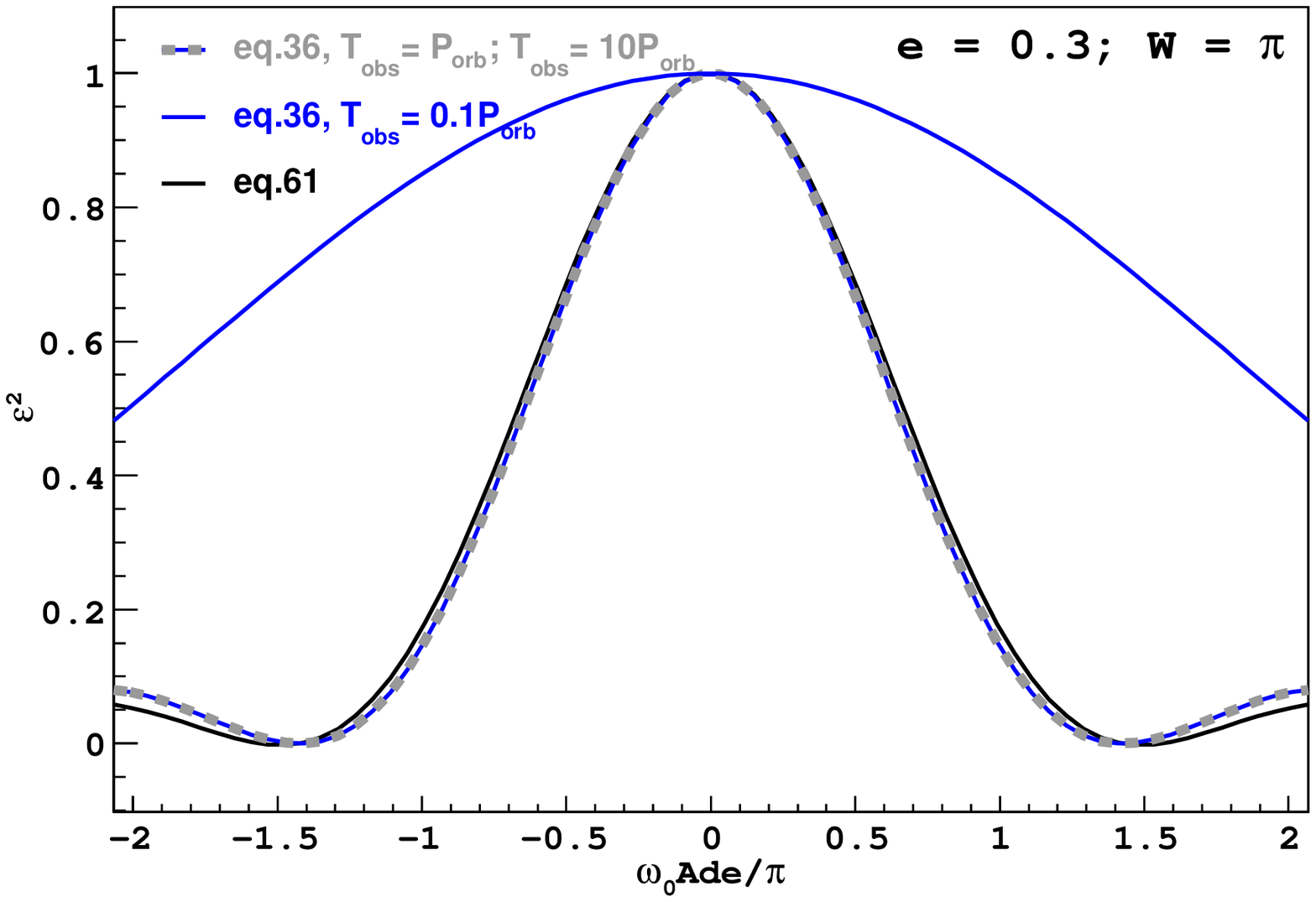}
\end{minipage}
\begin{minipage}[b]{0.5\textwidth}
\centering
\includegraphics[width=1\textwidth]{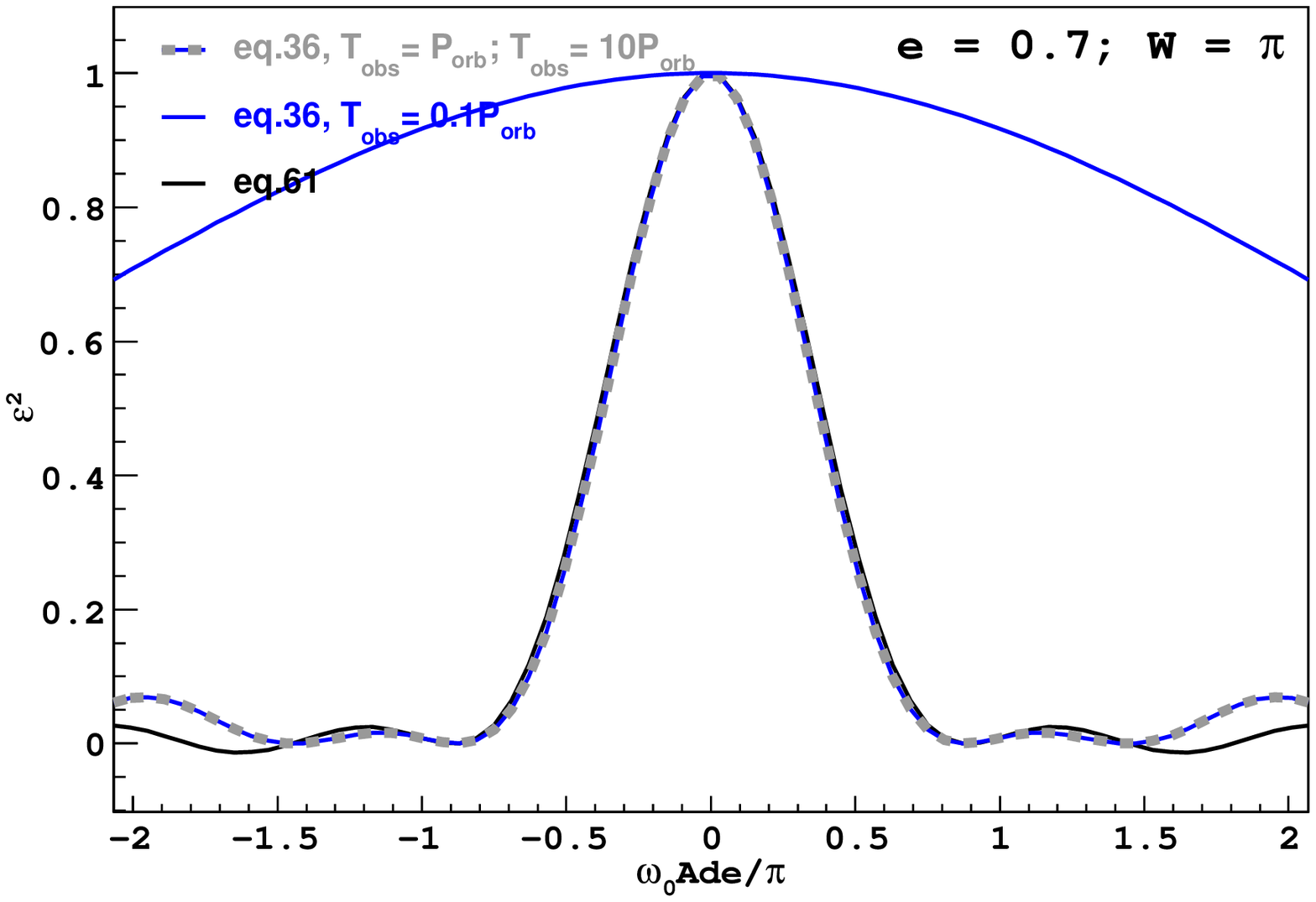}
\end{minipage}
\begin{minipage}[b]{0.5\textwidth}
\centering
\includegraphics[width=1\textwidth]{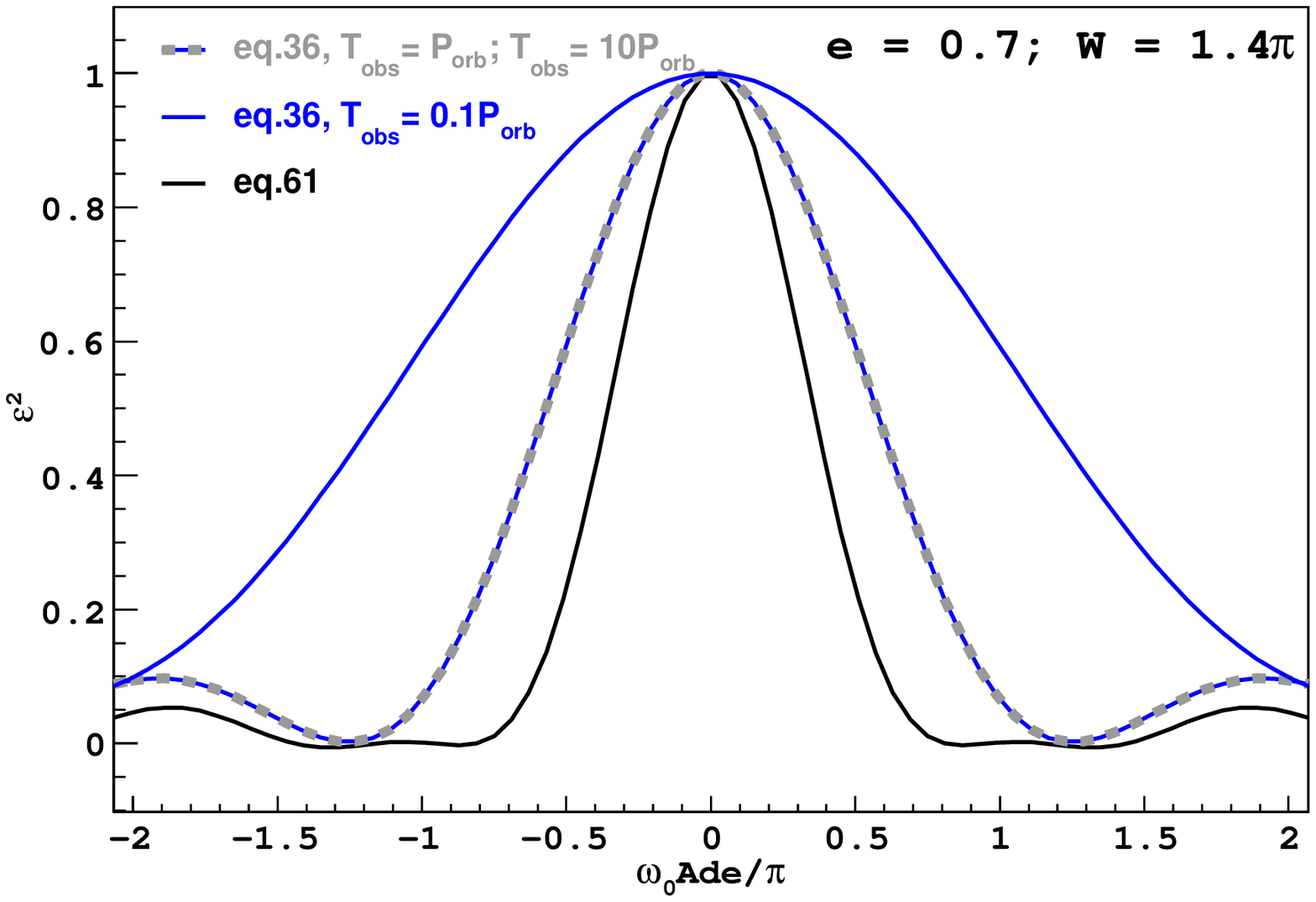}
\end{minipage}
\caption{$\varepsilon^2$ versus $\omega_0 A de/\pi$ as calculated using Eq.~(\ref{6.2.6}), and Eq.~(\ref{5.10}) for the case described in Section 6.2.1. Each panel assumes different values of $e$ and $W$. \textit{Top}: $e=0.3$, $W=\pi$. \textit{Middle}: $e=0.7$, $W=\pi$. \textit{Bottom}: $e=0.7$, $W=1.4\pi$. 
Three different observation times are assumed to compute Eq.~(\ref{5.10}): $T_{\rm obs}=10P_{\rm orb}; P_{\rm orb}$, and $0.1 P_{\rm orb}$.  }
\label{fig3}
\end{figure}

\begin{figure}
\begin{minipage}[b]{0.5\textwidth}
\centering
\includegraphics[width=1.0\textwidth]{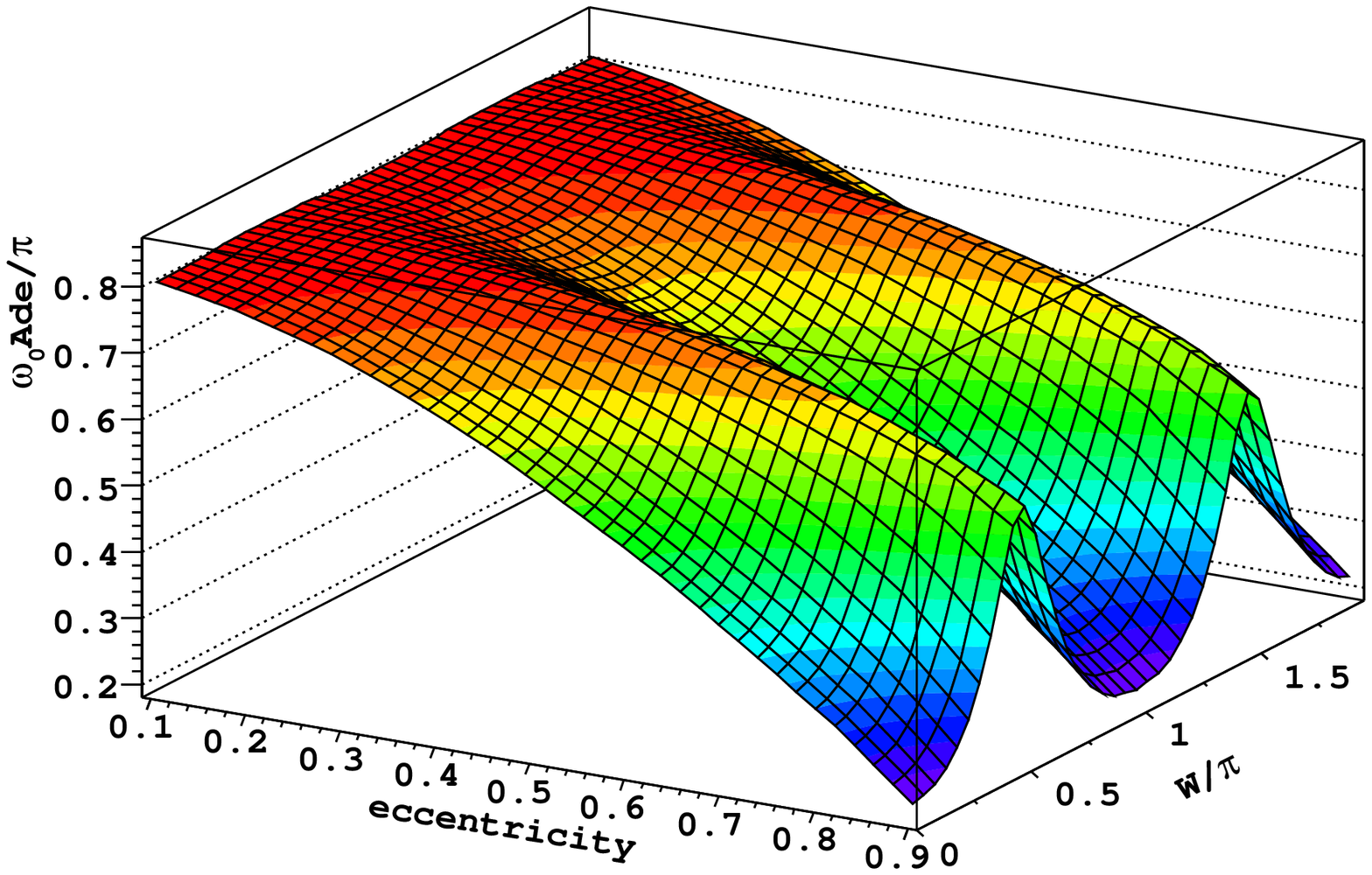}
\end{minipage}
\begin{minipage}[b]{0.5\textwidth}
\centering
\includegraphics[width=1.0\textwidth]{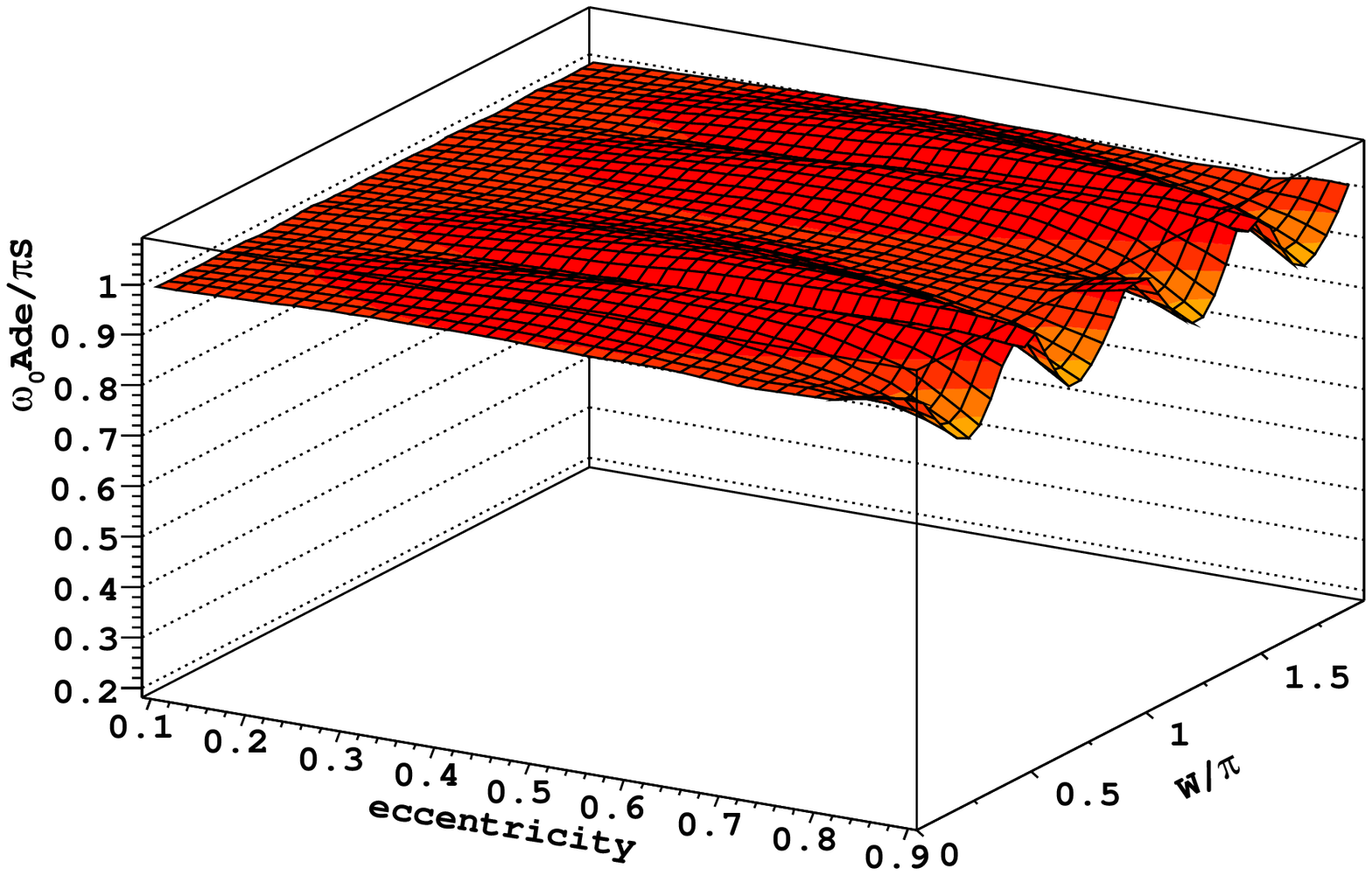}
\end{minipage}
\caption{\textit{Top}: Surface representing the values of $\omega_0Ade/\pi$ for which $\varepsilon^2 = 0.4$ versus $e$ and $W$. This surface is approximately fitted by the function $S(e,W)$ given in Eq.~(\ref{6.2.8}). \textit{Bottom}: The almost-flat surface in this plot represents the values of $\omega_0Ade/\pi S(e,W)$ for which $\varepsilon^2 = 0.4$. The flatness of the surface indicate that the term $\omega_0Ade/\pi S(e,W)$ absorbs the dependence of $\varepsilon^2$ on  $e$ and $W$.}
\label{fig4}
\end{figure}

\begin{figure}
\center
\includegraphics[width=0.5\textwidth]{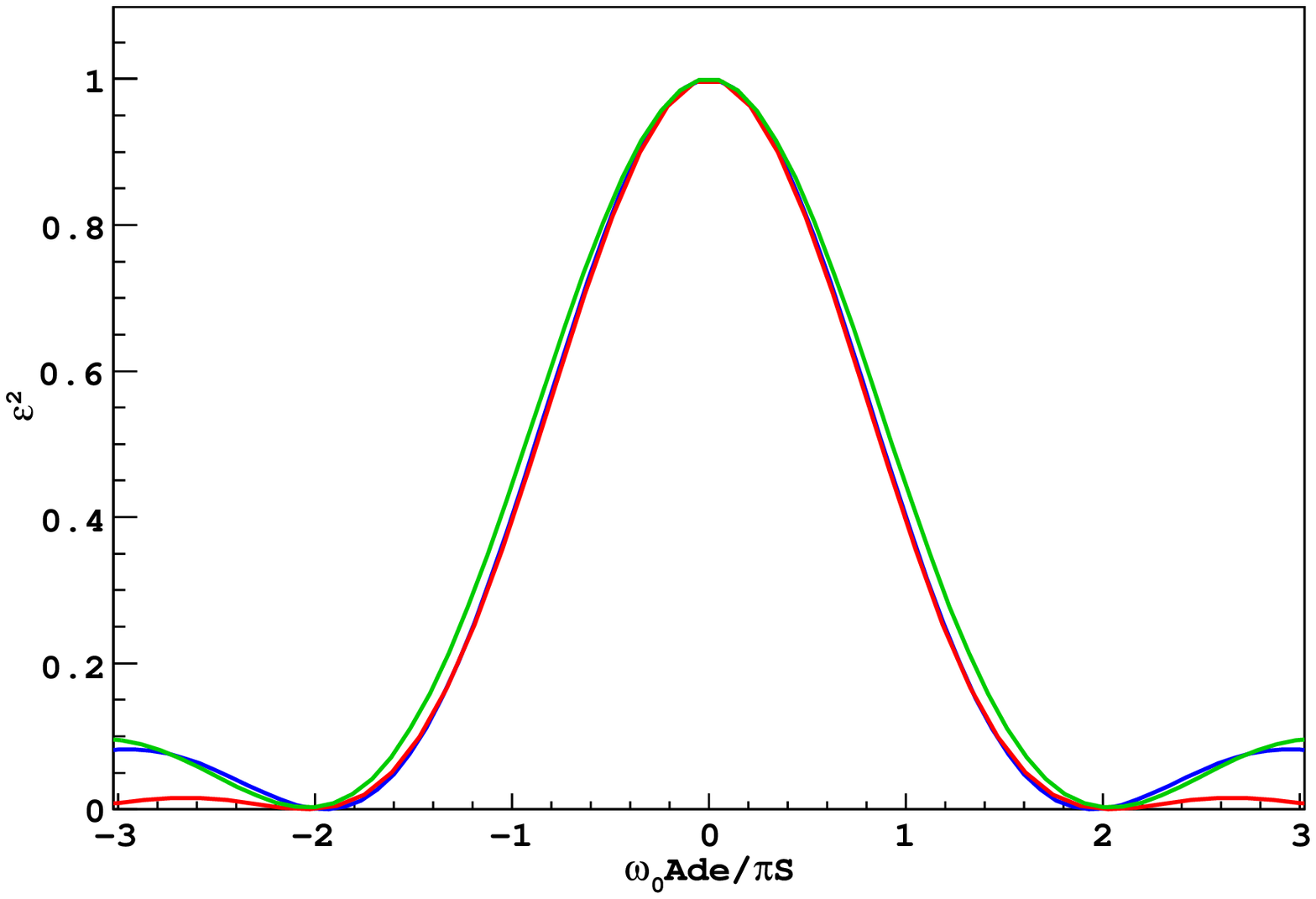}
\caption{$\varepsilon^2$ versus $\omega_0Ade/\pi S$, assuming for $e$ and $W$ the same values as for the three plots in Figure \ref{fig3}. The values of $\varepsilon^2$ are calculated by means of Eq.~(\ref{5.10}). \textit{Blue line}: $e = 0.3$, $W = \pi$. \textit{Red line}: $e = 0.7$, $W = \pi$. \textit{Green line}: $e = 0.7$, $W = 1.4\pi$. \label{EccSurfNorm}  }
\end{figure}

Also in this case, Eq.~(\ref{6.2.6}) does not depend on the duration of the observation ($T_{\rm obs}$), when it is greater than one orbital period ($P_{\rm orb}$). In order to check this, Figure \ref{fig3} shows Eq.~(\ref{6.2.6}) versus $\omega_0\, A\, de$ for different values of $e$ and $W$. They are compared with Eq.~(\ref{5.10}) assuming different observation times ($T_{\rm obs}=10P_{\rm orb}$, $P_{\rm orb}$, and
$0.1P_{\rm orb}$). 
From these plots we can notice that the agreement between Eq.~(\ref{5.10}) and Eq.~(\ref{6.2.6}) is not always good. Furthermore, the width of central the peak produced by Eq.~(\ref{5.10}), and its first zeroes have an evident dependence on $e$ and $W$. We investigated this dependence assuming a grid of values of $e$ and $W$, and numerically finding the values of $\omega_0 \, A\, de/\pi$ for which $\varepsilon^2 = 0.4$. They are represented by the surface plot of Figure \ref{fig4} (top panel), which is approximately fitted by the function 
\begin{eqnarray}
S(e,W) &=& U(e) - {\rm Ampl}(e) \cdot  | {\rm cos}W | , \label{6.2.8} \;\;\;\;\; {\rm with} \\
U(e) &=& U_0 + a \cdot {\rm Log}(b-e^{1.3}) , \nonumber \\
{\rm Ampl}(e) &=& {\rm Ampl}_0 \cdot (1-e/2)e^2, \nonumber 
\end{eqnarray}
and where $U_0 = 0.816 \pm 0.001$, $a = 0.345 \pm 0.005$, $b=1.08 \pm 0.01$, and Ampl$_0=0.84\pm0.01$. 
If we now plot the previous values of $\omega_0Ade$ divided by $S(e,W)$, we obtain an almost flat surface at $z=1$ with small fluctuations due to the approximations (see Figure \ref{fig4} bottom panel). 
The flatness of the surface indicate that the term $\omega_0Ade/\pi S(e,W)$ absorbs the dependence of $\varepsilon^2$ on $e$ and $W$. In other words, the factor $\varepsilon^2$ is approximately a function of the single variable $z=\omega_0Ade/\pi S$. To show this better, Figure \ref{EccSurfNorm} shows the values of $\varepsilon^2$ obtained using Eq.~(\ref{5.10}) versus $\omega_0Ade/\pi S$, assuming for $e$ and $W$ the same values as for the three plots in Figure \ref{fig3}.

\begin{figure}
\center
\includegraphics[width=0.5\textwidth]{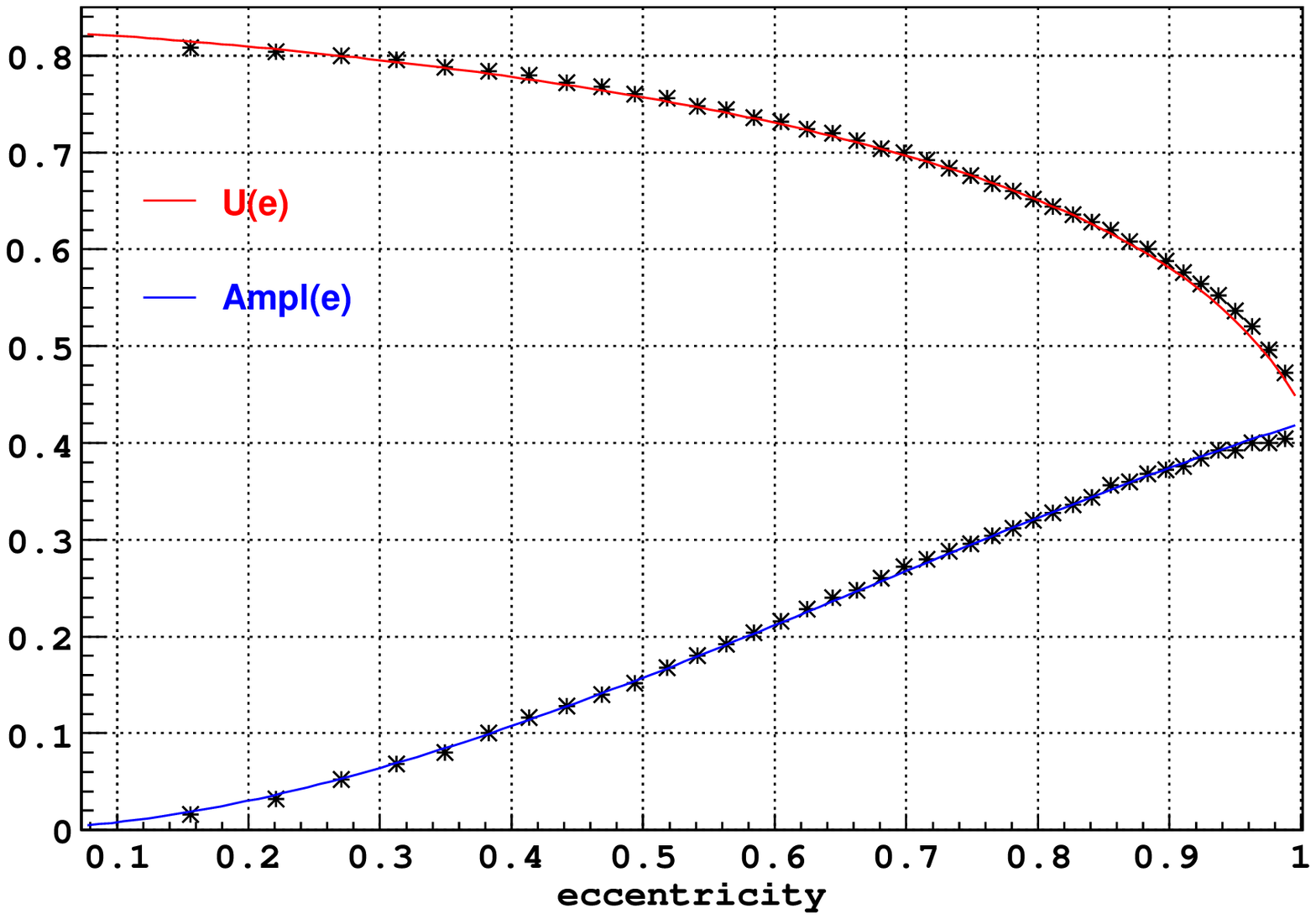}
\caption{$U(e)$ and Ampl$(e)$ evaluated for several values of the eccentricity, and fitted by the second (red line) and third  (blue line) functions of the group of Eqs.~(\ref{6.2.8}). \label{fig5}  }
\end{figure}

We can now state that in order to maintain the factor $\varepsilon^2$ higher than a certain level, the following condition must be satisfied 
\begin{eqnarray}
\frac{\omega_0Ade}{\pi} < [\varepsilon^2]^{-1} \cdot \left\lbrace U(e) + {\rm Ampl}(e) \cdot | {\rm cos}W | \right\rbrace  .
\label{6.2.9}
\end{eqnarray}
Figure \ref{fig5} shows the functions $U(e)$ and ${\rm Ampl}(e)$ fitted on a set of values of $e$. Looking at this figure, and from Eq.~(\ref{6.2.9}), we can conclude that for low eccentricities ($e<0.3$) ${\rm Ampl}(e)$ is negligible, and $U(e)$ acts like a weak scale factor $\sim 0.8$. In contrast, for high eccentricities ($e>0.6$) and especially for $W \sim 1$, the error $de$ is strongly constrained if we wish to maintain pulse-detection capability. Indeed, the right hand of Eq.~(\ref{6.2.9}) can reach values lower than 0.2 for $e>0.9$, implying that $de$ has to be lower than $0.2\pi/(\omega_0A)$ in order to avoid a prohibitive suppression of the pulsed signal ($\varepsilon^2 < 0.4$).

\begin{figure*}[ht]
\centering
\begin{minipage}[b]{0.9\textwidth}
\centering
\includegraphics[width=1.0\textwidth]{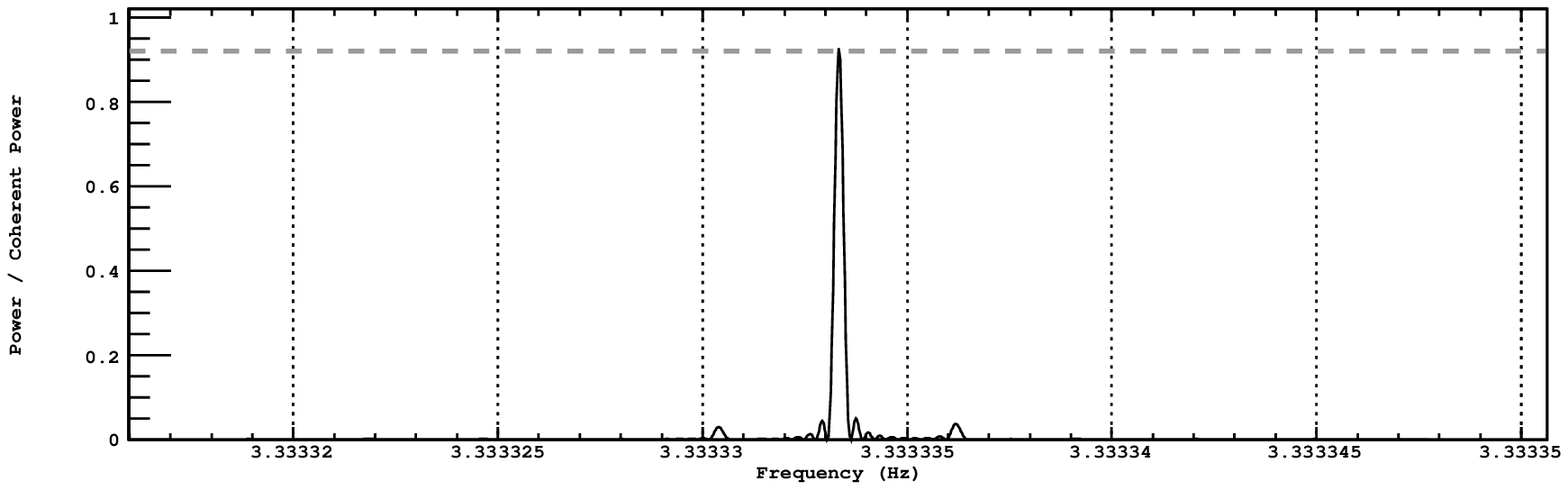}
\end{minipage}
\begin{minipage}[b]{0.9\textwidth}
\centering
\includegraphics[width=1.0\textwidth]{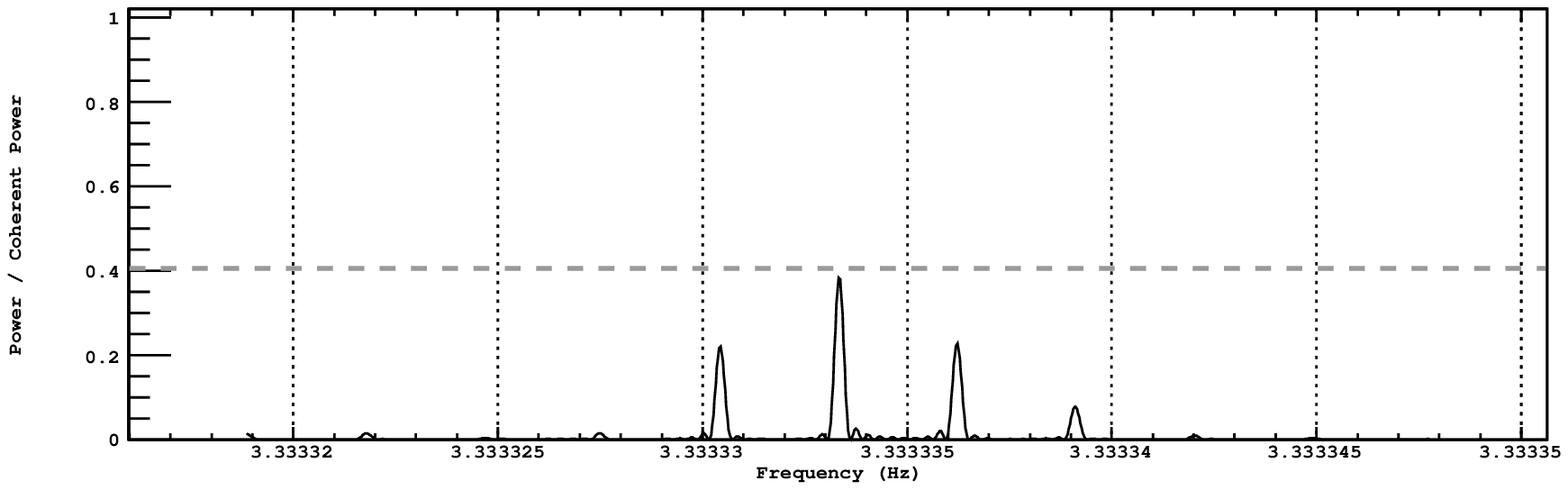}
\end{minipage}
\begin{minipage}[b]{0.9\textwidth}
\centering
\includegraphics[width=1.0\textwidth]{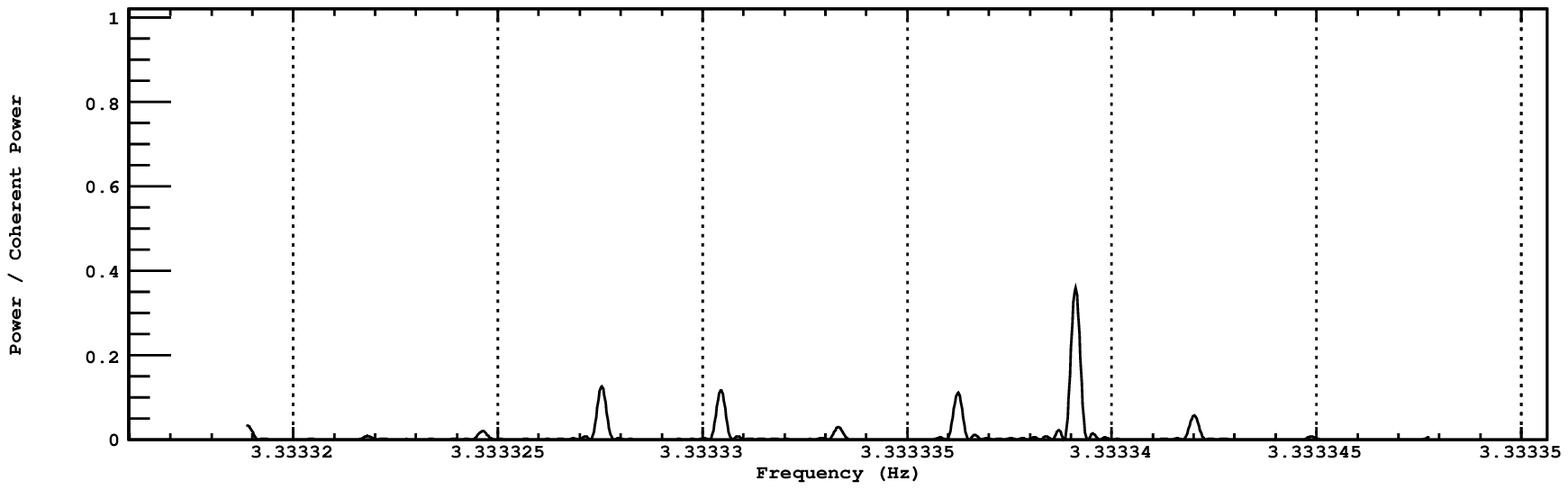}
\end{minipage}
\caption{Power spectra of the simulated time series, demodulated varying the eccentricity by three different amounts $de$. \textit{Top panel}: $de=0.0107$ corresponds to $\omega_0Ade/\pi S = 1/\pi$. The horizontal dashed line indicate the expected value of $\varepsilon^2$.
\textit{Middle panel}: $de=0.0337$, corresponds to $\omega_0Ade/\pi S = 1$, and $\varepsilon^2 = 0.4$ is expected (dashed line).
\textit{Bottom panel}: $de=0.0675$, implies $\omega_0Ade/\pi S = 2$, a null power is expected.
The \textit{Coherent Power} at the denominator of the $y$ axis is the power calculated for an unperturbed demodulation ($de=0$) at the pulsar frequency $\omega_0$. }
\label{fig.decc}
\end{figure*}

To check on all these results we used the arrival time series of the pulsar simulated in Section 6.1.1. The demodulation is performed varying the eccentricity by three different amounts $de$, and the corresponding power spectra are shown in Figure \ref{fig.decc}. In the top panel $de=0.0107$ correspond to $\omega_0Ade/\pi S = 1/\pi$.
In the middle panel $de=0.0337$, corresponding to $\omega_0Ade/\pi S = 1$, and $\varepsilon^2 = 0.4$ as discussed above.
In the bottom panel $de=0.0675$, implying $\omega_0Ade/\pi S = 2$, for which a null power is expected as we can deduce from Figure \ref{EccSurfNorm}.
The ratio of the power to the coherent power of the central peak in the plots of Figure \ref{fig.decc} can again be directly compared with the $\varepsilon^2$ as computationally evaluated in Figure \ref{EccSurfNorm} at the corresponding values of $\omega_0Ade/\pi S$. The very good agreement between the analytical and the simulation results validate the procedure adopted for the eccentricity.

\subsubsection{Epoch of the periastron $T_0$ and simulations}

The treatment of the uncertainty of the epoch of the periastron is similar to the previous case. 
From Eq.~(\ref{Ete}), the partial derivative of the eccentric anomaly $E$ with respect to $T_0$ is
\beq
\frac{\partial E}{\partial T_0} = \frac{-\Omega_{\rm orb}}{1-e \, {\rm cos}E}.
\label{6.3.1}
\enq
This is periodic, and implies that also $\delta k$ and $\psi$ are. 
In particular, the terms $\delta k_i$ and $\psi_i$ in Eq.~(\ref{6.2.2})
contain the partial derivative of $E$ as function of $(2\pi i + \theta)/\omega_0$, 
which is periodic with period $\Omega_{\rm orb}/\omega_0$. 
Since this is the same feature we have found in the case of the eccentricity, we can follow all the steps done
in the previous Section until Eq.~(\ref{6.2.6}).

Also in this case the values $K_{min}$ and $K_{max}$ are not symmetric. Hence the expression outside the integral can not be further simplified to a \textit{sinc} function. However, the function $K_i$ is equal to:
\begin{eqnarray}
 K_i = \delta k_i \cdot {\rm sin}\left(  \frac{\Omega_{\rm orb}}{\omega_0}2\pi i +\phi+\psi_i \right) = 
\nonumber \\
\-\frac{\Omega_{\rm orb} A \sqrt{1-e^2 {\rm cos}^2 W }} {1-e \, {\rm cos} x } dT_0 \cdot {\rm sin}(x+\phi-\pi/2) =
\nonumber \\
\-\frac{\Omega_{\rm orb} A}{1+e^2} \cdot (1+e^2) \sqrt{1-e^2 {\rm cos}^2 W } dT_0 \cdot \frac{{\rm cos}(x+\phi)} {1-e \, {\rm cos} x } =
\nonumber \\
\-\alpha dT_0 \cdot g(x; e, W)
\label{6.3.2}
\end{eqnarray}
where we have set $x=({\Omega_{\rm orb}}/{\omega_0}) 2\pi i$, $\alpha = \Omega_{\rm orb} A/(1+e^2)$, and $g(x; e, W) = (1+e^2) \sqrt{1-e^2 {\rm cos}^2 W } {\rm cos}(x+\phi) / (1-e \, {\rm cos} x)$. We have found useful to introduce the term $(1-e^2)$ in Eq.~(\ref{6.3.2}) for the comparison of the analytical results with several of the simulations we performed. 
Substituting $K_{max}$ and $K_{min}$ in Eq.~(\ref{6.2.6}) with the corresponding maximum and minimum of Eq.~(\ref{6.3.2}), it is clear that $\omega_0 \alpha dT_0$ has a key role in this case, from which we can conclude that the maximum error of the epoch of the periastron ($T_0$) that is allowed to avoid washing out the pulsed signal is inversely proportional to the pulsar frequency ($\omega_0$), and to the projection of the semi-major axis ($A$), while it is directly proportional to the orbital period $P_{\rm orb} = 2\pi / \Omega_{\rm orb}$.
\begin{figure}
\begin{minipage}[b]{0.5\textwidth}
\centering
\includegraphics[width=1\textwidth]{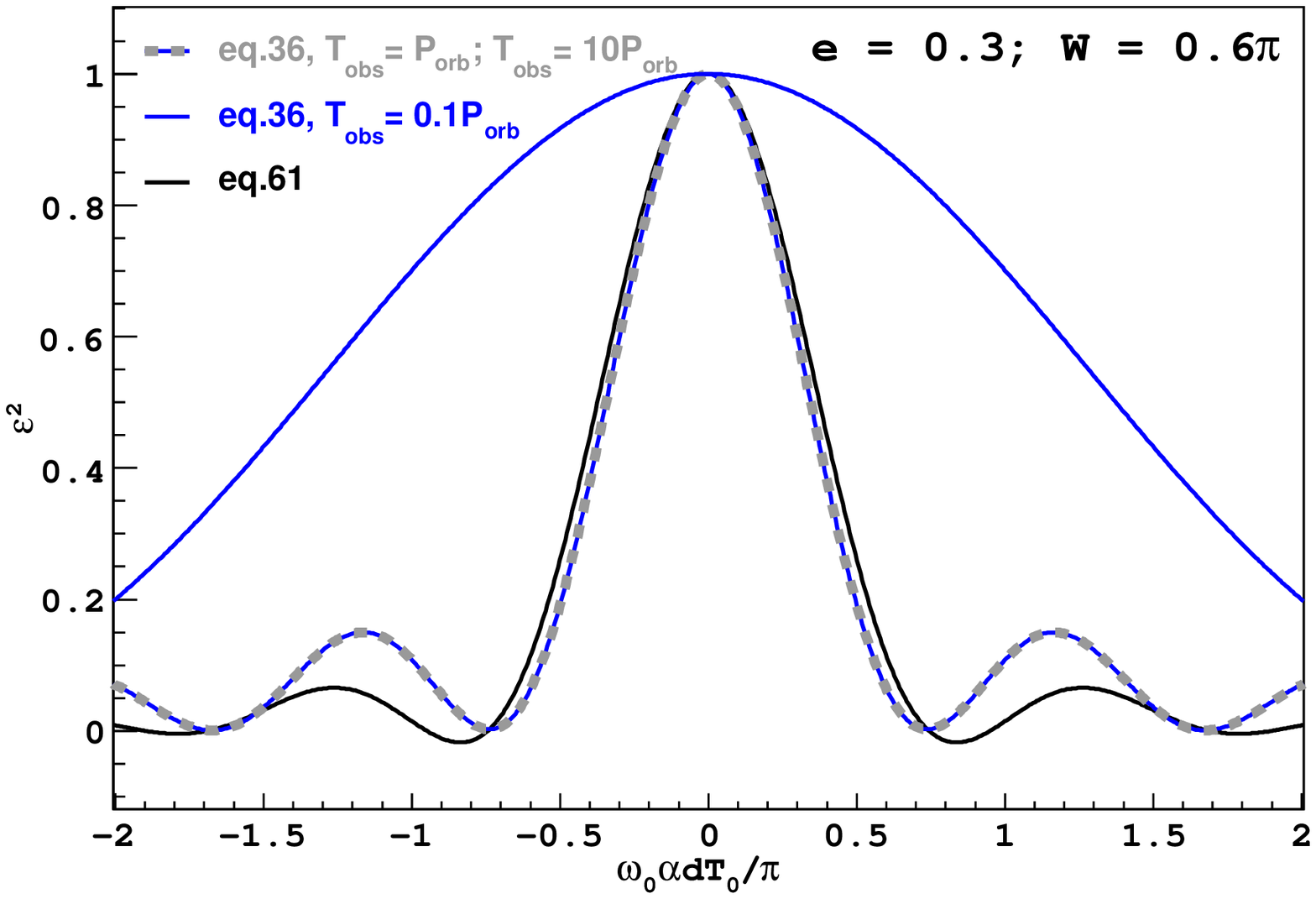}
\end{minipage}
\begin{minipage}[b]{0.5\textwidth}
\centering
\includegraphics[width=1\textwidth]{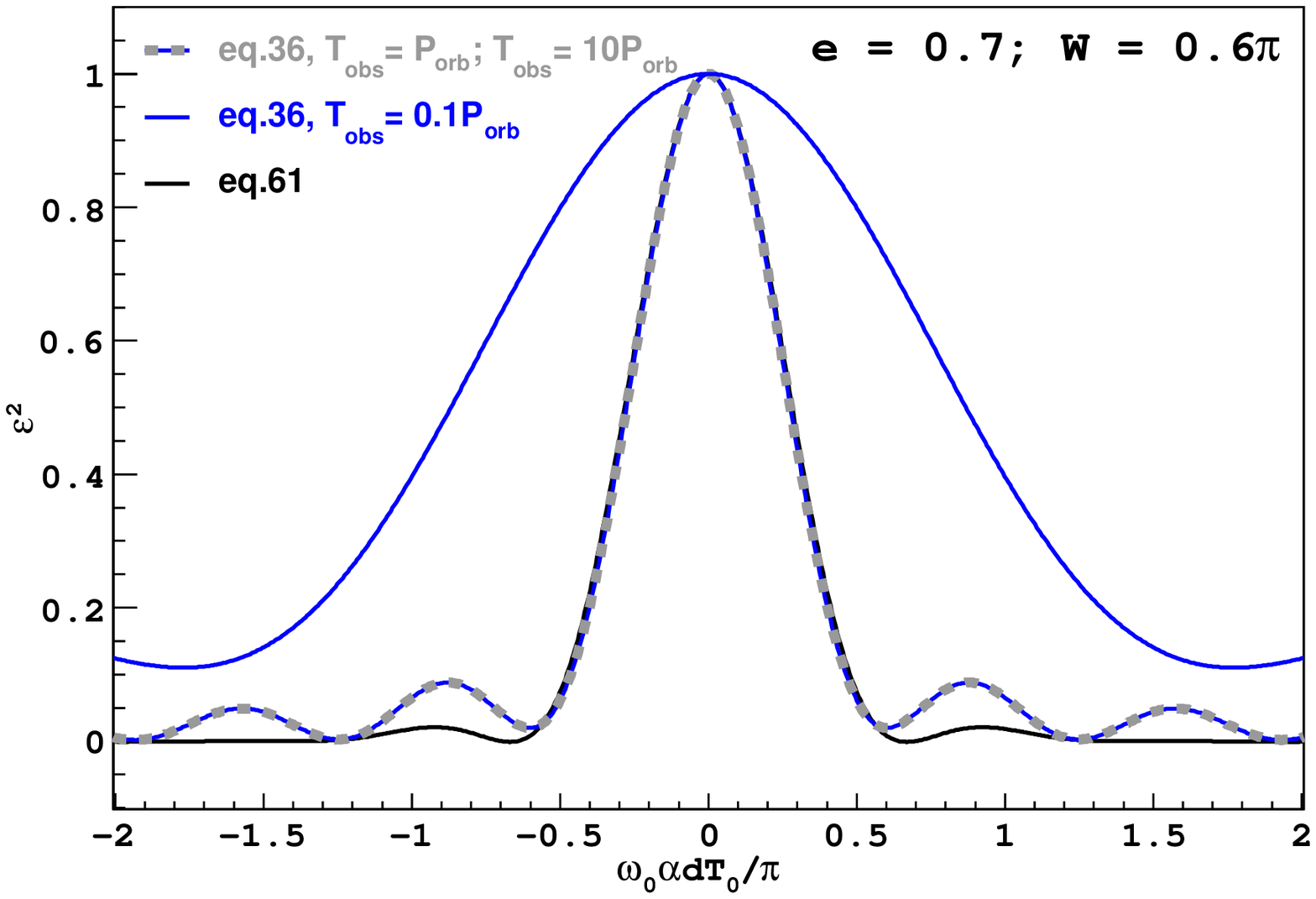}
\end{minipage}
\begin{minipage}[b]{0.5\textwidth}
\centering
\includegraphics[width=1\textwidth]{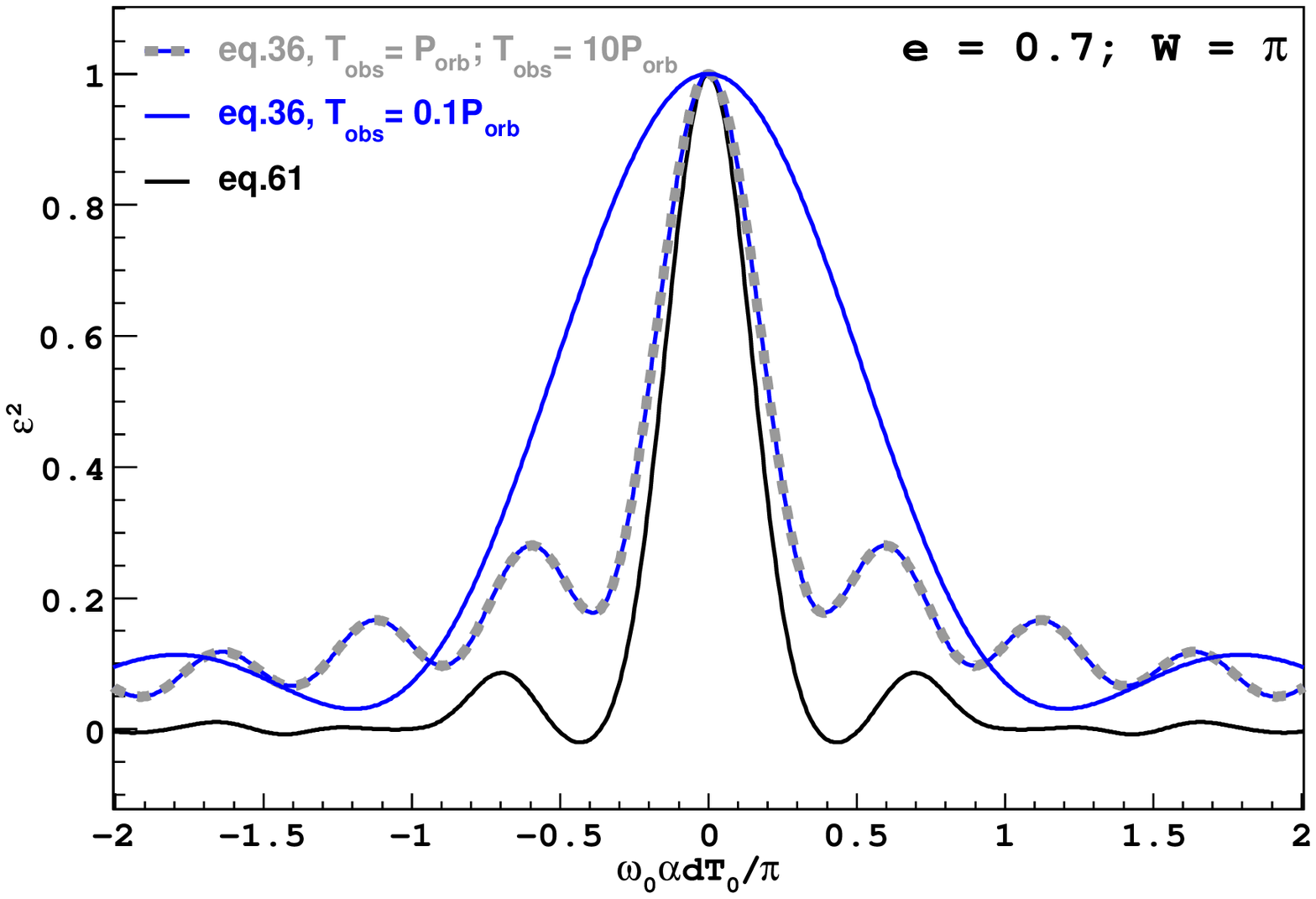}
\end{minipage}
\caption{$\varepsilon^2$ versus $\omega_0 \alpha dT_0/\pi$ as calculated using Eq.~(\ref{6.2.6}), and Eq.~(\ref{5.10}) for the case described in Section 6.2.2. Each panel assumes different values of $e$ and $W$. \textit{Top}: $e=0.3$, $W=0.6\pi$. \textit{Middle}: $e=0.7$, $W=0.6\pi$. \textit{Bottom}: $e=0.7$, $W=1.0\pi$. 
Three different observation times are assumed to compute Eq.~(\ref{5.10}): $T_{\rm obs}=10P_{\rm orb}; P_{\rm orb}$, and $0.1 P_{\rm orb}$. }
\label{Fig-T0.1}
\end{figure}

\begin{figure}
\center
\includegraphics[width=0.5\textwidth]{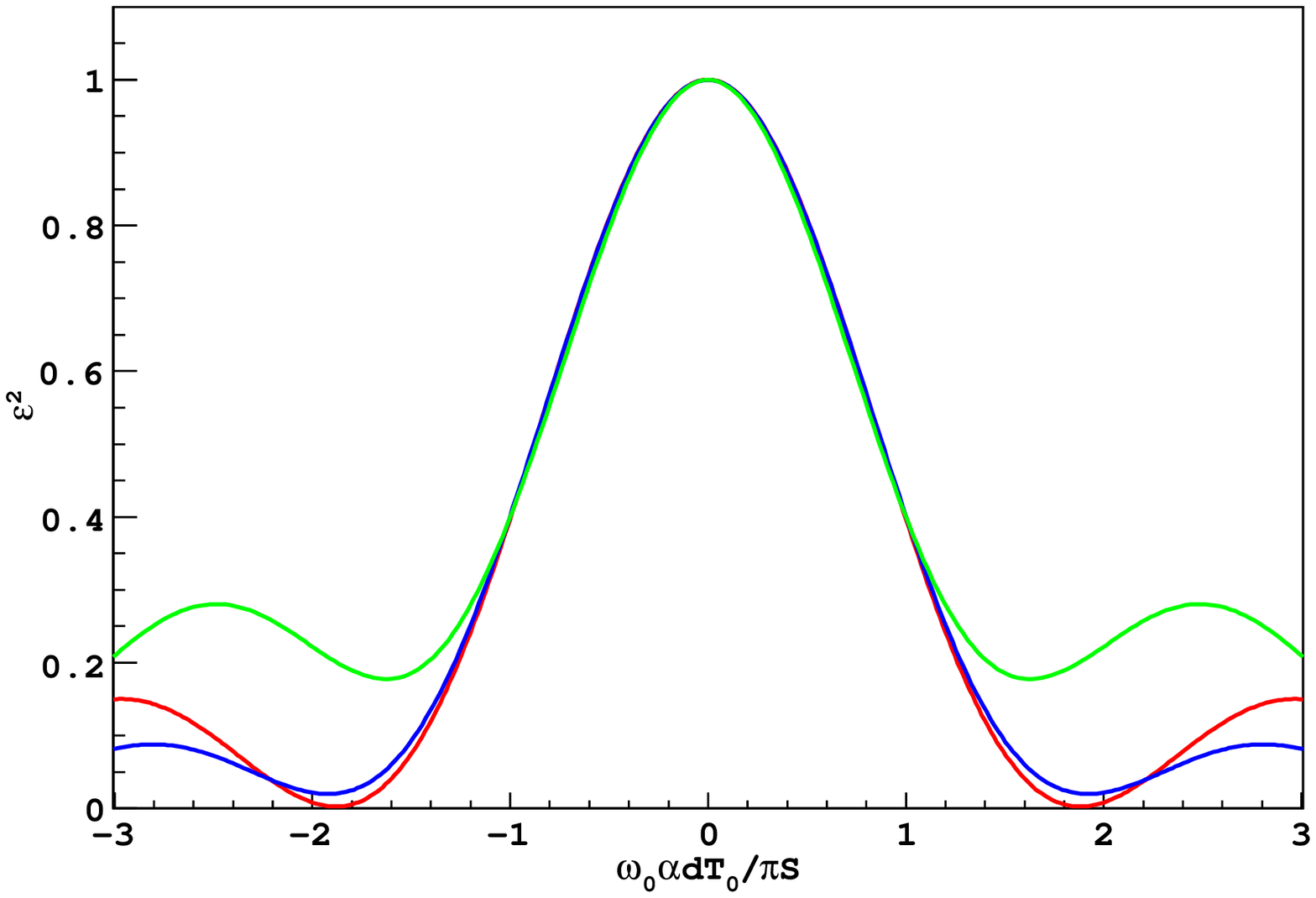}
\caption{$\varepsilon^2$ versus $\omega_0\alpha dT_0/\pi S$, assuming for $e$ and $W$ the same values as for the three plots in Figure \ref{Fig-T0.1}. The values of $\varepsilon^2$ are calculated by means of Eq.~(\ref{5.10}). \textit{Red line}: $e = 0.3$, $W = 0.6\pi$. \textit{Blue line}: $e = 0.7$, $W = 0.6\pi$. \textit{Green line}: $e = 0.7$, $W = \pi$. \label{Fig-T0.2} }
\end{figure}

The panels in Figure \ref{Fig-T0.1} show $\varepsilon^2$ versus $\omega_0 \alpha dT_0$ for different values of $e$ and $W$.
The values of $\varepsilon^2$ are calculated by Eq.~(\ref{6.2.6}), and by Eq.~(\ref{5.10}) for different observation periods . The plots show that there is a good agreement between Eq.~(\ref{6.2.6}) and Eq.~(\ref{5.10}). We note again that $\varepsilon^2$ does not depend on the duration of the observation ($T_{\rm obs}$) when it is greater than one orbital period ($P_{\rm orb}$); but the width of the central peak still has a dependence on $e$ and $W$. To normalize it, we have used the same approach adopted in the case of the eccentricity. Namely, we have found the surface of $\omega_0 \alpha dT_0$ for which $\varepsilon^2 = 0.4$ varying $e$ and $W$, and we have searched for an approximate fitting function, that in this case is
\begin{eqnarray}
S(e,W) &=& U(e) - \frac{{\rm Ampl}(e)}{2} \cdot (1 + {\rm cos}(2W)) , \label{6.3.3} \;\;\;\;\; {\rm with} \\
U(e) &=& U_0 + a \cdot {\rm Log}(b-e^{1.3}) , \nonumber \\
{\rm Ampl}(e) &=& {\rm Ampl}_0 \cdot (1-\frac{e}{2-c})e^{2+c}, \nonumber 
\end{eqnarray}
where $U_0 = 0.410 \pm 0.003$, $a = 0.25 \pm 0.01$, $b=1.07 \pm 0.01$, Ampl$_0=0.45 \pm 0.01$, and $c = 0.68 \pm 0.01$.
In this way we have found that the factor $\varepsilon^2$ is approximately a function of the single variable $z = \omega_0 \alpha dT_0 / \pi S$. This is evident in Figure \ref{Fig-T0.2}, where $\varepsilon^2$ obtained using Eq.~(\ref{5.10}) is plotted versus $\omega_0 \alpha dT_0/\pi S $, assuming for $e$ and $W$ the same values as  in Figure \ref{Fig-T0.1}.

In Figures \ref{Fig-T0.1} (bottom panel) and \ref{Fig-T0.2} (green curve) we can notice that for $W \sim \pi$ the curves of $\varepsilon^2$ obtained by Eq.~(\ref{5.10}) do not reach zero in the neighbourhood of $\omega_0 \alpha dT_0/\pi S = 0$. In particular, when $e \gtrsim 0.9$, the first local minimum closer to the peak of the curve has a value $\varepsilon^2 > 0.4$.
From the simulations we performed we observed that this feature is true also for values of $W \neq \pi$.
%
%
Unfortunately, this feature is lost in the approximations done to pass from Eq.~(\ref{5.10}) to Eq.~(\ref{6.2.6}), as shown in Figure \ref{Fig-T0.1} (black curves). Neglecting this, we can conclude that in order to maintain the factor $\varepsilon^2$ higher than a certain level, the following condition must be satisfied 
\begin{eqnarray}
\frac{\omega_0 \alpha dT_0}{\pi} < [\varepsilon^2]^{-1} \cdot \left\lbrace U(e) - \frac{{\rm Ampl}(e)}{2} \cdot (1 + {\rm cos}(2W)) \right\rbrace  ,
\label{6.3.4}
\end{eqnarray}
where the inverse function $[\varepsilon^2]^{-1}$ can be deduced by Figure \ref{Fig-T0.2}.

\begin{figure*}
\centering
\begin{minipage}[b]{0.9\textwidth}
\centering
\includegraphics[width=1.0\textwidth]{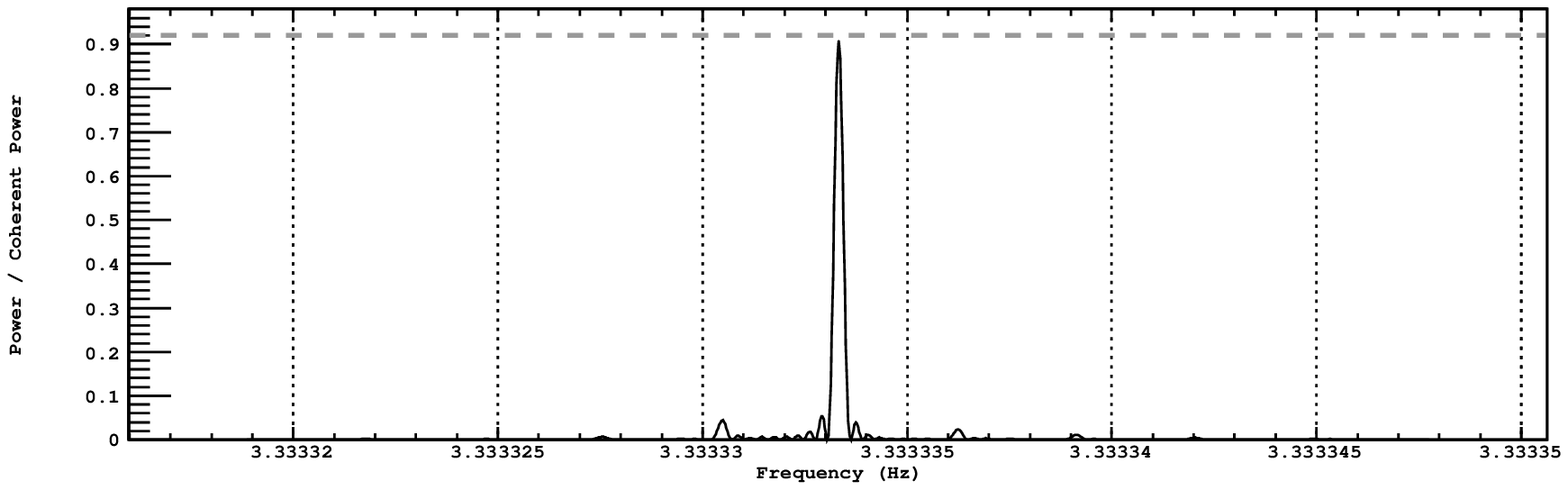}
\end{minipage}
\begin{minipage}[b]{0.9\textwidth}
\centering
\includegraphics[width=1.0\textwidth]{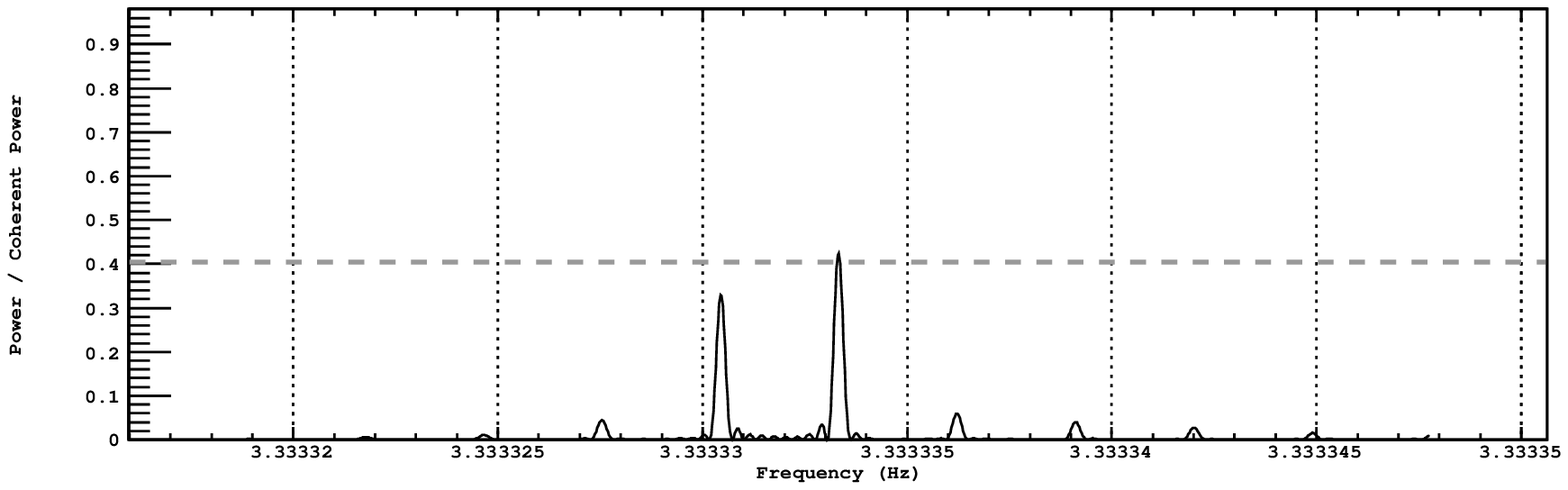}
\end{minipage}
\begin{minipage}[b]{0.9\textwidth}
\centering
\includegraphics[width=1.0\textwidth]{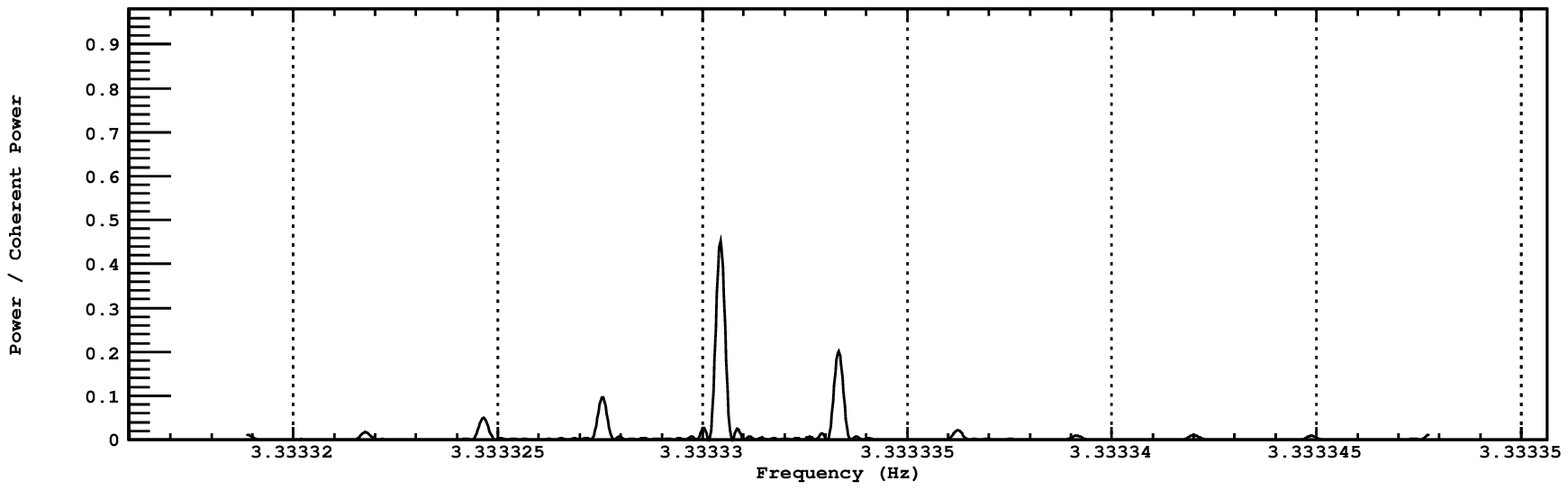}
\end{minipage}
\caption{Power spectra of the simulated time series, demodulated varying the epoch of the periastron by three different amounts $dT_0$. \textit{Top panel}: $dT_0 = 5.2 \times 10^{-3}$ days corresponds to $\omega_0 \alpha dT_0/\pi S = 1/\pi$. The horizontal dashed line indicate the expected value of $\varepsilon^2$.
\textit{Middle panel}: $dT_0 = 1.63 \times 10^{-2}$ days, corresponds to $\omega_0 \alpha dT_0/\pi S = 1$, and $\varepsilon^2 = 0.4$ is expected (dashed line).
\textit{Bottom panel}: $dT_0 = 3.26 \times 10^{-2}$ days, implies $ \omega_0 \alpha dT_0/\pi S= 2$.
The \textit{Coherent Power} at the denominator of the $y$ axis is the power calculated for an unperturbed demodulation ($dT_0=0$) at the pulsar frequency $\omega_0$.}
\label{fig.dT0}
\end{figure*}

To check on these results we have also used the arrival time series of the pulsar simulated in Section 6.1.1. The demodulation is performed varying the epoch of the periastron by three different amounts $dT_0$, and the corresponding power spectra are shown in Figure \ref{fig.dT0}. In the top panel $dT_0 = 5.2 \times 10^{-3}$ days corresponds to $\omega_0 \alpha dT_0/\pi S = 1/\pi$.
In the middle panel $dT_0 = 1.63 \times 10^{-2}$ days, corresponding to $ \omega_0 \alpha dT_0/\pi S = 1$, and $\varepsilon^2 = 0.4$ as discussed above.
In the bottom panel $dT_0 = 3.26 \times 10^{-2}$ days, implying $ \omega_0 \alpha dT_0/\pi S = 2$, for which a null power is expected.
The ratio of the power to the coherent power of the central peak in the plots of Figure \ref{fig.dT0} is in good agreement with the values of $\varepsilon^2$ as computationally evaluated in Figure \ref{Fig-T0.2} at the corresponding values of $\omega_0 \alpha dT_0/\pi S$, with the exception of the bottom panel in Figure \ref{fig.dT0} where the central peak has a power of $\sim 0.2$ against the null value expected. This discrepancy is due to the neglected dependence of the level of the first local minimum of $\varepsilon^2$ by the eccentricity, as explained above.

\subsubsection{Solution for orbital period $P_{\rm orb}$ and simulations}

\begin{figure}
\center
\includegraphics[width=0.5\textwidth]{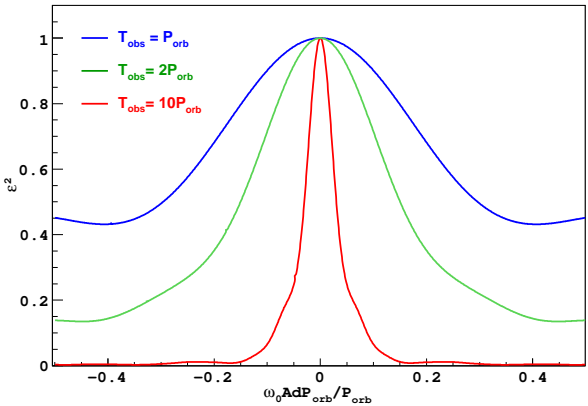}
\caption{$\varepsilon^2$ versus $\omega_0AdP_{\rm orb}/P_{\rm orb}$, assuming $e = 0.5$ and $W = \pi/3$. The values of $\varepsilon^2$ are calculated using Eq.~(\ref{5.10}). Three different observation times are assumed: $T_{\rm obs} = P_{\rm orb}$ (blue line), $T_{\rm obs} = 2P_{\rm orb}$ (green line),  $T_{\rm obs} = 10P_{\rm orb}$ (red line).  }
\label{Pfig1}
\end{figure}

\begin{figure*}
\centering
\begin{minipage}[b]{0.9\textwidth}
\centering
\includegraphics[width=1.0\textwidth]{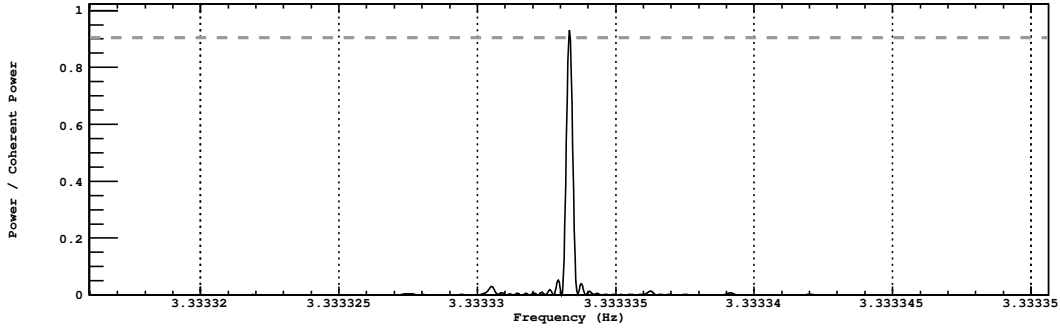}
\end{minipage}
\caption{Power spectra of the simulated time series, demodulated varying the orbital period by $dP_{\rm orb} = 7.6 \times 10^{-4}$ days. The expected value of $\varepsilon^2 = 0.9$ is marked by the horizontal dashed line.
The \textit{Coherent Power} at the denominator of the $y$ axis is the power calculated for an unperturbed demodulation ($dP_{\rm orb}=0$) at the pulsar frequency $\omega_0$.}
\label{fig.dP}
\end{figure*}

From Eq.~(\ref{Ete}), the partial derivative of the eccentric anomaly $E$ with respect to $P_{\rm orb}$ is
\begin{eqnarray}
\frac{\partial E}{\partial P_{\rm orb}} = \frac{\partial E}{\partial \Omega_{\rm orb}} \frac{d \Omega_{\rm orb}}{d P_{\rm orb}} =
- \frac{\Omega_{\rm orb}^2 (t_e-T_0)}{2\pi(1-e \, {\rm cos}E)} = \nonumber \\
\sim - \frac{\Omega_{\rm orb} E}{2\pi(1-e \, {\rm cos}E)}.
\label{6.4.1}
\end{eqnarray}
Eq.~(\ref{6.4.1}) is not periodic. Therefore, the procedure followed in the previous cases can not be applied for the orbital period. 
Eq.~(\ref{6.4.1}) has instead a linear dependence on $t_e-T_0$. As we will show, this implies that the effect of the error $d P_{\rm orb}$ does depend on the duration of the observation $T_{\rm obs}$, in contrast with the other parameter errors.
Substituting Eq.~(\ref{6.4.1}) in Eqs.~(\ref{3.6}), (\ref{fE}), and (\ref{5.4}), we obtain the perturbation function due to the uncertainty of the orbital period
\beq
f_i = -A \frac{dP_{\rm orb}}{P_{\rm orb}} \sqrt{1-e^2 {\rm cos}^2 W} \frac{x \, {\rm cos}(x+\phi)}{1-e \, {\rm cos}x} ,
\label{6.4.2}
\enq
where $x = \frac{\Omega_{\rm orb}}{\omega_0} 2\pi i$. 
The factor $\varepsilon^2$ can be calculated substituting Eq.~(\ref{6.4.2}) in Eq.~(\ref{5.10}). Figure \ref{Pfig1} shows $\varepsilon^2$ versus $\omega_0 A dP_{\rm orb}/P_{\rm orb}$ calculated for three different observation times ($T_{\rm obs} = 1 P_{\rm orb}$, $T_{\rm obs} = 2 P_{\rm orb}$, and $T_{\rm obs} = 10 P_{\rm orb}$), and random values of the $e$ and $W$ ($e = 0.5$, and $W = \pi/3$). The dependence of $\varepsilon^2$ on $T_{\rm obs}$ is evident in Figure \ref{Pfig1}.
We found that close to the peak, the factor $\varepsilon^2$ is roughly approximated by
\beq
\varepsilon^2 = 1 - 10 n^2 \left( \omega_0 A \frac{dP_{\rm orb}}{P_{\rm orb}} \right)^2 ,
\label{6.4.3}
\enq
where $n = T_{\rm obs}/P_{\rm orb}$ is the number of orbital periods observed. 
Applying Eq.~(\ref{6.4.3}) to the pulsar simulated in Section 6.1.1 (see Table \ref{Tab1}), a decrease of the power by a factor $\varepsilon^2=0.9$ is expected 
if the arrival time series are demodulated varying the orbital period by $dP_{\rm orb} = 7.6 \times 10^{-4}$ days.
This is confirmed in Figure \ref{fig.dP}.

In conclusion, the condition to maintain $\varepsilon^2$ higher than a certain level (say $\varepsilon^2>0.9$)
can be derived by Eq. ~(\ref{6.4.3}) as
\beq
\frac{dP_{\rm orb}}{P_{\rm orb}} < \frac{1}{n \omega_0 A} \sqrt{\frac{1-\varepsilon^2}{10}}.
\label{6.4.4}
\enq
Eq.~(\ref{6.4.4}) shows that the maximum error $dP_{\rm orb}$ useful to maintain the loss in power $\varepsilon^2$ above a certain value is inversely proportional to the pulsar frequency ($\omega_0$), to the projection of the semi-major axis $A$, and also to the lenght of the observation $T_{\rm obs} = n P_{\rm orb}$.

\section{Discussion and conclusions}

We have presented an analytical study aimed 
to understand the impact of the uncertainties of the orbital parameters on the pulsation searches. We  validated the analytical study with numerical simulations. 
We especially focused on the cases where the observations span a time longer than the orbital period of the system. This is the usual case, for instance, in GeV observations with the {\it Fermi}-LAT gamma-ray satellite.

The search for pulsations is performed by calculating the power spectrum of the emission times, which are evaluated demodulating the arrival time series. But the errors on the orbital parameters unavoidably leads to a wrong estimation of the R\"{o}mer delay, and consequently to an uncertain demodulation. 
Here, we have studied the relation between the exact and the imperfect emission time series considering that the latter are a perturbation of the former. Within this frame, we have defined the perturbation function (Eq. \ref{fte}), and have found how it affects the probability density distribution of the phases attributed to each event time assuming a generic frequency (Eq. \ref{4.11}). 

The power spectrum calculated at the pulsar frequency is reduced by a factor $\varepsilon^2$ when the demodulation is not correct. Eq.~(\ref{5.10}) describes the relation among the perturbation function and the factor $\varepsilon^2$.
Starting from Eq.~(\ref{5.10}), we have analyzed, in a  case by case basis, the impact of the uncertainties of each of the orbital parameters. Figure \ref{fig2}, and Eq.~(\ref{6.1.13}) show the behavior of the power loss, $\varepsilon^2$, in the cases of the semi-major axis $A$ and the longitude of the periastron $W$. The maximum value of the error on the semi-major axis $A$ in order to maintain the factor $\varepsilon^2$ larger than a certain amount is explicitly given by Eq.~(\ref{6.1.1.2}), while in the case of $W$ it can be deduced from Eq.~(\ref{6.1.16}).
The maximum error for the eccentricity and the epoch of the periastron are given by Eq.~(\ref{6.2.9}) and Eq.~(\ref{6.3.4}), respectively. Whereas for the orbital period, the maximum error can be estimated by Eq.~(\ref{6.4.4}). All this is summarized in Table \ref{stepTab}.

\begin{table*}
\begin{minipage}{\textwidth}
\centering
  \caption{Derived formulae to calculate the maximum value of the error on the parameters in order to maintain the loss factor $\varepsilon^2$ larger than a certain amount. For the maximum error $de$ ($dT_0$) the functions Ampl$(e)$ and $U(e)$ are defined in Eqs. (\ref{6.2.8}) (Eqs. (\ref{6.3.3}), respectively).  For the maximum error on $P_{\rm orb}$, we use $n=T_w/P_{\rm orb}$. In the first column $\left[ \varepsilon^2\right]^{-1}$ means the inverse function of $\varepsilon^2$ given in the second column. For the cases we do not found an explicit formula for $\varepsilon^2$ (labeled with $-$) the inverse value can be deduced by the figure listed in the third column. As explained in Section 7.1, in sampling the space of the parameters, the maximum step is defined as twice the maximum error given in the first column.} 
  \begin{tabular}{lll}
  \hline
  Maximum error & $\varepsilon^2$ function & $\varepsilon^2$ plot \\
  \hline
   $dA < \frac{\pi \left[ \varepsilon^2\right]^{-1}}{\omega_0 \sqrt{1-e^2 \cos(W)^2}}$  &  
$\varepsilon^2 = \left[ \frac{ {\rm sin}(\pi x) }{\pi x}\right]^2 \left[1-2\left( \pi x \right) ^2 \right]$ & Figure \ref{fig2} \\

   $ dW < \frac{\pi \left[ \varepsilon^2\right]^{-1}}{\omega_0 A  \sqrt{\frac{1-e^2+e^4{\rm cos}^2W {\rm sin}^2W}{1-e^2 \cos(W)^2}}} $  & $\varepsilon^2 = \left[ \frac{ {\rm sin}(\pi x) }{\pi x}\right]^2 \left[1-2\left( \pi x \right) ^2 \right]$ & Figure \ref{fig2}\\

   $ de < \frac{\pi [\varepsilon^2]^{-1} \cdot \left\lbrace U(e) + {\rm Ampl}(e) \cdot | {\rm cos}W | \right\rbrace}{\omega_0 A} $  & $-$ & Figure \ref{EccSurfNorm}\\

   $ dT_0 < \frac{\pi [\varepsilon^2]^{-1} \cdot \left\lbrace U(e) - \frac{{\rm Ampl}(e)}{2} \cdot (1 + {\rm cos}(2W)) \right\rbrace}{\omega_0 \Omega_{\rm orb} A / (1+e^2)}$   & $-$ & Figure \ref{Fig-T0.2}\\

   $ dP_{\rm orb} < \frac{P_{\rm orb}}{n \omega_0 A} \sqrt{\frac{1-\varepsilon^2}{10}} $& $ \varepsilon^2 = 1 - 10 n^2 ( x )^2 $ & Figure \ref{Pfig1}\\
\hline
\end{tabular}
\label{stepTab}
\end{minipage}
\end{table*}

The results discussed so far concern  the case in which only a single parameter is affected by a significant uncertainty. 
The total loss factor that reduce the power of the coherent signal is expected to be the product of the factors calculated for each single parameter:
\begin{equation}
\varepsilon^2 = \varepsilon^2_A \cdot \varepsilon^2_W \cdot \varepsilon^2_e \cdot \varepsilon^2_{T_0} \cdot \varepsilon^2_{P_{\rm orb}}.
\label{7.1}
\end{equation}
To understand this suppose  that only the value of $A$ differs from the true one. The measured power at the pulsar frequency is then reduced by $P(\omega_0, dA) =  \varepsilon^2_A P(\omega_0, 0)$. If from this starting condition the value of $W$ is varied by $dW$, we expect a further reduction of the power, such that $P(\omega_0, dA, dW) =  \varepsilon^2_W P(\omega_0, dA) = \varepsilon^2_W \cdot \varepsilon^2_A P(\omega_0, 0)$. This reasoning leads to Eq.~(\ref{7.1}). 
Anyway, being this only an intuitive demonstration, we can not exclude that combinations of the uncertainties on different parameters can deviate from the expected behavior of the total factor $\varepsilon^2$. In this sense, Eq.~(\ref{7.1}), and the results that follow in this discussion have to be considered as conservative.
In particular, we have not taken into account possible correlations among the parameter uncertainties. 
For example, the uncertainties of the epoch and the longitude of the periastron ($dT_0$ and $dW$) are likely correlated, and the correlation is probably regulated by the orbit eccentricity.

\subsection{The case of LS 5039}

\begin{table*}
  \caption{Maximum steps of the sampling of the orbital parameters calculated for LS 5039. In the last row the total number of steps needed to cover the $\pm 1 \sigma$ uncertainty ranges are given.}
  \begin{tabular}{l l ccc}
\multicolumn{5}{c}{LS 5039} \\
  \hline
   \multicolumn{2}{c}{Most updated parameters}  &   \multicolumn{3}{c}{Maximum steps to have the total factor $\varepsilon^2>0.36$}  \\ 
\cline{3-5}
  \multicolumn{2}{c}{\citep{Aragona2009}}  & $P_{psr}=300$ ms & $P_{psr}=30$ ms & $P_{psr}=3$ ms \\
 \hline
 $P_{\rm orb}$ (d) &   3.90608 $\pm$ 0.0001          & $-$  & $-$  & 1.1$\times 10^{-4}$  \\
 $T_0$ (MJD) &   52825.985 $\pm$ 0.053    & 1.2$\times 10^{-2}$ & 1.2$\times 10^{-3}$ & 1.2$\times 10^{-4}$  \\
 $A   $ (lt-s)     &  3.33 $\pm$ 0.15     & 6.1$\times 10^{-2}$ & 6.1$\times 10^{-3}$  & 6.1$\times 10^{-4}$  \\ 
 $W   $ (deg)      &  236.0 $\pm$ 5.8     & 1.1                 & 1.1$\times 10^{-1}$  & 1.1$\times 10^{-2}$  \\ 
 $e   $            & 0.337 $\pm$ 0.036    & 3.5$\times 10^{-2}$ & 3.5$\times 10^{-3}$ & 3.5$\times 10^{-4}$ \\ 
 \multicolumn{2}{l}{Total number of trials} &  990 & 9.89$\times 10^{6}$ & 9.73$\times 10^{10}$ \\

\hline
\end{tabular}
\label{LS5039tab}
\end{table*}

The most recent estimate of the orbital parameters of LS 5039 \citep{Aragona2009} are reported in the first column of Table \ref{LS5039tab}. With these uncertainties, we find that the detection of a pulsar with a period faster than some seconds is, regrettably, still very unlikely. 
A search for pulsations implies sampling the space of the orbital parameters in order to scan the range given by the current uncertainties. The results found in this work are useful to define the maximum step 
in the parameter sampling that guarantees 
a loss factor $\varepsilon^2$ greater than a pre-defined value $\varepsilon^2_{\rm min}$. 
The maximum step for each parameter is defined as twice the maximum error calculated by Eqs.~(\ref{6.1.1.2}), (\ref{6.1.16}), (\ref{6.2.9}), (\ref{6.3.4}), (\ref{6.4.4}), for $A$, $W$, $e$, $T_0$, $P_{\rm orb}$, respectively. 
Indeed, the maximum error $dp$ is defined as the distance from the true value of the parameter, 
where the worst case that can happen with the sampling is that the true value of a parameter is right at the center of a step. Thus, the full length of the step is equal to $2 dp$, and at both edges of the step, the loss factor is equal to the pre-defined value $\varepsilon^2_{\rm min}$.  
In all the other cases one of the two edges of the step measure a loss factor $\varepsilon^2 > \varepsilon^2_{\rm min}$.

Table \ref{LS5039tab} shows the maximum steps that should be taken to search for a young pulsar with slow (300 ms), and fast (30 ms) period, as well as for a millisecond pulsar (with period 3 ms). 
The steps in the table are calculated such that the loss factors for the parameters $A$, $e$, $W$, and $T_0$ are $\varepsilon^2_p > 0.8$. 
We will suppose to perform a FFT analysis with a time windows equal to the orbital period ($T_w = P_{\rm orb}$). Therefore,
the maximum steps for $P_{\rm orb}$ are calculated setting $n=1$ in Eq.~(\ref{6.4.4}), and such that its loss factor is $\varepsilon^2_{P_{\rm orb}} > 0.9$. Indeed, we can be more tight with the orbital period, because commonly it is the parameter with the smallest uncertainty with respect to the others. 
With the steps defined in this way, using Eq.~(\ref{7.1}) we see that the total loss factor is $\varepsilon^2 \gtrsim 0.36$.

The last row of Table \ref{LS5039tab} gives the total number of trials needed to cover the $\pm 1 \sigma$ uncertainty ranges on the orbital parameters.
For instance, for $P_{\rm psr} = 300$ ms the uncertainty range is covered by 9 steps of $T_0$, 5 steps of $A$, 11 steps of $W$, and 2 steps of $e$. No trials are needed for $P_{orb}$, because its $\pm 1 \sigma$ uncertainty ranges is shorter than its maximum step.
The total number of trials of orbital parameters is then the multiplication of the former steps, $N_{\rm op} = 990$. 
For all the parameters, the steps have an inverse dependence with the pulsar frequency.

\subsection{The case of LS I +61 303}

\begin{table*}
\begin{minipage}{\textwidth}
\centering
  \caption{Maximum steps of the sampling of the orbital parameters calculated for LS I 61 +303. (The uncertainty of the orbital period is from \citet{Gregory2002}). In the last row the total number of steps needed to cover the $\pm 1 \sigma$ uncertainty ranges are given.}
  \begin{tabular}{l l cccc}
\multicolumn{6}{c}{LS I +61 303} \\
  \hline
   \multicolumn{2}{c}{Most updated parameters}  &   \multicolumn{4}{c}{Maximum steps to have the total factor $\varepsilon^2>0.36$}  \\ 
\cline{3-6}
  \multicolumn{2}{c}{\citep{Aragona2009}} & $P_{psr}>2$ s & $P_{psr}=300$ ms & $P_{psr}=30$ ms & $P_{psr}=3$ ms \\
 \hline
$P_{\rm orb}$ (d) &   26.4960 $\pm$ 0.0028  &  $-$    & $-$    & 0.0013  & 1.3$\times 10^{-4}$ \\
$T_0$ (MJD) &   51057.39 $\pm$ 0.23         &   0.090 & 1.3$\times 10^{-2}$   & 1.3$\times 10^{-3}$  & 1.3$\times 10^{-4}$ \\
$A   $ (lt-s)     &  20.0 $\pm$ 1.2         &   0.44  & 6.6$\times 10^{-2}$  & 6.6$\times 10^{-3}$  & 6.6$\times 10^{-4}$   \\
$W   $ (deg)      &  40.5 $\pm$ 5.7         &    1.2  & 1.8$\times 10^{-1}$  & 1.8$\times 10^{-2}$  & 1.8$\times 10^{-3}$  \\
$e   $            & 0.537 $\pm$ 0.034       &   0.032 & 4.8$\times 10^{-3}$  & 4.8$\times 10^{-4}$  & 4.8$\times 10^{-5}$  \\
\multicolumn{2}{l}{Total number of trials} &  450 & 1.06$\times 10^{6}$ & 4.41$\times 10^{10}$ & 4.86$\times 10^{15}$\\
\hline
\end{tabular}
\label{LSItab}
\end{minipage}
\end{table*}

We calculated the maximum steps of the orbital parameters also for LS I +61 303,
as reported in Table \ref{LSItab}. For this system the steps of $W$, and $e$  are much smaller than for LS 5039, because the orbit of LS I +61 303 is much larger.
We calculated here too the possibility to detect a very slow pulsar with period greater than 2 s (see, e.g., \citet{TorresRea2012}). 

\subsection{Simulations} 

To have a feeling on how the calculated steps of the orbital
parameters can allow (or not) the plausible detection of pulsations in
gamma-ray energies, 
we simulated LS I +61 303 as it was observed by {\it Fermi}-LAT for 2 years, 
assuming a 300 ms pulsar as the compact object.
We suppose to analyse the LAT data
with an aperture photometry technique --collecting all the events within an
angular separation-- of 2.4 deg from the source, as done in \citet{LSIfermi}. 
This implies an event rate of 30 counts
per day, and the background level due to the diffuse gamma-ray
emission and the nearby sources approximately equal to 2/3 of the
total counts. We also conservatively assume that the pulse fraction is 50\%.
The barycentered arrival time series was simulated following the same procedure described in Section 6.1.1, with the only difference that instead using Eq.~(\ref{6.1.1.3}) for the distribution of the emitted times in the pulsar reference frame, we used 
\begin{equation}
 P_t(t) = 1 + \eta \cdot {\rm sin}(\omega_0 t),
 \label{7.2}
\end{equation}
where the factor $\eta$ take into account both the background level, and the pulsed fraction of the source, so that in this case we have $\eta = (1 - 2/3) \times 50\% = 1/6$. The pulsar frequency is $\omega_0 = 3.33$ Hz.
The simulated arrival time series have been analysed using the PRESTO
software. In order to demodulate the time series, and to fold the
photons, this software need the orbital parameters as input, as well
as an initial guess of the pulsar frequency and its derivative. 

\begin{figure*}
\centering
\begin{minipage}{0.33\textwidth}{\tt a) Correct parameters\\}
\includegraphics[width=0.98\textwidth]{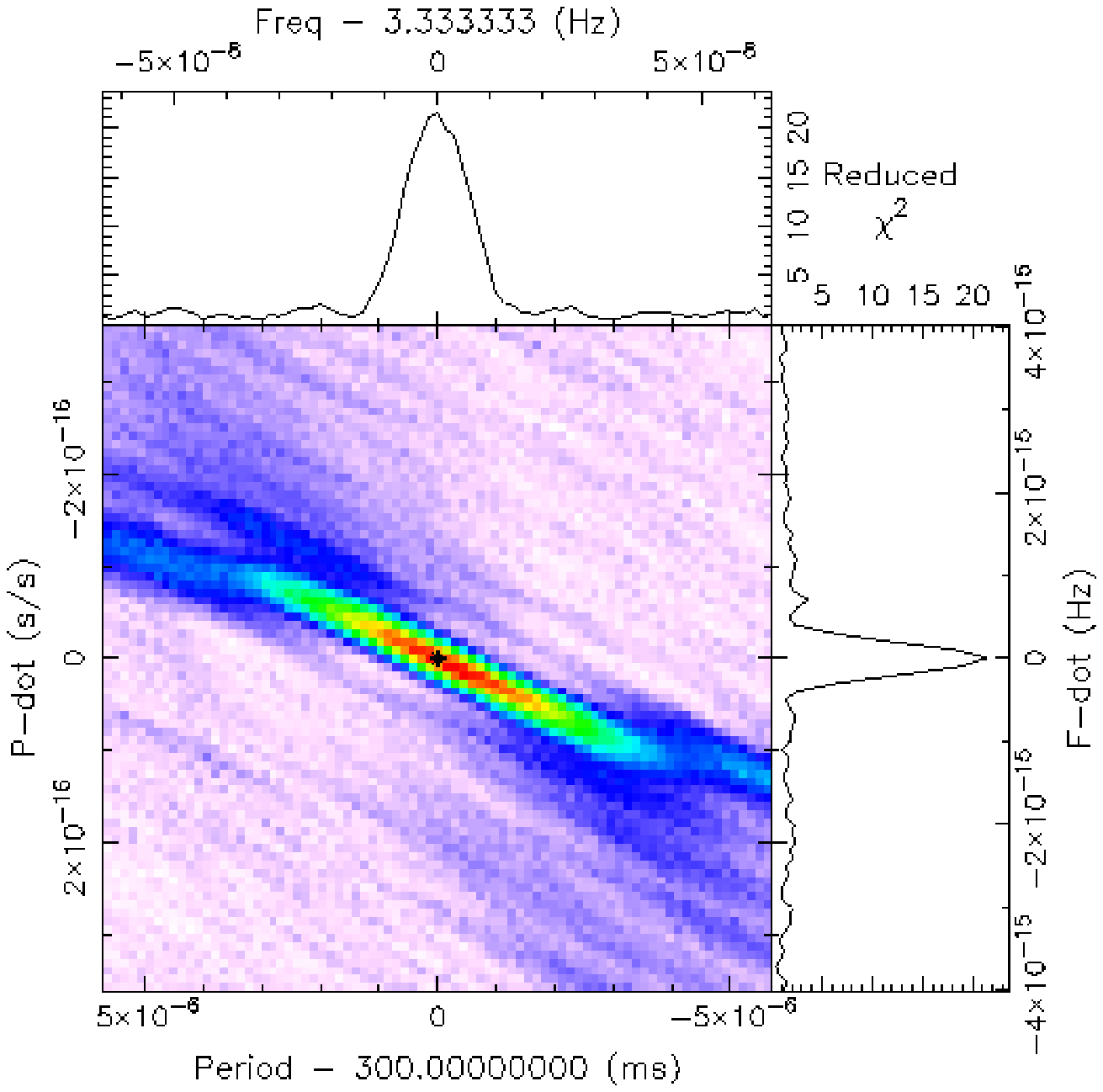}
\end{minipage}
\begin{minipage}{0.33\textwidth}{\tt b) Current uncertainties added\\}
\includegraphics[width=0.98\textwidth]{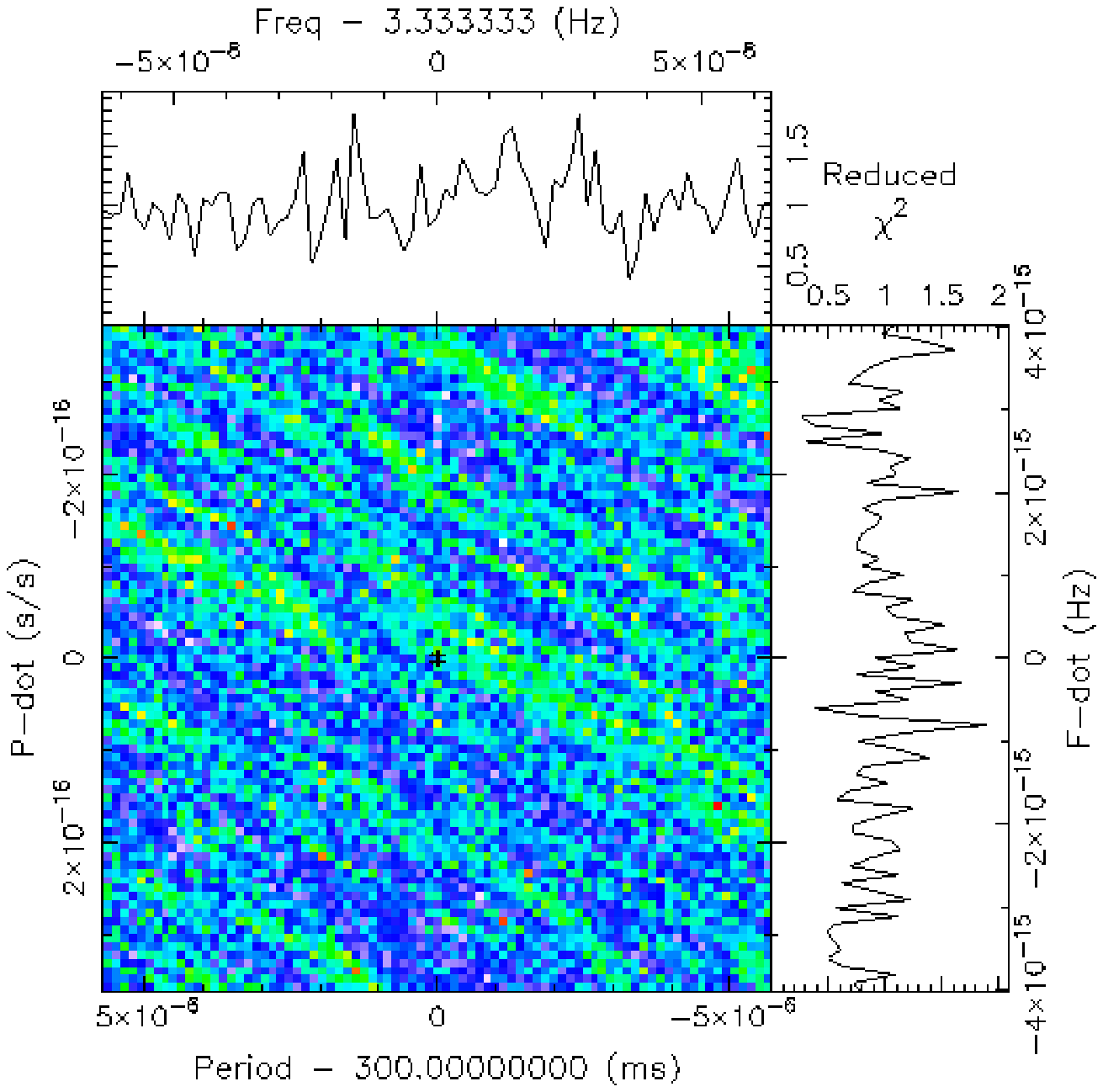}
\end{minipage}
\begin{minipage}{0.33\textwidth}
\hspace{1.0\textwidth}%
\end{minipage}
\vspace{0.3 cm}\\
\begin{minipage}{0.33\textwidth}{\tt c) 1$\times$dp$_{\tt max}$ added\\}
\includegraphics[width=0.98\textwidth]{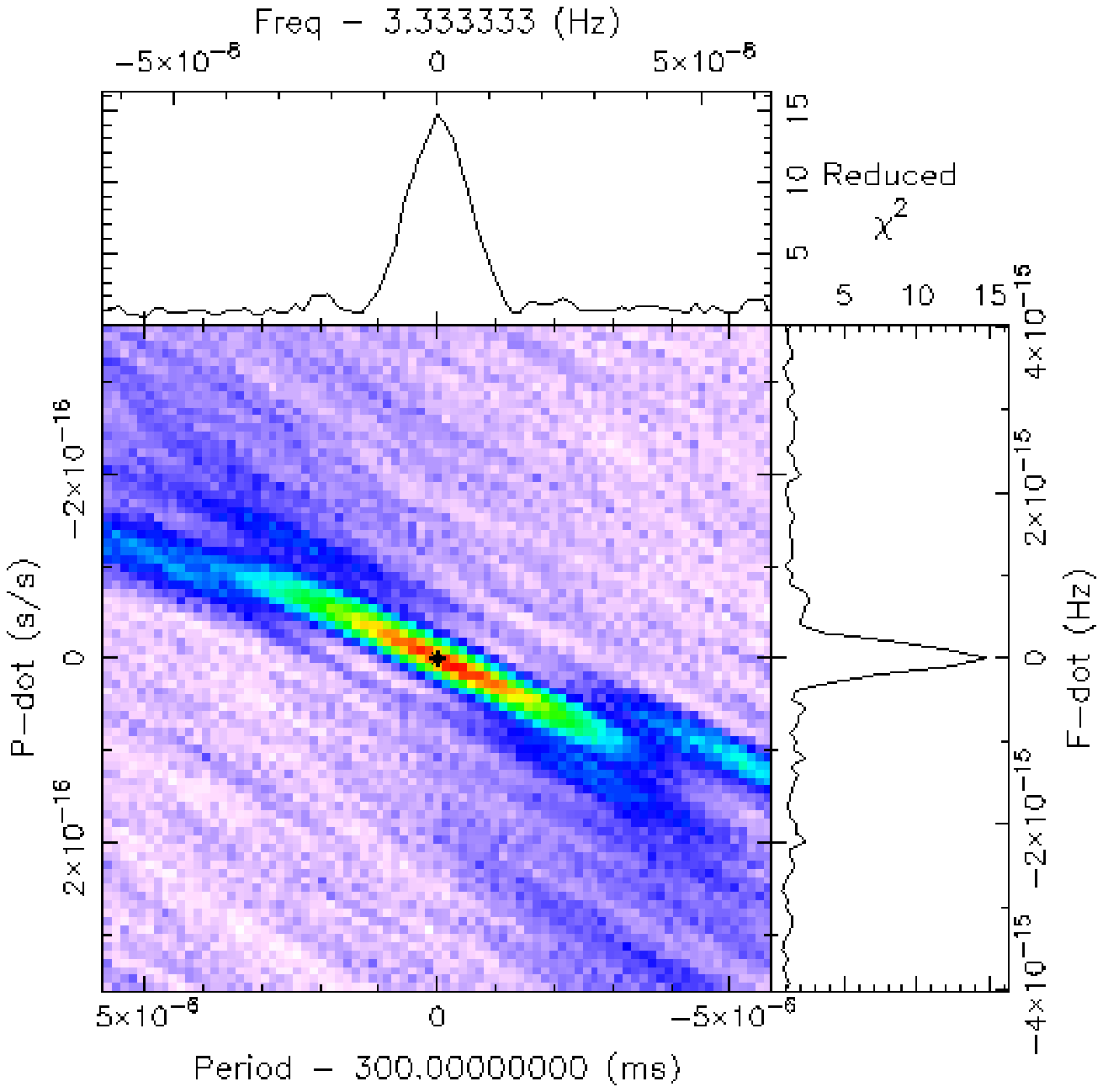}
\end{minipage}
\begin{minipage}{0.33\textwidth}{\tt d) 2$\times$dp$_{\tt max}$ added\\}
\includegraphics[width=0.98\textwidth]{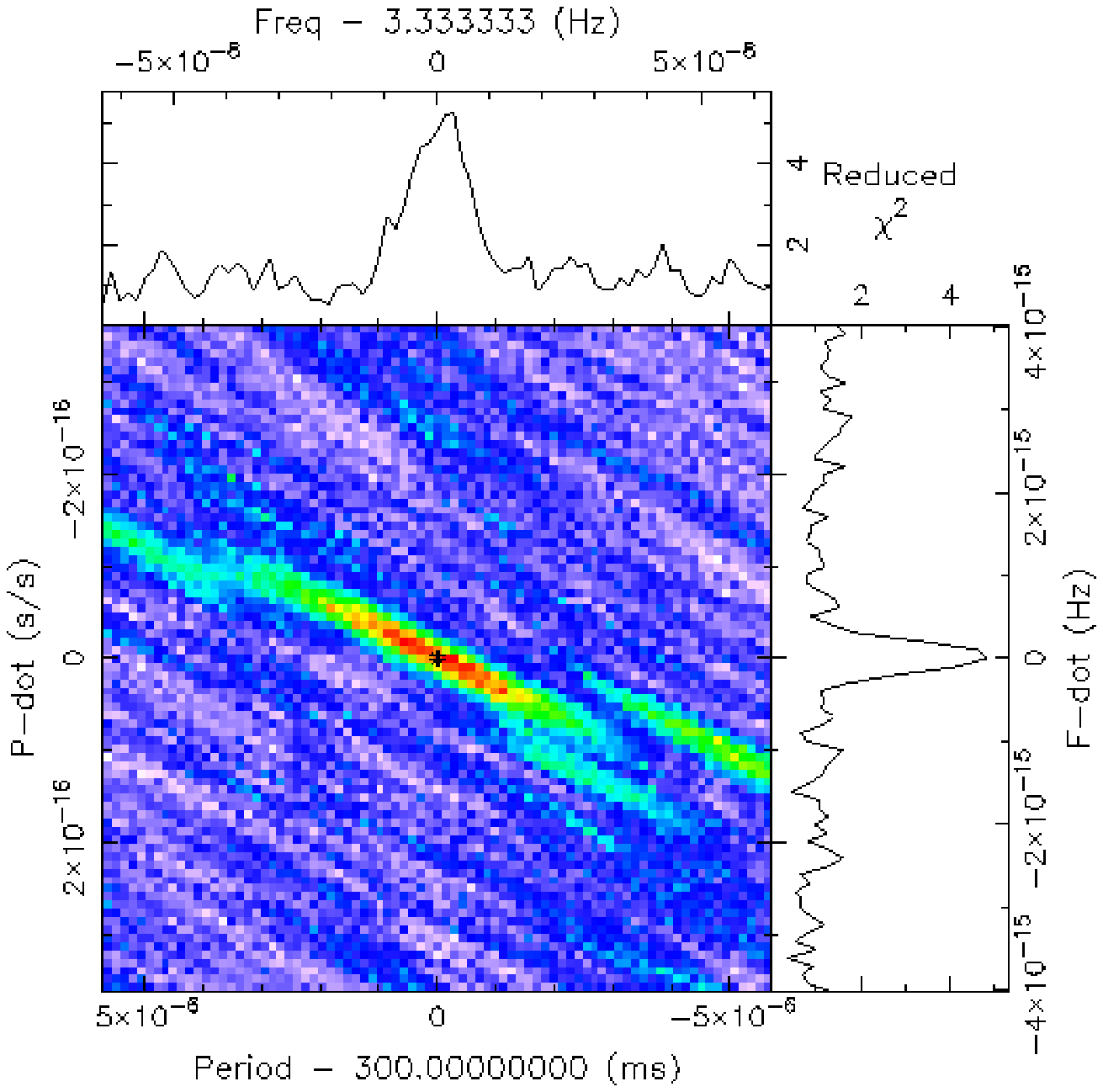}
\end{minipage}
\begin{minipage}{0.33\textwidth}{\tt e) 3$\times$dp$_{\tt max}$ added\\}
\includegraphics[width=0.98\textwidth]{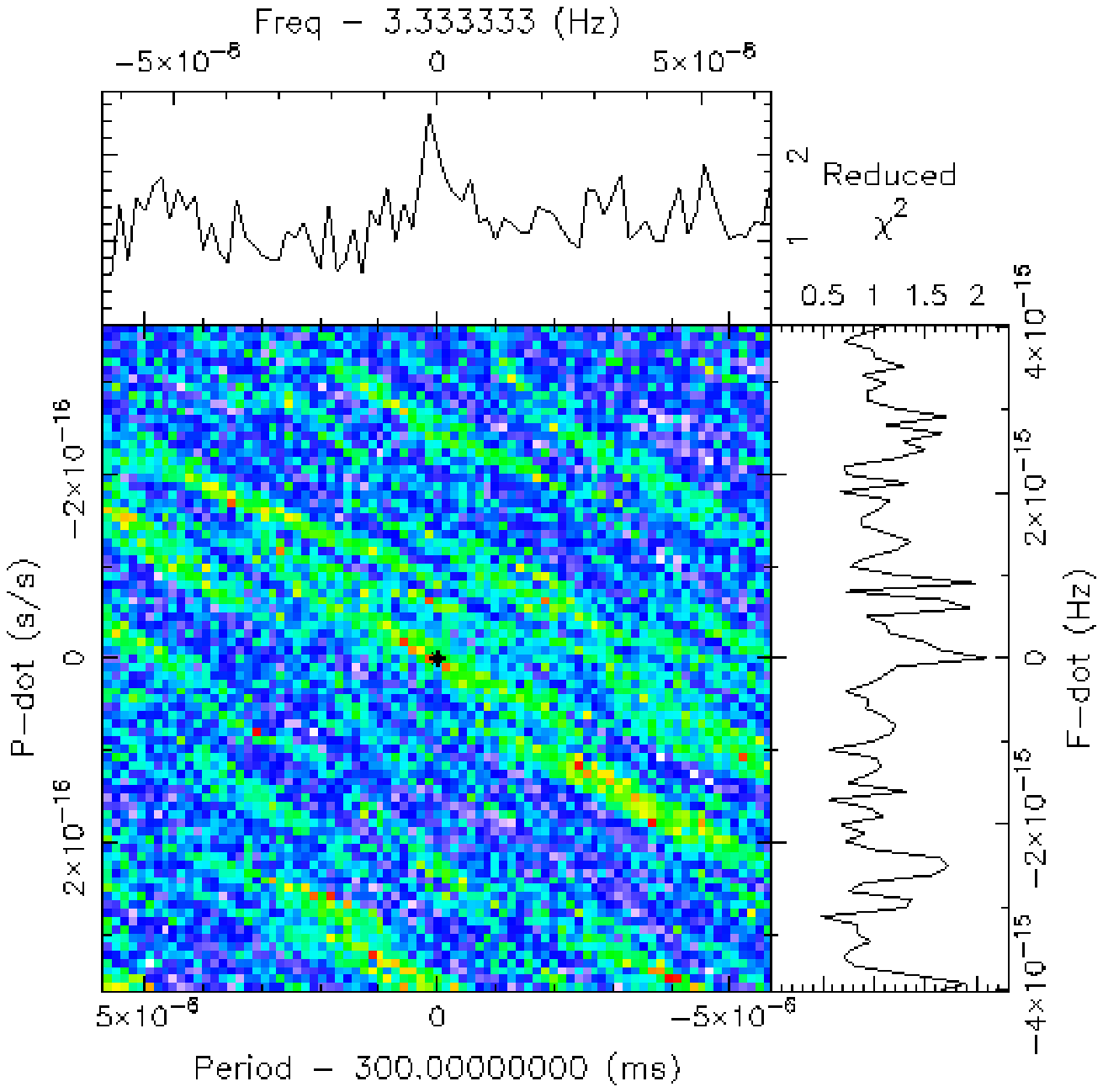}
\end{minipage}
\caption{Detection of the simulated pulsar in LS I +61 303. A 300 ms pulsar with 50\% pulse fraction and 2 years of {\it Fermi}-LAT data have been assumed in the simulation. {\it Panel a)}: Search for the most probable values of the pulsar period and its first derivative with PRESTO, after demodulating the arrival time series with the correct orbital parameters. {\it Panel b)}: The same but after demodulating the arrival time series using parameters that deviate from the correct ones by an amount equal to the current uncertainties in \citet{Aragona2009}. {\it Panel c)}: The same but with the demodulation performed using the orbital parameters $A$, $e$, and $W$ that deviate from the correct ones by half of the maximum steps  defined in table \ref{LSItab} (${\rm step}_{\rm max}$), or equivalently by their maximum errors $dp_{\rm max} = {\rm step_{max}/2}$. {\it Panel d)}: The same but the orbital parameters $A$, $e$, and $W$ deviate by 2 times the maximum errors $dp_{\rm max}$. {\it Panel d)}: and 3 times the maximum errors $dp_{\rm max}$. \label{Pplots}}
\end{figure*}

Figure \ref{Pplots} 
shows the 
typical plots produced by PRESTO. They show the most probable 
value of the pulsar period and its first derivative searched over a grid. 
In panel $a)$ of Figure \ref{Pplots} the demodulation of the arrival time series is performed using the correct values of the orbital parameters, while in panel $b)$ the current uncertainties from \citet{Aragona2009} are added. Whereas the first panel shows that the signal is clearly visible after a correct demodulation, the second regrettably shows that it is
completely lost with the current level of uncertainty. 
In panel $c)$ the demodulation is performed using the orbital parameters $A$, $W$, and $e$ that deviate from the correct ones just by half of the maximum steps defined in Table \ref{LSItab}, or equivalently by their maximum errors ($dp_{\rm max} = {\rm step}_{\rm max}/2$). For simplicity we kept $T_0$ at its true value, while the uncertainty on $P_{\rm orb}$ is negligible in this case, as showed in Table \ref{LSItab}. We enlarged the error having the maximum errors ($dp_{\rm max}$) as units, and show the results we obtain in the case they deviate
2 (in panel d) and 3 times (in panel e) the maximum errors.
We can notice that the signal is still significant ($\sim 6\sigma$) in panel $d)$, even if the total factor is $\varepsilon^2 \sim 0.1$. This depends on the fact that the signal to noise ratio in the power spectrum is proportional to the total number of counts observed. Specifically, $P(\omega_0)/P(\omega \neq \omega_0) = N_0 \varepsilon^2/2$. The two years-long simulated observation makes possible a detection at low $\varepsilon^2$, even if the data rate and the total pulsed fraction are relatively small.
The plots in Figure \ref{Pplots} (panels $c)$, $d)$, and $e)$) show that detection is possible only if the deviations of the orbital parameters are within the bell shape in Figure \ref{fig2} and Figure \ref{EccSurfNorm}, depending on the overall pulse fraction (factor $\eta$ in Eq. \ref{7.2}). In contrast, outside that bell, the pulsation detection is washed out (as in panel $b$ of Figure \ref{Pplots}).

\subsection{Comparing the maximum steps with current uncertainties}

The comparison of the maximum steps in Tables \ref{LS5039tab} and \ref{LSItab} with the current uncertainties of the orbital parameters points out that a fine sampling is needed in order to detect pulsations with long observations. In the case of a magnetar-like period, we can still hope that an improvement of the uncertainties with further optical observations can allow a detection, but for young or millisecond pulsars, 
sampling is unavoidable. 
For instance, for pulsars with a period of $300$ ms orbiting LS I +61 303, the uncertainty on the semi-major axis has to be smaller than $0.03$ lt-s, that is $\sim 40$ times smaller than the current one. The larger the pulsar period the closer current optical observations are with respect to the maximum steps. 
The situation is slightly more optimistic for the consideration of a young pulsar in LS 5039, which has a semi-major axis of $3.33 \pm 0.15$ lt-s. We calculated that in order to have $\varepsilon^2>0.36$ for a pulsar with a period of $300$ ms the uncertainties should be improved by a factor of 10 
for $T_0$ and $W$, a factor of 5 for $A$, and a factor of 2 for $e$.

The study presented here is conducted assuming a pure sinusoidal pulse shape for the pulsar. In contrast, the gamma-ray profiles show commonly two narrow peaks, or a single broad one \citep{FermiPC}. Therefore, the total coherent power is shared in several harmonics.
In the periodicity search, the significance of the signal can be improved  summing the power of different harmonics. 
The extension of the results in this work to higher harmonic is immediate, except for the fact that in the sampling of the orbital parameters one should optimize the maximum steps with respect to the highest harmonic to sum.

In conclusion, the analytical study presented in this paper 
clarifies a number of issues regarding the possible detection of pulsations from binary systems. It is likely the basis for the
development of a software to perform a blind search for pulsations in binary systems, applying an optimized sampling of the orbital parameters searching in different frequency ranges.

\section*{Acknowledgements}

This work was supported by the  grants AYA2009-07391 and SGR2009-811, as well as the Formosa program
TW2010005 and iLINK program 2011-0303. We acknowledge an anonymous referee for suggestions.

\appendix
\section{} \label{App0}
We here briefly 
introduce a method to evaluate the expectation value of the power spectrum at the signal and nearby frequencies, based on a statistical approach.

Following the definition by \cite{Scargle1982}, the power spectrum of a 
sample data set $\{X(t_i), i=1,2,....,N_0\}$ calculated at a frequency $\omega$ is given by
\begin{equation}
\! \! 
P(\omega) \!=\! \frac{1}{N_0} \!
\left[ \left( \sum_{i=1}^{N_0} X_i {\rm cos}(\omega t_i)\right) ^2
\! \!  + \!
\left( \sum_{i=1}^{N_0} X_i {\rm sin}(\omega t_i)\right) ^2\right] . \!
\label{A0.1}
\end{equation} 
Here, we consider the series of the arrival times of single events on a detector, i.e. the sample data set $X$ is such that $X_i=1$ for each $i$.
Attributing a phase value to each event $\theta_i = \omega t_i$ we have for the power spectrum
\begin{equation}
P(\omega) = \frac{1}{N_0}\left[ \left( \sum_{i=1}^{N_0} {\rm cos}\theta_i\right) ^2+\left( \sum_{i=1}^{N_0} {\rm sin}\theta_i\right) ^2\right].
\label{pow}
\end{equation} 
For a large number of events ($N_0 \gtrsim 100$) in Eq. (\ref{pow}), 
the sums of the trigonometric functions of the phases, 
cos$(\theta_i)$ and sin$(\theta_i)$, are well approximated by their mean values times $N_0$. Thus,
\begin{equation}
\sum_{i=1}^{N_0} {\rm cos}\theta_i \longrightarrow N_0  \left\langle {\rm cos}\theta_i \right\rangle 
= N_0\int_{-1}^1 {\rm cos}\theta \cdot P_{\rm cos}(\theta) d{\rm cos}\theta,
\label{A0.3}
\end{equation}
where $P_{\rm cos}(\theta)$ is the distribution of the values of ${\rm cos}\theta$ expressed as function of $\theta$, and 
\begin{equation}
\sum_{i=1}^{N_0} {\rm sin} (\theta_i) \longrightarrow N_0  \left\langle {\rm sin}\theta_i \right\rangle  = N_0\int_{-1}^1 {\rm sin}\theta \cdot P_{\rm sin} (\theta) d{\rm sin}\theta  ,
\label{A0.4}
\end{equation}
where $P_{\rm sin}(\theta)$ is the distribution of the values of ${\rm sin}\theta$.
The expectation value of the power spectrum is obtained substituting these in Eq. (\ref{pow})
\begin{equation}
P(\omega) = N_0\left[ \left\langle {\rm cos}\theta_i \right\rangle^2 + \left\langle {\rm sin}\theta_i \right\rangle^2\right] .
\label{pow2}
\end{equation} 
The two distributions $P_{\rm cos}(\theta)$, and $P_{\rm sin}(\theta)$ can be expressed as function of the probability density distribution (pdf) of the phases attributed to each time stamp of the time series $P_{\theta}(\theta)$. 
We use that the pdf of a variable $z=f(x)$ that is function of a random variable $x$ whose pdf $P_x(x)$ is known can be calculated as (see e.g. \citealt{Rotondi})
\begin{equation}
P_{z}(z) = \sum_i \frac{P_x(x_{2\_i})}{f^{\prime}(x_{2\_i})} - \frac{P_x(x_{1\_i})}{f^{\prime}(x_{1\_i})} .
\label{rotondi}
\end{equation}
Here, the prime (as in $f^{\prime}$) represents a derivative with respect to $x$, and the intervals $\left[ x_{1\_i}, x_{2\_i}\right] $ are those for which for a given $z=z_0$, $f(x)<z_0$.
In our case, $P_z$ in Eq. (\ref{rotondi}) corresponds to $P_{\rm cos}$ or $P_{\rm sin}$, while $P_x$ is the phase distribution $P_{\theta}$.
A detailed calculation of $P_{\rm cos}$ and $P_{\rm sin}$ leads to the result 
that both $P_{\rm cos}$ and $P_{\rm sin}$ can be expressed as a linear function of the sums
\beq
\sum_{i=0}^{N-1} P_{\theta}(2\pi(i+k) \pm \theta),
\label{A0.5} 
\enq
where $k = 0, 1/2, 1$, 
and $N$ is the number of rotations made by the plausible neutron star during the whole observation $T_{\rm obs}$.
Since in Eq.~(\ref{A0.5}), $0 \leq \theta<2\pi$, it corresponds  to the distribution of the phases folded by $2 \pi$.
Substituting the solutions of $P_{\rm cos}$ and $P_{\rm sin}$ in Eqs. (\ref{A0.3}) and (\ref{A0.4}) respectively, and these in Eq. (\ref{pow2}), we obtain the direct dependence of the power spectrum by the pdf of the phases $P_{\theta}(\theta)$.

\section{} \label{App1}

In Section 3, we approximated the eccentricity with $E \sim \Omega_{\rm orb} t_e$ within the perturbation functions 
in Eq.~(\ref{Ete}).  
In this Appendix  we shall evaluate the level of accuracy of this approximation. 

\begin{figure}
\begin{minipage}[b]{0.5\textwidth}
\centering
\includegraphics[width=1.0\textwidth]{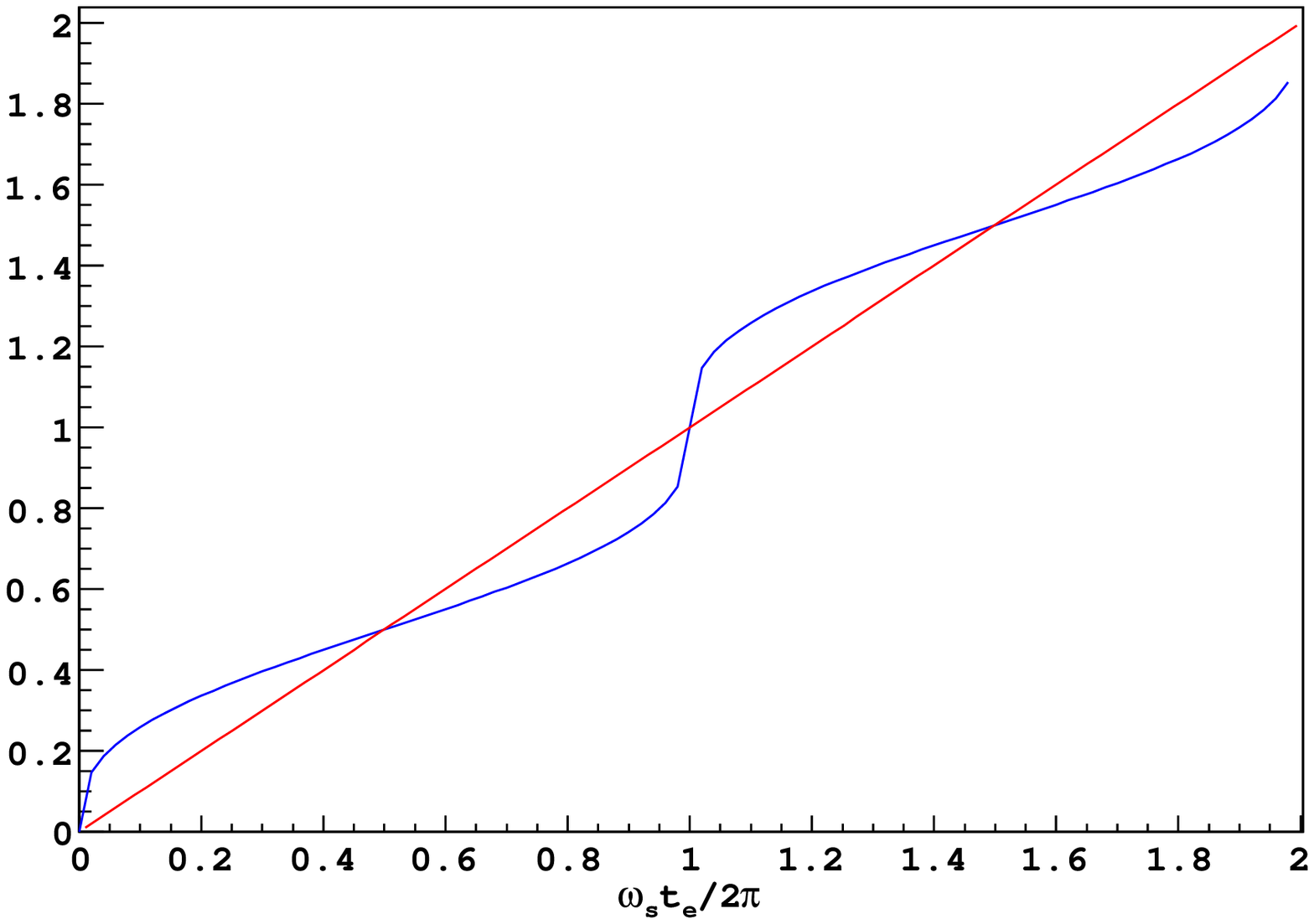}
\end{minipage}
\begin{minipage}[b]{0.5\textwidth}
\centering
\includegraphics[width=1.0\textwidth]{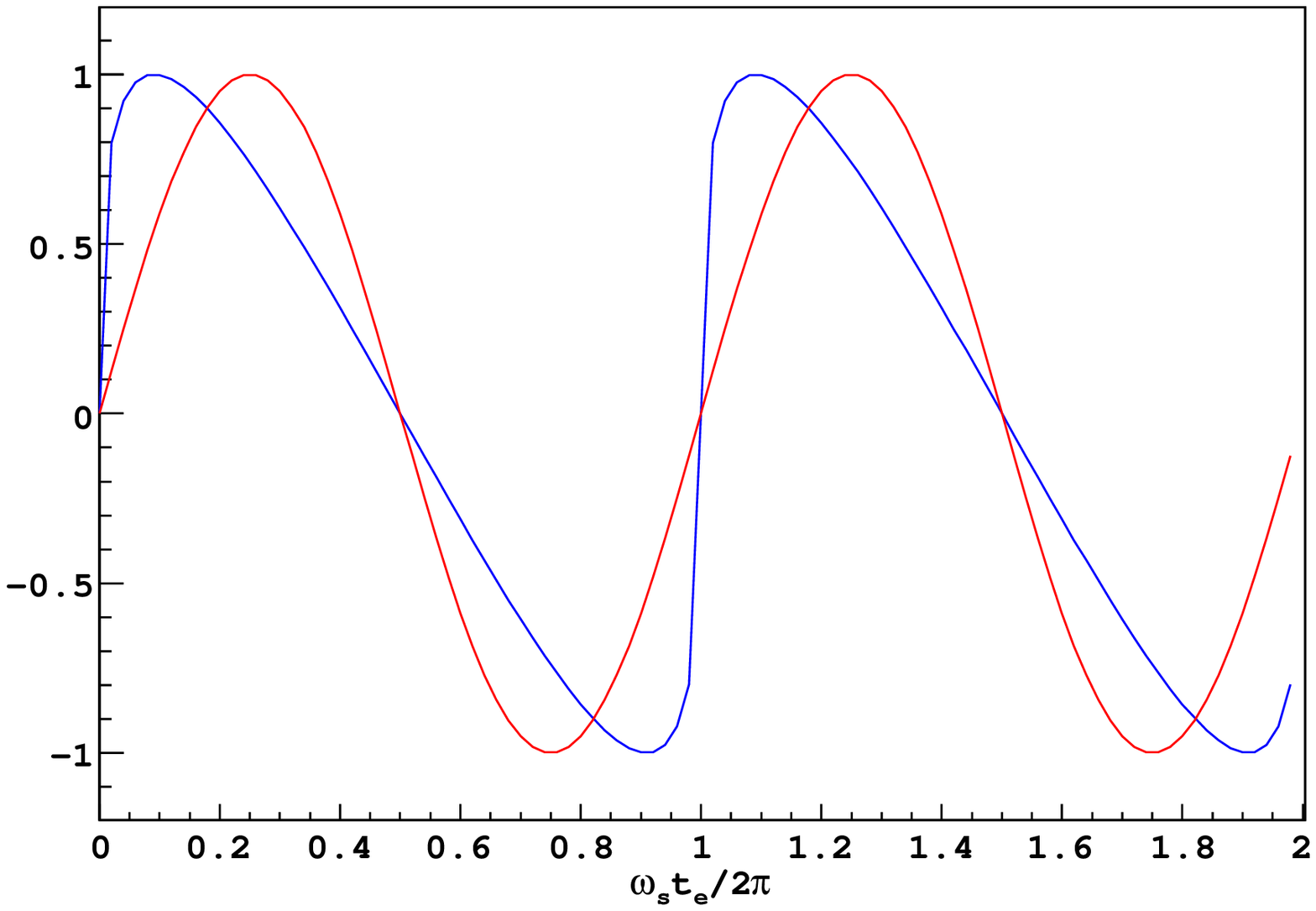}
\end{minipage}
\caption{\textit{Top}: Numerically computed eccentric anomaly $E/2\pi$ (in blue), compared with its approximation $\widetilde{E}/2\pi = \Omega_{\rm orb} t_e/2\pi$ (in red). \textit{Bottom}: Numerically computed sine of the eccentric anomaly (in blue), compared with the sine of the approximation adopted (sin $\widetilde{E}$, in blue). Magnitudes in both panels are shown as a function of $\Omega_{\rm orb} t_e/2\pi$.}
\label{FA1}
\end{figure}

\begin{figure}
\centering
\includegraphics[width=0.5\textwidth]{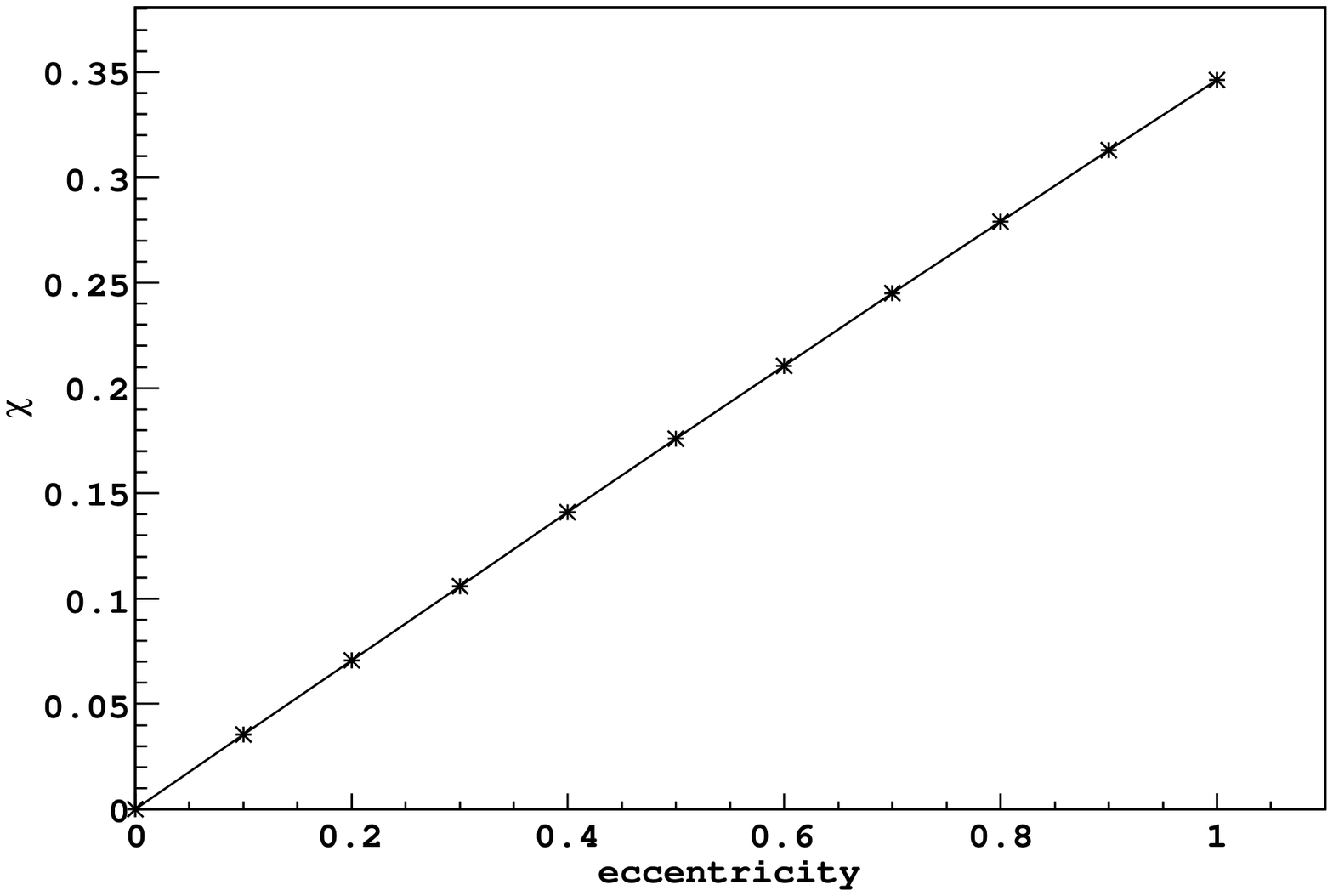}
\caption{Mean squared difference $\chi$ between the sine of the eccentric anomaly 
and the sine of its approximation (Eq. \ref{A1.1}), calculated for several values of the eccentricity.}
\label{FA2}
\end{figure}

The worst possible case for the approximation is when the orbit has an eccentricity close to 1. Figure \ref{FA1} shows how different are the correct values of $E(t_e)$ computed numerically inverting the Eq.~(\ref{Ete}), with respect to its approximation. 
In the same figure we show the differences between the sine of $E(t_e)$ and the sine of its approximation (hereafter $\widetilde{E}$). The linear and the sine dependencies on $E$ are the two instances concerned by the perturbation function. The former appears only in the case the perturbed function is due to the uncertainty on the orbital period.
In the case of the linear dependence, the larger is $\Omega_{\rm orb} t_e$, the smaller is the relative difference between $E(t_e)$ and $\widetilde{E}$. So the approximation improves  for long observations, and the difference becomes negligible by setting the zero of the time stamps $t_e$ far enough from the observation time
In the case of the sine dependence, we estimated the mean squared difference using the following formula
\beq
\chi = \sqrt{\frac{\left[ {\rm sin}\widetilde{E}-{\rm sin}E \right]^2 }{N_s}}
\label{A1.1}
\enq
using a sampling $N_s = 1000$ of $\Omega_{\rm orb} t_e$ in the range 0, $2\pi$.
In the case showed in figure \ref{FA1} ($e=1$) $\chi = 0.346$. Figure \ref{FA2} plots $\chi$ 
as calculated for several values of the eccentricity. All in all, these approximations do not seem, a priori, to introduce a significant deviation. This is later confirmed by simulations. 

\section{} \label{App2}

In this Appendix we find the conditions under which the function $t_{ew} = t_e - f(t_e)$ increases monotonically. We are going to demonstrate that these conditions are satisfied by all the binary systems that are already observed. To increase monotonically,  the first derivative of $t_{ew}$ with respect to $t_e$ should satisfy
\beq
\frac{df}{dt_e} < 1.
\label{A.1}
\enq
In Eq.~(\ref{2.6}) the perturbation function is defined as 
\beq
f(E(t_e)) = \frac{\partial \Delta_R(E(t_e))}{\partial p} dp ,
\label{A.2}
\enq
where $p$ is the orbital parameter to which the perturbation function is referred to. We stress that in Eq.~(\ref{A.2}) both $f$ and $\Delta_R$ are functions of the eccentric anomaly $E$, which in turn is function of $t_e$. Taking this into account, the first derivative of $f$ can be calculated as
\beq
\frac{df(E(t_e))}{dt_e} = \frac{\partial^2 \Delta_R}{\partial t_e \partial p} dp = \frac{\partial}{\partial p} \left( \frac{\partial \Delta_R}{\partial t_e}\right) dp ,
\label{A.3}
\enq
where the exchange of the order of the derivatives in the last passage is allowed because $t_e$ and the orbital parameters $p$ are independent variables.
Since $\Delta_R$ is only indirectly a function of $t_e$, its derivative with respect to it is
\beq
\frac{\partial \Delta_R}{\partial t_e} = \frac{\partial \Delta_R}{\partial E} \frac{dE}{dt_e} .
\label{A.4}
\enq
From Eq.~(\ref{3.2}), the derivative of $\Delta_R$ respect to $E$ is equal to
\beq
\frac{\partial \Delta_R}{\partial E} = M(A,e,W) {\rm cos}(E + \phi(e, W)) ,
\label{A.5}
\enq
where here we explicitly write the dependence of $M$ and $\phi$ (defined in Eq. \ref{3.3}) on the canonical orbital parameters.
To evaluate the derivative of the eccentric anomaly $E$ with respect to $t_e$, we can differentiate Eq.~(\ref{Ete})
\beq
\frac{dE}{dt_e} - e \, {\rm cos}E \frac{dE}{dt_e} = \Omega_{\rm orb} ,
\label{A.6}
\enq
from which we obtain 
\beq
\frac{dE}{dt_e} = \frac{\Omega_{\rm orb}}{1-e \, {\rm cos}E} .
\label{A.7}
\enq
Substituting Eqs.~(\ref{A.4}, \ref{A.5}, \ref{A.7}) in Eq.~(\ref{A.3}), we get
\begin{eqnarray}
\frac{df(E(t_e))}{dt_e} = \hspace{5cm}
 \nonumber \\
\frac{\partial}{\partial p} \left[ \frac{\Omega_{\rm orb}}{1-e \, {\rm cos}E} M(A,e,W) {\rm cos}(E + \phi(e, W)) \right] dp .
\label{A.8}
\end{eqnarray}
We will find useful to indicate the expression within the square brackets in Eq. \ref{A.8} as the function $\frac{\partial \Delta_R}{\partial t_e} = g(p,E)$.
To calculate the derivative in Eq.~(\ref{A.8}), and to evaluate the inequality $df/dt_e < 1$, we need to analyze 
case by case each orbital parameter $p$. 

\subsection{${\bf [p=A]}$}

For the projection of the semi-major axis ($A$), Eq.~(\ref{A.8}) becomes
\begin{eqnarray}
\frac{df(E(t_e))}{dt_e} = \hspace{5cm}
\nonumber \\
 \frac{\partial}{\partial A} \left[ \frac{\Omega_{\rm orb}}{1-e \, {\rm cos}E} M(A,e,W) {\rm cos}(E + \phi(e, W)) \right] dA = \nonumber \\
 \left[ \frac{\Omega_{\rm orb}}{1-e \, {\rm cos}E} \sqrt{1-e^2 {\rm cos}^2W}  \right] {\rm cos}(E + \phi) dA < 1,
\label{A.9}
\end{eqnarray}
where the last inequality is what we need to prove.
We eliminate the dependence of the uncertainty $dA$ by the eccentric anomaly $E$, because the former is the uncertainty 
on a parameter that describe the system, and is fixed in time.
Since ${\rm cos}(E + \phi)$ has as maximum and minimum values 1 and $-1$, respectively, and the term in square bracket is positive, $dA$  is still well constrained by
\beq
\left[ \frac{\Omega_{\rm orb}}{1-e \, {\rm cos}E} \sqrt{1-e^2 {\rm cos}^2W}  \right] |dA| < 1 .
\label{A.10}
\enq
In order to be as conservative as possible with respect the uncertainty $dA$, we can maximize the term in square brackets in Eq.~(\ref{A.10}). With this purpose the denominator ($1-e \, {\rm cos}E$) is minimized setting ${\rm cos}E\,=1$, while the root square at the nominator is maximized setting ${\rm cos}W\,=0$. We so obtain 
\beq
|dA| < \frac{1-e}{\Omega_{\rm orb}}.
\label{A.11}
\enq
We can multiply and divide by $(1+e)$, obtaining
\beq
|dA| < \frac{1-e^2}{\Omega_{\rm orb}(1+e)}.
\label{A.12}
\enq
We now minimize the right hand side, so that the constraints on the uncertainty are the most conservative possible, getting,
\beq
|dA| < \frac{P_{\rm orb} (1-e^2)}{4 \pi} ,
\label{A.13}
\enq
where we considered that $\Omega_{\rm orb} = 2\pi / P_{\rm orb}$. 

It may seems at this point that the inequality would not hold, since for the two body problem with a gravitational potential the eccentricity could be as close as possible to 1. But tidal forces would come to help. 

Tidal forces tend to circularize the orbit of a binary system, affecting the period and the eccentricity at once. They strongly depend by the separation ($r$) of the objects in the system, varying as $1/r^6$ \citep{Lecar1976}. Thus, in systems with short periods, tidal forces are strong. On the other hand, in systems with long periods, they are still not negligible when high eccentricities make the two objects very close at periastron. Systems affected by tidal forces modify in short time their orbit until they reach a balanced status in which either these forces are no longer important, or the orbit become circular. 
In \citet{Halbwachs2008}, the maximum eccentricity that a system can have without being significantly affected by tidal forces is expressed as function of its period by
\beq
e_{\rm Max} = \sqrt{1-(P_{\rm Coff}/P_{\rm orb})^{2/3}} , 
\label{A.14}
\enq
where $P_{\rm Coff}$ is the circularization limit, meaning that systems with periods that are shorter than the latter inevitably fall in a circular orbit. Using the sample of the observed spectroscopic binaries, \citet{Halbwachs2008} (and references therein) estimate $P_{\rm Coff} = 5 \div 10$ days. These values are compatible with the sample of the HMXBs analyzed by \citet{Townsend2011} (compare their figure 6 with Eq. \ref{A.14}). To be conservative, in this work we will consider $P_{\rm Coff} = 5$ days. 

The right term in Eq.~(\ref{A.13}) is further minimized if the maximum eccentricity expressed by Eq.~(\ref{A.14}) is taken into account. So that we have
\begin{eqnarray}
\frac{|dA|}{A} &<& \left( \frac{P_{\rm Coff}}{P_{\rm orb}} \right)^{2/3} \frac{P_{\rm orb}}{4 \pi A}, \hspace{.5cm} {\rm for} \hspace{0.2cm} P_{\rm orb} >P_{\rm Coff} ; 
\label{A.15} \\
\frac{|dA|}{A} &<& \frac{P_{\rm orb}}{4 \pi A}, \hspace{.5cm} {\rm for} \hspace{0.3cm} P_{\rm orb} \le P_{\rm Coff} ,
\label{A.16}
\end{eqnarray}
where we have also divided both the left and right side by $A$, to take into account the relative uncertainty $|dA|/A$. In order to express the right hand side of the inequalities as function only of the orbital period, we can invoke the third Kepler's law 
\beq
A = \left( \frac{G}{4 \pi^2 c^3} \frac{M_s^3}{(M_s+M_{psr})^2} P_{\rm orb}^2 \right)^{1/3} {\rm sin}\,i , 
\label{A.17}
\enq
where $G$ is the gravitational constant, $c$ the speed of light, $M_{psr}$ the mass of the compact object, $M_s$ the mass of the companion star, and ${\rm sin}\,i$ is the inclination of the orbit plane respect to the observer. When the companion star is very massive, the ratio $M_s^3/(M_s+M_{psr})^2 \sim M_s$, while when the masses are similar $M_s^3/(M_s+M_{psr})^2 \sim M_s/4$. Since $A$ is at the denominator of the terms in the right hand side of Eqs.~(\ref{A.15}, \ref{A.16}), the most stringent constraint on $dA$ can be found by setting ${\rm sin}\,i = 1$, and $M_s^3/(M_s+M_{psr})^2 = M_s$ in all the cases.
Finally we have
\begin{eqnarray}
&& \hspace{-.7cm} \frac{|dA|}{A} < 9.54\times10^2 \left[ \frac{M_s}{2M_{\odot}}\right]^{-1/3}  \left[ \frac{P_{\rm Coff}}{5{\rm days}}\right]^{2/3}  \left[ \frac{P_{\rm orb}}{5{\rm days}}\right]^{-1/3} ,  
 \nonumber \\
  && {\rm for} \hspace{0.2cm} P_{\rm orb} >P_{\rm Coff} ;
\label{A.18} \\
&&\frac{|dA|}{A} <  9.54\times10^2 \left[ \frac{M_s}{2M_{\odot}}\right]^{-1/3}  \left[ \frac{P_{\rm orb}}{5{\rm days}}\right]^{1/3} , 
\nonumber \\
 && {\rm for} \hspace{0.3cm} P_{\rm orb} \le P_{\rm Coff} .
\label{A.19}
\end{eqnarray}
From Eqs.~(\ref{A.18}, \ref{A.19}), we can clearly see that for low mass systems with orbital periods of few days, the uncertainty $dA$ is totally unconstrained, since it is only required that the relative uncertainty $|dA|/A \lesssim 10^3 $, which is obviously true for uncertainties that make sense. In general, we can say that a constrain on $dA$ is effectively worrisome when $|dA|/A < 1 $. This happens only for systems with very large masses ($M_s > 50 M_{\odot}$), and unrealistic periods as long as $10^6$ years (from Eq. \ref{A.18}), or as short as few milliseconds (from Eq. \ref{A.19}). 
In other words the monotonically-increasing-function condition for the function $t_{ew} = t_e - f(t_e)$ is always satisfied for all the observed binary systems when the perturbation function is due to the uncertainty on the projection of the semi-major axis.

\subsection{${\bf [p=W]}$}

For the longitude of the periastron ($W$), Eq.~(\ref{A.8}) (with the imposition of the searched inequality) reads
\begin{eqnarray}
\frac{df(E(t_e))}{dt_e} = \hspace{4cm}  \\
\frac{\partial}{\partial W} \left[ \frac{\Omega_{\rm orb}}{1-e \, {\rm cos}E} M(A,e,W) {\rm cos}(E + \phi(e, W)) \right] dW = \nonumber \\
 \frac{\Omega_{\rm orb}}{1-e \, {\rm cos}E} 
 \left[ \frac{\partial M}{\partial W} \cdot {\rm cos}(E+\phi(e,W)) - \right. 
  \nonumber \\
  \left.
  M(A,e,W) \frac{\partial \phi}{\partial W} {\rm sin}(E+\phi(e,W))\right] dW < 1. \nonumber
\label{A.W0}
\end{eqnarray}
Within the square brackets in the last equation,
we can apply the trigonometric identity described in Eq.~(\ref{3.1}) 
to ${\rm cos}(E+\phi)$ and ${\rm sin}(E+\phi)$, obtaining
\begin{eqnarray}
\frac{\Omega_{\rm orb}}{1-e \, {\rm cos}E} \sqrt{\left( \frac{\partial M}{\partial W}\right) ^2 + M^2\left( \frac{\partial \phi}{\partial W} \right)^2 } \times
\nonumber \\
\hspace{2cm} {\rm sin}(E + \phi + \beta) dW < 1,
\label{A.W1}
\end{eqnarray}
where $\beta$ is a function of $A$, $e$, and $W$. Since the factor multiplying ${\rm sin}(E + \phi + \beta)$ is positive, and the sine oscillates between 1 and $-1$, the constraints on $dW$ can be written as
\beq
|dW| \frac{\Omega_{\rm orb}}{1-e \, {\rm cos}E} \sqrt{\left( \frac{\partial M}{\partial W}\right) ^2 + M^2\left( \frac{\partial \phi}{\partial W} \right)^2 } < 1 .
\label{A.W2}
\enq
Maximizing the factors that multiply $|dW|$ we obtain
\begin{eqnarray}
\frac{\Omega_{\rm orb}}{1-e \, {\rm cos}E} < \frac{\Omega_{\rm orb}}{1-e} = \Omega_{\rm orb} \frac{1+e}{1-e^2} < \frac{2\Omega_{\rm orb}}{1-e^2} ,
\nonumber \\
M(A,e,W) = A\sqrt{1-e^2{\rm cos}^2W} < A ,
\label{A.W3} \\
\frac{\partial M}{\partial W} = \frac{Ae^2{\rm sin}W\,{\rm cos}W}{\sqrt{1-e^2{\rm cos}^2W}} < \frac{A}{2\sqrt{1-e^2}} ,\nonumber \\
\frac{\partial \phi}{\partial W} = \frac{\left( 1+ {\rm tan}(W)^2\right) \sqrt{1-e^2}}{1-e^2+{\rm tan}(W)^2} < \frac{1}{\sqrt{1-e^2}} .
\nonumber
\end{eqnarray}
Eq.~(\ref{A.W2}) then becomes
\beq
|dW| < \frac{P_{\rm orb}}{2\pi \sqrt{5}} \frac{(1-e^2)^{(3/2)}}{A} ,
\label{A.W5}
\enq
where we have substituted $\Omega_{\rm orb} = 2\pi/P_{\rm orb}$.
Substituting the eccentricity $e$ by $e_{\rm Max}$ as defined in Eq.~(\ref{A.14}) in Eq. \ref{A.W5} we get
\begin{eqnarray}
|dW| &<&  \frac{P_{\rm Coff}}{2\pi \sqrt{5}}  \frac{1}{A},  \hspace{1cm} {\rm for} \hspace{0.2cm} P_{\rm orb} >P_{\rm Coff} ;
\label{A.W6} \\
|dW| &<&  \frac{P_{\rm orb}}{2\pi \sqrt{5}}  \frac{1}{A} , \hspace{1cm} {\rm for} \hspace{0.3cm} P_{\rm orb} \le P_{\rm Coff}.
\label{A.W7}
\end{eqnarray}
Finally, introducing the third Kepler's law (Eq. \ref{A.17}) with ${\rm sin}\,i = 1$, and $M_s^3/(M_s+M_{psr})^2 = M_s$, we get
\begin{eqnarray}
|dW| <  8.54\times10^2 \left[ \frac{M_s}{2M_{\odot}}\right]^{-1/3}  \left[ \frac{P_{\rm Coff}}{5{\rm days}}\right]   \left[ \frac{P_{\rm orb}}{5{\rm days}}\right]^{-2/3} ,
\nonumber \\
\hspace{.5cm} {\rm for} \hspace{0.1cm} P_{\rm orb} >P_{\rm Coff} ; \hspace{.5cm} 
\label{A.W8} \\
|dW| <  8.54\times10^2 \left[ \frac{M_s}{2M_{\odot}}\right]^{-1/3} \left[ \frac{P_{\rm orb}}{5{\rm days}}\right]^{1/3} ,
\nonumber \\
\hspace{.5cm} {\rm for} \hspace{0.1cm} P_{\rm orb} \le P_{\rm Coff} . \hspace{.5cm} 
\label{A.W9}
\end{eqnarray}
Since $W$ is an angle, the constraints on its uncertainty $dW$ are effective if $dW < 2\pi$ rad is required. A greater uncertainty means that we do not have any knowledge of $W$ whatsoever. From the equations above, we can calculate that $dW$ is effectively constrained only for binary systems with periods longer than 7 years and masses $M_s \gtrsim 20 M_{\odot}$. A more constraining but still very large uncertainty $dW<1$ rad satisfy the monotonic condition of the function $t_{ew}$ for all the binary systems with $M_s < 50 M_{\odot}$, and period shorter than 70 years. Thus, it is safe to consider that the monotonically-increasing-function condition for the function $t_{ew} = t_e - f(t_e)$ is always satisfied for the systems of interest.

\subsection{${\bf [p=e]}$}

Since the eccentric anomaly depends by the eccentricity ($e$), for this parameter Eq.~(\ref{A.8}) (with the imposition of the searched inequality) reads
\begin{eqnarray}
\frac{df(E(t_e))}{dt_e} = \frac{\partial}{\partial e} \left[ g(e,E) \right] de = 
\frac{\partial g}{\partial e} \cdot 1 + \frac{\partial g}{\partial E} \frac{\partial E}{\partial e} de < 1
\label{A.e0}
\end{eqnarray}
where we defined with $g(e,E)$ the expression in the square brackets of Eq.~(\ref{A.8}).
Term by term we have
\begin{eqnarray}
\frac{\partial g}{\partial e} = \frac{\Omega_{\rm orb}}{1-e \, {\rm cos}E}\left\lbrace \left[ \frac{M \, {\rm cos}E}{1-e \, {\rm cos}E} + \frac{\partial M}{\partial e}\right] {\rm cos}(E+\phi) - \right. 
  \nonumber \\
  \left.
 M\frac{\partial \phi}{\partial e} {\rm sin}(E+\phi)\right\rbrace 
\label{A.e1}
\end{eqnarray}
\begin{eqnarray}
\frac{\partial g}{\partial E} = -\frac{\Omega_{\rm orb}}{1-e \, {\rm cos}E}\left\lbrace \left[ \frac{e \, M \, {\rm sin}E}{1-e \, {\rm cos}E} \right] {\rm cos}(E+\phi) - \right. 
  \nonumber \\
  \left.
 M \, {\rm sin}(E+\phi) \right\rbrace 
\label{A.e2}
\end{eqnarray}
\begin{eqnarray}
\frac{\partial E}{\partial e} = \frac{{\rm sin}E}{1-e \, {\rm cos}E}
\label{A.e3}
\end{eqnarray}
Substituting Eqs.~(\ref{A.e1}) (\ref{A.e2}) and (\ref{A.e3}) in Eq.~(\ref{A.e0}), and applying the trigonometric identity described in Eq.~(\ref{3.1}) to ${\rm cos}(E+\phi)$ and ${\rm sin}(E+\phi)$, we obtain
\begin{eqnarray}
\frac{\Omega_{\rm orb}}{1-e \, {\rm cos}E} \left\lbrace \left[ \frac{M \, {\rm cos}E}{1-e \, {\rm cos}E} + \frac{\partial M}{\partial e} - \frac{e \, M \, {\rm sin}^2 E}{(1-e \, {\rm cos}E)^2} \right]^2 + \right. 
  \nonumber \\
  \left.
\left[ M\frac{\partial \phi}{\partial e} + \frac{M \, {\rm sin}E}{1-e \, {\rm cos}E}\right]^2  \right\rbrace^{\frac{1}{2}} {\rm sin}(E+\phi+\beta) de < 1
\label{A.e4}
\end{eqnarray}
where $\beta$ is a function of $A$, $e$, and $W$. The terms in the brace brackets can be maximised using the following inequalities, in addition to those in Eq.~(\ref{A.W3})
\begin{eqnarray}
\left( \frac{\partial M}{\partial e} \right)^2  = \left[ -\frac{Ae \, {\rm cos}^2W}{\sqrt{1-e^2\, {\rm cos}^2W}} \right]^2 <   \left( \frac{Ae}{\sqrt{1-e^2}}\right)^2 , 
\label{A.e5} \\
\frac{\partial \phi}{\partial e} = \frac{e \, {\rm tan}(W)}{( 1-e^2+{\rm tan}(W)^2 ) \sqrt{1-e^2}} < \frac{e}{2(1-e^2)}, 
\nonumber \\
\end{eqnarray}
The constraint on $de$ is then reduced to
\beq
|de| < \frac{P_{\rm orb}}{2 \pi \sqrt{61}} \frac{(1-e^2)^2}{A}, 
\label{A.e6}
\enq
where we have already substituted $\Omega_{\rm orb} = 2\pi/P_{\rm orb}$.
Taking into account Eq.~(\ref{A.14}), and the third Kepler's law (Eq. \ref{A.17}), we finally get
\begin{eqnarray}
|de| <  2.44\times10^2 \left[ \frac{M_s}{2M_{\odot}}\right]^{-1/3}  \left[ \frac{P_{\rm Coff}}{5{\rm days}}\right]^{4/3}   \left[ \frac{P_{\rm orb}}{5{\rm days}}\right]^{-1} , \!\!\!
\nonumber \\
\hspace{1.5cm}  {\rm for} \hspace{0.1cm} P_{\rm orb} >P_{\rm Coff} ;\hspace{1.5cm} 
\label{A.e7} \\
|de| <  2.44\times10^2 \left[ \frac{M_s}{2M_{\odot}}\right]^{-1/3} \left[ \frac{P_{\rm orb}}{5{\rm days}}\right]^{1/3},
\nonumber \\
\hspace{1.5cm} {\rm for} \hspace{0.1cm} P_{\rm orb} \le P_{\rm Coff} .\hspace{1.5cm} 
\label{A.e8}
\end{eqnarray}
Since the eccentricity is in the range $0 \leq e < 1$, its uncertainty $de$ is really constrained if $de < 1$. Taking this into account, we can calculate from the equations above that also in this case the monotonically-increasing-function condition for $t_{ew} = t_e - f(t_e)$ is satisfied for all realistic binary systems.

\subsection{${\bf [p=T_0]}$}

Since the eccentric anomaly depends by the epoch of the periastron ($T_0$), for this parameter Eq.~(\ref{A.8}) (with the imposition of the searched inequality) reads
\begin{eqnarray}
\frac{df(E(t_e))}{dt_e} = \frac{\partial}{\partial T_0} \left[ g(e,E) \right] dT_0 = \nonumber \\
\frac{\partial g}{\partial T_0} \cdot 1 + \frac{\partial g}{\partial E} \frac{\partial E}{\partial T_0} dT_0 < 1
\label{A.T0}
\end{eqnarray}
where we defined with $g(T_0,E)$ the expression in the square brackets of Eq.~(\ref{A.8}).
Term by term we have
\begin{eqnarray}
\frac{\partial g}{\partial T_0}=0
\label{A.T1}
\end{eqnarray}
\begin{eqnarray}
\frac{\partial E}{\partial T_0} = -\frac{\Omega_{\rm orb}}{1-e \, {\rm cos}E}
\label{A.T2}
\end{eqnarray}
while $\frac{\partial g}{\partial E}$ is given in Eq.~(\ref{A.e2}).
Substituting in Eq.~(\ref{A.T0}), and applying the trigonometric identity described in Eq.~(\ref{3.1}) to ${\rm cos}(E+\phi)$ and ${\rm sin}(E+\phi)$, we obtain
\begin{eqnarray}
\frac{\Omega_{\rm orb}^2}{(1-e \, {\rm cos}E)^2} \left\lbrace \left[ \frac{e \, M \, {\rm sin}E}{1-e \, {\rm cos}E} \right]^2 + M^2  \right\rbrace^{\frac{1}{2}} \times 
\nonumber \\
{\rm sin}(E+\phi+\beta) dT_0 < 1
\label{A.T4}
\end{eqnarray}
where $\beta$ is a function of $A$, $e$, and $W$. Maximising the terms in the brace brackets using the inequalities in Eq.~(\ref{A.W3}) the constraint on $de$ is then reduced to
\beq
\frac{|dT_0|}{P_{\rm orb}} < \frac{P_{\rm orb}}{16 \pi^2 \sqrt{5}} \frac{(1-e^2)^3}{A}, 
\label{A.T5}
\enq
where we have already substituted $\Omega_{\rm orb} = 2\pi/P_{\rm orb}$.
Taking into account Eq.~(\ref{A.14}), and the third Kepler's law (Eq. \ref{A.17}), we finally get
\begin{eqnarray}
|dT_0| <  3.4\times10^1 \left[ \frac{M_s}{2M_{\odot}}\right]^{-1/3}  \left[ \frac{P_{\rm Coff}}{5{\rm days}}\right]^{2}   \left[ \frac{P_{\rm orb}}{5{\rm days}}\right]^{-5/3} , \!\!\!
\nonumber \\
\hspace{1.5cm}  {\rm for} \hspace{0.1cm} P_{\rm orb} >P_{\rm Coff} ;\hspace{1.5cm} 
\label{A.T6} \\
|dT_0| <  3.4\times10^1 \left[ \frac{M_s}{2M_{\odot}}\right]^{-1/3} \left[ \frac{P_{\rm orb}}{5{\rm days}}\right]^{1/3},
\nonumber \\
\hspace{1.5cm} {\rm for} \hspace{0.1cm} P_{\rm orb} \le P_{\rm Coff} .\hspace{1.5cm} 
\label{A.T7}
\end{eqnarray}
A constrain on $dT_0$ is effective when  $|dT_0|/P_{\rm orb} < 1$. This happens only for systems with very large masses, and unrealistic very long periods. In conclusion, the monotonically-increasing-function 
condition $t_{ew} = t_e - f(t_e)$ is  satisfied.

\subsection{${\bf [p=P_{\rm orb}]}$}

Since the eccentric anomaly depends by the orbital period ($P_{\rm orb}$), for this parameter Eq.~(\ref{A.8}) (with the imposition of the searched inequality) reads
\begin{eqnarray}
\frac{df(E(t_e))}{dt_e} = \frac{\partial}{\partial P_{\rm orb}} \left[ g(P_{\rm orb},E) \right] dP_{\rm orb} = 
\nonumber \\
\left[ \frac{\partial g}{\partial \Omega_{\rm orb}} \cdot 1 + \frac{\partial g}{\partial E} \frac{\partial E}{\partial \Omega_{\rm orb}} \right] \frac{d \Omega_{\rm orb}}{dP_{\rm orb}} dP_{\rm orb}< 1
\label{A.P0}
\end{eqnarray}
where we defined with $g(\Omega_{\rm orb},E)$ the expression in the square brackets of Eq.~(\ref{A.8}), and we take into account that $\Omega_{\rm orb} = 2 \pi / P_{\rm orb}$
Term by term we have
\begin{eqnarray}
\frac{d \Omega_{\rm orb}}{dP_{\rm orb}} = -\frac{2\pi}{P_{\rm orb}^2}
\label{A.P1}
\end{eqnarray}
\begin{eqnarray}
\frac{\partial g}{\partial \Omega_{\rm orb}} = \frac{M \, {\rm cos}(E+\phi)}{1-e \, {\rm cos}E}
\label{A.P2}
\end{eqnarray}
\begin{eqnarray}
\frac{\partial E}{\partial \Omega_{\rm orb}} = \frac{t_e-T_0}{1-e \, {\rm cos}E}
\label{A.P3}
\end{eqnarray}
while $\frac{\partial g}{\partial E}$ is given in Eq.~(\ref{A.e2}).
Substituting in Eq.~(\ref{A.P0}), and applying the trigonometric identity described in Eq.~(\ref{3.1}) to ${\rm cos}(E+\phi)$ and ${\rm sin}(E+\phi)$, we obtain
\begin{eqnarray}
-\frac{2\pi M}{(1-e \, {\rm cos}E)P_{\rm orb}} \left\lbrace \left[ 1 - \frac{\Omega_{\rm orb} \, e \, {\rm sin}E \, (t_e-T_0) }{(1-e\, {\rm cos}E)^2}\right]^2 + \right.
\nonumber \\
\left.
\left[ \frac{\Omega_{\rm orb} (t_e - T_0)}{1-e \, {\rm cos}E} \right]^2  \right\rbrace^{\frac{1}{2}} {\rm sin}(E+\phi+\beta) < 1 
\label{A.P4}
\end{eqnarray}
where $\beta$ is a function of $A$, $e$, and $W$. Maximising the terms in the brace brackets using the inequalities in Eq.~(\ref{A.W3}) the constraint on $dP_{\rm orb}$ is then reduced to
\beq
\frac{|dP_{\rm orb}|}{P_{\rm orb}} < \frac{P_{\rm orb}}{16 \pi \sqrt{1+4\pi^2 (t_e-T_0)^2/P_{\rm orb}^2}} \frac{(1-e^2)^3}{A}, 
\label{A.P5}
\enq
where we have already substituted $\Omega_{\rm orb} = 2\pi/P_{\rm orb}$.
If $T_0$ is set in order to be as close as possible to the beginning of the observation or better within it, 
then  $(t_e-T_0) \leq T_{\rm obs}$. The inequality in Eq.~(\ref{A.P5}) is still valid if we substitute 
$(t_e-T_0)$ with  $T_{\rm obs} = n P_{\rm orb}$.
Taking into account Eq.~(\ref{A.14}), and the third Kepler's law (Eq. \ref{A.17}), we finally get
\begin{eqnarray}
\frac{|dP_{\rm orb}|}{P_{\rm orb}} <  \frac{2.4 \times 10^2}{\sqrt{1+4\pi^2 n^2}} \left[ \frac{M_s}{2M_{\odot}}\right]^{-1/3}  \left[ \frac{P_{\rm Coff}}{5{\rm days}}\right]^{2}   \left[ \frac{P_{\rm orb}}{5{\rm days}}\right]^{-5/3} , \!\!\!
\nonumber \\
\hspace{1.5cm}  {\rm for} \hspace{0.1cm} P_{\rm orb} >P_{\rm Coff} ;\hspace{1.5cm} 
\label{A.P6} \\
\frac{|dP_{\rm orb}|}{P_{\rm orb}} <  \frac{2.4 \times 10^2}{\sqrt{1+4\pi^2 n^2}} \left[ \frac{M_s}{2M_{\odot}}\right]^{-1/3} \left[ \frac{P_{\rm orb}}{5{\rm days}}\right]^{1/3},
\nonumber \\
\hspace{1.5cm} {\rm for} \hspace{0.1cm} P_{\rm orb} \le P_{\rm Coff} .\hspace{1.5cm} 
\label{A.P7}
\end{eqnarray}
Commonly, the uncertainty on $P_{\rm orb}$ is such that $|dP_{\rm orb}|/P_{\rm orb} < 10^{-3}$. This level of constrain is reached in Eq.~(\ref{A.P6}) and (\ref{A.P7}) only for $n>10^4$. On the other hand, in the analysis proposed use time windows $T_w = P_{\rm orb}$ for long period systems ($n=1$), or for orbital periods of few hours a time window of a week implies $n$ to be of the order of 100.
A part for $n$, is not satisfied only for systems with very large masses, and unrealistic very long periods. In conclusion, the monotonically-increasing-function condition $t_{ew} = t_e - f(t_e)$ is  satisfied.

\subsection{The most general case}

So far we have considered the case in which only a single parameter is estimated with a given uncertainty, while all the others are known,
 so that their contribution is null to the non-factorised perturbation function, as defined by Eq.~(\ref{2.5})
\begin{equation}
f = \sum_p \frac{\partial \Delta_R}{\partial p} dp .
\label{B.g1}
\end{equation}
In practice, however,  all  orbital parameters have an uncertainty, and the condition over the function $t_{we}$ must be satisfied when the full derivative of Eq.~(\ref{B.g1}) with respect to $t_e$ (that is equal to the sum of the terms in Eq. \ref{A.3} relative to each one of the parameters) is lower than 1, i.e., 
\begin{equation}
\frac{df(E(t_e))}{dt_e} = \sum_p \frac{\partial}{\partial p} \left( \frac{\partial \Delta_R}{\partial t_e}\right) dp < 1 .
\label{B.g2}
\end{equation}
This inequality is satisfied if we divide by 5 each of the formerly derived constraints, i.e.,  Eqs.~(\ref{A.18}, \ref{A.19}, \ref{A.W8}, \ref{A.W9}, \ref{A.e7}, \ref{A.e8},  \ref{A.T6}, \ref{A.T7},  \ref{A.P6}, \ref{A.P7}). This introduce no significant worry for any of the parameters, since the 
conditions were easily satisfied in all cases of realistic binary systems.

\label{lastpage}

\end{document}